\documentclass[a4paper]{CMNIEN}

\renewcommand\refname{REFERENCES}
\usepackage{graphicx}
  \usepackage{amsmath}
    \usepackage{amssymb}
    \usepackage{bm}
    \usepackage{amstext}
    \usepackage{amsthm}

\title{MULTIBODY MINIMUM-ENERGY TRAJECTORY WITH APPLICATIONS TO PROTEIN FOLDING}

\author{Carlos Leandro$^{\bf {1*}}$ and Jorge Ambr\'osio$^{\bf 2}$}

\heading{Carlos Leandro, Jorge Ambr\'osio}

\address{1: Departamento de Matem\'{a}tica\\
Instituto Superior de Engenharia de Lisboa\\
Rua Conselheiro Emidio Navarro 1, 1959-007 Lisboa,\\
e-mail: miguel.melro.leandro@gmail.com\\
\
\\
2: LAETA, IDMEC, Instituto Superior T\'ecnico, University of Lisbon\\
Av. Rovisco Pais, P-1049-001 Lisboa,\\
e-mail: {jorge@dem.ist.utl.pt}
}

\keywords{Multibody dynamics, optimal control, protein dynamics, statistical potentials}

\abstract{This work addresses the optimal control of multibody systems being actuated with control forces in order to find a dynamically feasible minimum-energy trajectory of the system. The optimal control problem and its constraints are integrated in a discrete version of the equation of motion allowing the minimization of system energy with respect to a discrete state and control trajectory. The work is centred on a specific type of open-chain multibody system, with strong local propensity, where the overall system kinematics is described essentially by the torsion around the links that connect rigid bodies. The coupling between the rigid body motion, and the optimal conformation is described as an elastic band of replicas of the original system with different conformations. The band forces are used to control system's motion directly, reflecting the influence of the system energy field on its conformation, using for that the Nudged-Elastic Band method.  Here the equation of motion of the multibody grid are solved by using the augmented Lagrangean method. In this context, if a feasible minimum-energy trajectory of the original system exists it is a stationary state of the extended system.
This approach is applied to the folding of a single chain protein. Protein chains can be seen as multibody systems described as open-loop chains in which the rigid bodies represent clusters of atoms, that are linked together by bounds modelled as kinematics joints. Since peptide planes remain relatively rigid during protein motion and its overall dynamics is described by the local propensities of their backbone dihedral angles. In this contribution the presented method to find a transition state pathway between two conformations is tested. The use of the method in two types of protein coerce-grained models are discussed, i.e., when the  system components  linked together by revolute joints, for the second case we allowed some flexibility on the bodies structure. In both study cases a statistical potential field defined using a database with protein native conformations is used.}

\begin{document}

\section{INTRODUCTION}

\noindent  Multibody conformation and deformation are essential on the simulation of materials, chemical systems, and biological matter. A multibody system may take different conformations defined by its degrees of freedom, and some conformations may be more favourable than others depending on its energy. However in many cases this is a multiscale process occurring on several length and time scales. For example biomolecular systems such as proteins in aqueous solution exhibit nuclear motion on a multitude of time scales, ranging from localized bond vibrations within tens of femtoseconds dynamics to nanoseconds and local conformational transitions and large conformational arrangements occurring on a microsecond time scale or even larger. Multiscale dynamics is often illustrated by a ``hierarchical'' free energy landscape, characterized by its minima, which represent the metastable states of the system, and its barriers, which connect these states \cite{Buchenberg 2015}. Its dynamics drives the system to a stable conformations, and conformation changes can be seen as rare events, having its dynamic characterized by long waits periods around stable states followed by sudden jumps from one state to another produced by external phenomena like impact or thermal fluctuations. In this sense a conformation fluctuations have high probability of be found near a stable state (metastable sets), and its dynamic is characterized by rare conformation jumps between metastable sets. This changes are assumed to occur on a time scale that is much larger than the microtime scale in the system dynamics. The reason is that these processes require an unusually large force fluctuations to drive the system over energy barriers separating the conformations. And because of the wide separation of time scales, it is impossible to study this processes by conventional dynamics simulations. When the goal is to analyse a trajectory between two fixed conformations, one must introduce an appropriate bias on the dynamics which greatly enhances the probability of observing the portion of the trajectories during which they perform a transition from one designated metastable set to another. 

How to integrated this bias in a classic multibody formulation is the focus of this work. Our approach is centred on a specific type of multibody system where the overall system dynamic is described essentially by bodies torsion, with strong local conformation propensity and multiple metastable conformations. Typically, these are open-chain systems defined using relatively rigid components linked together using joints. The torsion motion is predominantly local in nature, but the predominant factor on its potential energy. Typical examples of this type of mechanism can be found in nature as nano systems like peptide chain of a protein.  

Peptide chain are biomolecules defined by chains of peptides with its local conformation propensity described by the chain dihedral angles. The systems local propensity, in the context of a  multibody system is presented bellow, as an optimal joint conformation, or an energy constrain in the system energy surface, and is assumed to be the main factor to explain the system dynamics as a transition paths between metastable conformations. To define these paths it is reasonable to introduce the bias that each of the best transition paths be the center of a \textit{channel reaction} through which the system conformation is likely to pass, made of highly probable, low entropic or low energetic local conformations.

Probably the most widely known and used bias for study the dynamic of rare events is the \textit{minimum energy path} (MEP) \cite{Weinan 2005}. MEPs are paths in configuration space that connects the metastable states along which the potential force is parallel to the tangent vector. The MEP can be justified as being the most likely path by which a reaction occurs under specific assumption, depending only on local characteristics of the energy landscape.  It is however quite clear that they often fail for well characterized applications. The reaction channels, of nano systems like peptide chains, is strongly influenced by more global features of the underlying landscape, usually associated with the long range electrostatic potentials and solvent iterations.

In this contribution kinematic analysis of multibody systems is applied, to find a \textit{Minimum Energy Path} (MEP), between two stable conformations,  describing the way through dynamical bottlenecks such as energetic or entropic barriers which separate metastable sets in system configuration space. This is done by reducing a time dependent dynamic problem to a steady problem, having by solution the relaxation of an elastic surface, with drive the system conformation change.  For that conformations are assumed to depend only on the local structure of system energy landscape, defined by the system relative degrees of freedom. For the must interesting cases of application, the energy landscape are usually highly irregular. To overcame this shape irregularities and possibly energetic barriers here the path bias is constructed as an elastic band with extremities fixed in two stable conformations, having its shape modelled progressively by minimizing the work done by the system to full fill the trajectory. Different computational methods can be found in the literature to calculate this minimum energy paths. For this work we used a version of the classic \textit{Nudged-Elastic Band method} (NEB).

The \textit{configurational state} of an open-chain mutibody system is defined by a set of collective variables, charactering its joint angles, in the system phase space. Different model parametrizations correspond to different system conformations, with a specific potential energy, and in this approach the energy only depend of them. A conformation path can be well described by a continuous assignment of values to those collective variables.  In the NEB method a discritization of this path is defined as a set of replicas of the system coupled together by sets of springs, driving the joint angles amplitude. The algorithm iterates upon the following two-step procedure \cite{Jónsson 1998}: (i) compute the gradient of the energy along a discretized curve (or string) while keeping the string fixed, then (ii) utilize this data to update the position of the string by solving the motion equation, i.e. move every point along the string in the direction of gradient component perpendicular to the string, and reparametrize the string nodes using the component of the springs forces parallel to the curve. The spring force helps the reparametrization by spreading the points along the path. In the framework of the multibody dynamics this corresponds to evaluate the dynamic, using the \textit{Direct Integration Method} (DIM) \cite{Flores 2004}, on an extension to the original mechanical model defined by a set of its replicas coupled by an elastic bands. However the spring forces are modified so that when converged, the system configurations  are all aligned along the minimum-energy path between conformations.  Convergence is based on how far the conformation changes after each full iteration. Here the original system timescale is lost, and the system evolution is parametrized by the path discretization. The use of the \textit{Nudged-Elastic Band} for this problem  is not new, however here we describe our attempt to integrate it in the framework of multibody systems kinematic analysis \cite{Jónsson 1998}.

In the typical applications of the method of \textit{elastic multibody systems} (EMBS) the modelling of the elastic bodies use finite elements methods introducing into the model a large number of elastic degrees of freedom. One essential step for an efficient simulation of generic EMBS is the reduction of the linear elastic, procedures for this are well documented and  extensively discussed, see \cite{Ambrosio 1996}. Here however, we are interested in models where deformations are descried by few elastic degrees of freedom, only allowing linear deformation along the joint axis and to control the separation between system components. Therefore our simulation were performed using the DIM, selecting a formulation of the models known to be computationally accurate, when convergent. It is a Cartesian coordinate based formulations proven to ensure the stability and accuracy in many mechanical simulations, the work presented here is developed in this framework, with the help of floating frames \cite{Nikravesh 1988}. The choice  of the coordinate has a direct influence in the structure of the equations of motion that describe the extended multibody kinematic and on the energy landscape description.

This paper is organized as follows: In Section \ref{II} we describe our case of study and discuss the methodology used to create a coarse-grain models for peptide chains. In Section \ref{III} we review the formulation for mulibody dynamic with kinematic constrains defined by cylindrical and revolute joints. Defining what we mean by conformation propensity in a kinematic joint. In Section \ref{IV}  we describe the minimum-energy path as an equilibrium state for an elastic band constrained system and how to find a pair stable conformations. In Section \ref{V} this equilibrium state if interpreted as a steady version for the problem of finding an optimal trajectory using the motion equation. We begin Section \ref{VI} by describing the application of the presented method for the study of polypeptide chains dynamics and we present the results of numerical simulations giving emphasis to the method limitations. Conclusions and final discussion are given in Section \ref{VIII}.

\section{Application: Protein folding}\label{II}
\noindent Most of the macromolecules operating inside the living organisms are proteins. All of them consist of long polymers composed by twenty different types of monomer units called amino acids, differing each other by the three-dimensional conformation in which they fold after the biosynthesis and in which they are biologically active. During the protein biosynthesis the amino acids are assembled together in a specific sequence according to the rule of the genetic code. The final product of this process, the polypeptide chain, undergoes a folding process reaching the protein proper stable state.
Its final conformation varies and hence the functions the protein is able to perform, according to the sequence of amino acids proper of each protein.

The protein folding represents a process of fundamental importance. The remarkable rapidity and reproducibility of this process still present many elements which are not yet completely understood. It is commonly recognised that the folding of the protein chain is a thermodynamically driven process, and that the biologically active protein conformation represents the energetic balance of various kinds of interactions between protein groups, and between these groups and the surrounding medium \cite{McCammon 1984}. If on the one hand this justifies a physical approach on the other one the high complexity of the system seems to prevent any possibility of solution to the problem, from a microscopical point of view. Despite this open issues, many facts about the folding of small single-domain proteins are well established. The physical properties of most of them in the folded stable conformation do not change, or change very little, when the environment is altered by changes in temperature, pH, or pressure, until a threshold is overcome \cite{Buchenberg 2015}. When this point is reached the protein denaturation occurs, a sudden complete unfolding of the protein. The unfolding transition is a two-state cooperative phenomenon, with only the native fully folded and the denatured fully unfolded states present \cite{Buchenberg 2015}. Partially unfolded structures are unstable relative to both states. In order to understand the connection between amino acid sequence and folding kinetics, several studies involved concepts borrowed by the statistical mechanics of disordered systems. The main idea is that, the high complexity of such systems can be conveniently described as a stochastic process. A typical  feature is the presence of conflicting forces and geometric constraints, the system can not satisfy all the impositions at the same time. As a consequence of this fact an unique optimal state does not exist, and the system behaviour is governed by the rugged energy landscape, with huge number of minima separated by high barriers. This makes the folding trajectory of hard simulation using conventional multibody dynamics. Since the dynamic usually proceeds by long waits periods around optimal state followed by sudden jumps from one state to another. 

\subsection{Protein structure}
\begin{figure}[h]
\begin{tabular}{cc}
\includegraphics[width=100pt]{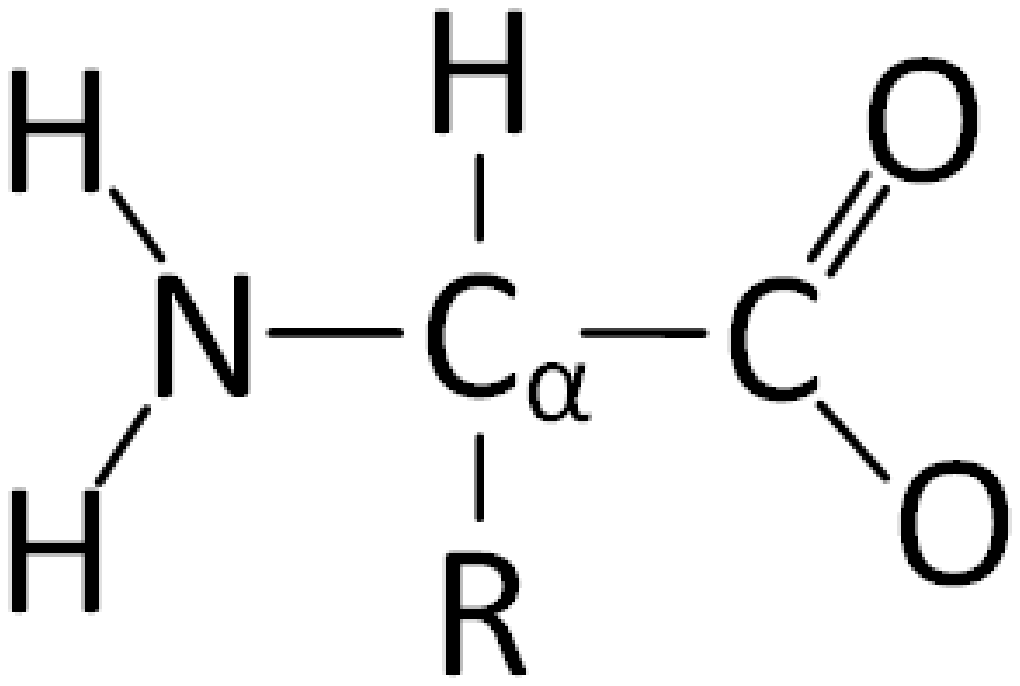}&
\includegraphics[width=300pt]{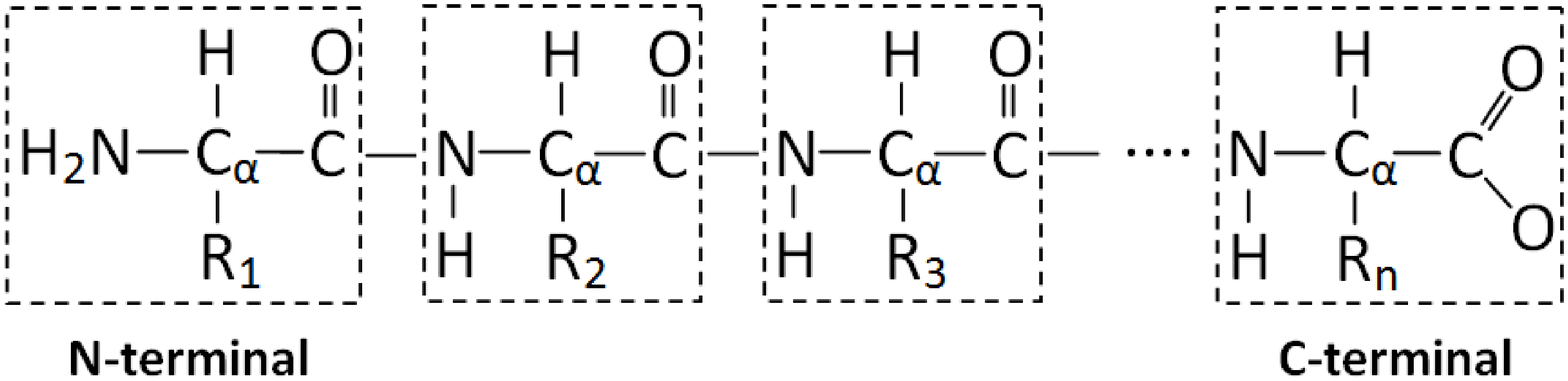}\\
\end{tabular}
\caption{(a) Amino acids are molecules containing an amine group ($H_2N-$), a carboxylic acid group ($-COOH$), and a side chain ($R$) that is pecific to each amino acide. The first carbon that attaches to a functional group is named apha-carbon ($C_\alpha$).(b) Every peptide has a $N$-terminus residue and a $C$-terminus residue on the ends of the peptide (Source Wikipedia).}\label{fig1}
\end{figure}

\noindent The method described in this work will be evaluated for the kinematic analysis of polypeptide chains. Biological proteins are polymeric chains build from amino acid monomers. These amino acids contain five chemical components: a central $\alpha$-carbon ($C_\alpha$), an $\alpha$-proton ($H$), an \textit{amino functional} group ($-NH$), a \textit{carboxylic acid functional} group ($-COOH$), a \textit{side chain} group ($R$) (see Figure \ref{fig1}) \cite{McCammon 1984}. The residual side chain group differentiates the common biological amino acids, and is the main factor of the peptide chains local stable conformations. These amino acids combine to become proteins through an energy-driven combination. This result in the creation of a peptide bond between the two amino acids, and repeating the process creates a polypeptide containing several peptide bonds. These peptide bonds behave like a partial double bond, which have restricted rotation about the bond. This restriction results in a stable peptide plane. These peptide planes are repeating units that exhibit constant structures in the protein and reduce the number of degrees of freedom. The polypeptide chain is intrinsically flexible because many of the covalent bonds that occur in its backbone and side chains are rotationally permissible. A protein can be defined by one or more polypeptide chains.
Geometric relationship involving atoms in the polypeptide fully define a thee-dimensional proteins stable conformation. The relationships consist of bond lengths, bond angles, dihedral angles and improper dihedral angles. The primary contributions from these parameters, which determine overall polypeptide structure, are the dihedral angles. Typically, the peptide plane remains relatively rigid during protein dynamics such that the bond lengths and bond angles remain constant, due the large energy cost for its deformation. As a result, the dihedral angles are the essential degrees of freedom that dictate the position of the polypeptide backbone atoms, defining the protein secondary structure \cite{McCammon 1984}. 

\begin{figure}[h]
\begin{center}
\includegraphics[width=200pt]{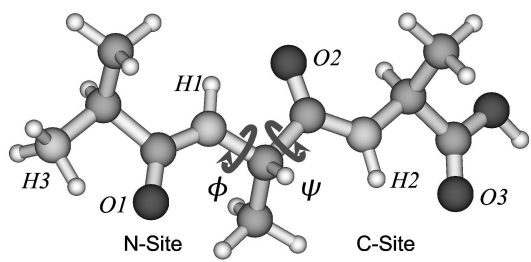}
\end{center}
\caption{Backbone dihedral angles in the molecular structure of trialanine (Altis 2008).}\label{fig2}
\end{figure}

\subsection{Coarse-grained models}
\noindent In a polypeptide chain the torsional motion is predominantly local in character. Therefore its model is here simplified as a constrained multibody system, and the overall dynamic is described by backbone dihedral angles, and possible linear elastic deformation allowing only covalent bonds lengths fluctuations. This type of model is not new, it have been used on molecular coarse-grained simulations, e.g. in systems like GROMACS \cite{Hess 2008},  widely used for long time simulations, where however the system dynamics is constrain by force fields like MARTINI force field \cite{Bekker 1990}. The most important feature of this type of approach is the possibility of modelling protein dynamics without explicitly treating every atom in the problem. Using this quasi-continuum approach, must degrees of freedom are eliminate, and force or energy calculations are largely expedited. Here however we are interested in analysing the possible advantages of imposing kinematic constrains on the system using kinematic joints. Since the definition of such constrains increase the simulation computational complexity, we are concern on its numerical stability for long time simulations \cite{Bekker 1990}. 

\begin{figure}[h]
\begin{center}
\includegraphics[width=350pt]{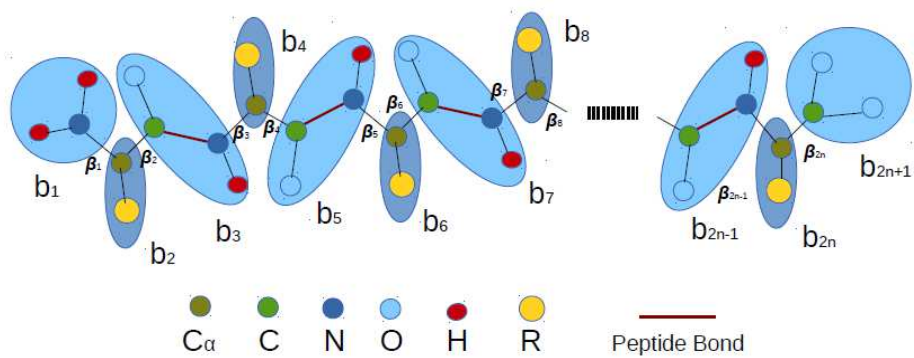}
\end{center}
\caption{Refinement for a protein with $n$ amino acids split in $2n+1$ bodies, with its $2n$ dihedral angles identified with $\beta_1,\cdots,\beta_{2n}$}\label{fig3}
\end{figure}

The methodology described bellow was evaluated on two types of  coarse-grained models, defined using different kinematic joints. The first is defined by an open chain of rigid bodies, linked together using revolute joints. For the second model we used the same open chain system but, to allow some fluctuations in the system structure, the components of the system are linked using cylindrical joints. In Fig. \ref{fig3}, presents the coarse-grained model for a generic polypeptide chain defined by $n$ amino acid, using a constrained multibody system with $N=2n+1$ bodies $b_1,b_2,\dots,b_{2n+1}$, with dihedral angles $\beta_1,\beta_2,\dots,\beta_{2n}$. Note that, each amino acid is split between three bodies, and each body is linked to the next by a covalent bonds. This gives great flexibility to them and its conformation can be characterized by dihedral angles on this bonds. The values of these angles are not uniformly distributed, they have strong local propensity  and are highly correlated \cite{Betancourt 2004}. 

The use of simplified models reduces the complexity of the interactions and hence reduces the amount of computation involved. This shouldn't be seen as a limitation. Software tools for reconstructing all-atoms structure from backbone structures (e.q. PHOENIX\footnote{http:$\backslash\backslash$cbsu.tc.cornell.edu/software/photarch}  and Maxsprout \footnote{http:$\backslash\backslash$www.ebi.ac.uk/maxsprout} ) are often employed to complete folding trajectory with side chain information.

\subsection{Protein local propensity}
\noindent Polypeptide chains are known to have strong local propensities, well characterized by the existence of specific amino acid patterns in chain with predefined ranges of dihedral angles. A commune strategy to study the angular propensity is through \textit{Ramachandran maps} used to produce distributions of the dihedral angles and their probabilities, extracted from statistical libraries with local propensities generated using the PDB. Studies of this plots \cite{McCammon 1984} show that they reflect the local interactions of free energy. The type of conformation are determined from a balance between local interactions (those closed to the sequence) and non-local ones. Effective statistical potentials can be extract from these populations have been studied for over 40 years. These potentials are based on the correlation of the observed frequency of a structure with its associated free energy. Thus, those potentials have a global minimum corresponding to the must frequency observed native informations (or collections of substructures found must often in native conformations). Such local potentials have been combined with non-local potential to predict protein folding structure, used  by the state-of-the-art systems for conformation perdition like ROSETTA.  

\subsection{Van Mises propensity distributions}
\begin{figure}[h]
\begin{tabular}{cc}
\includegraphics[width=200pt]{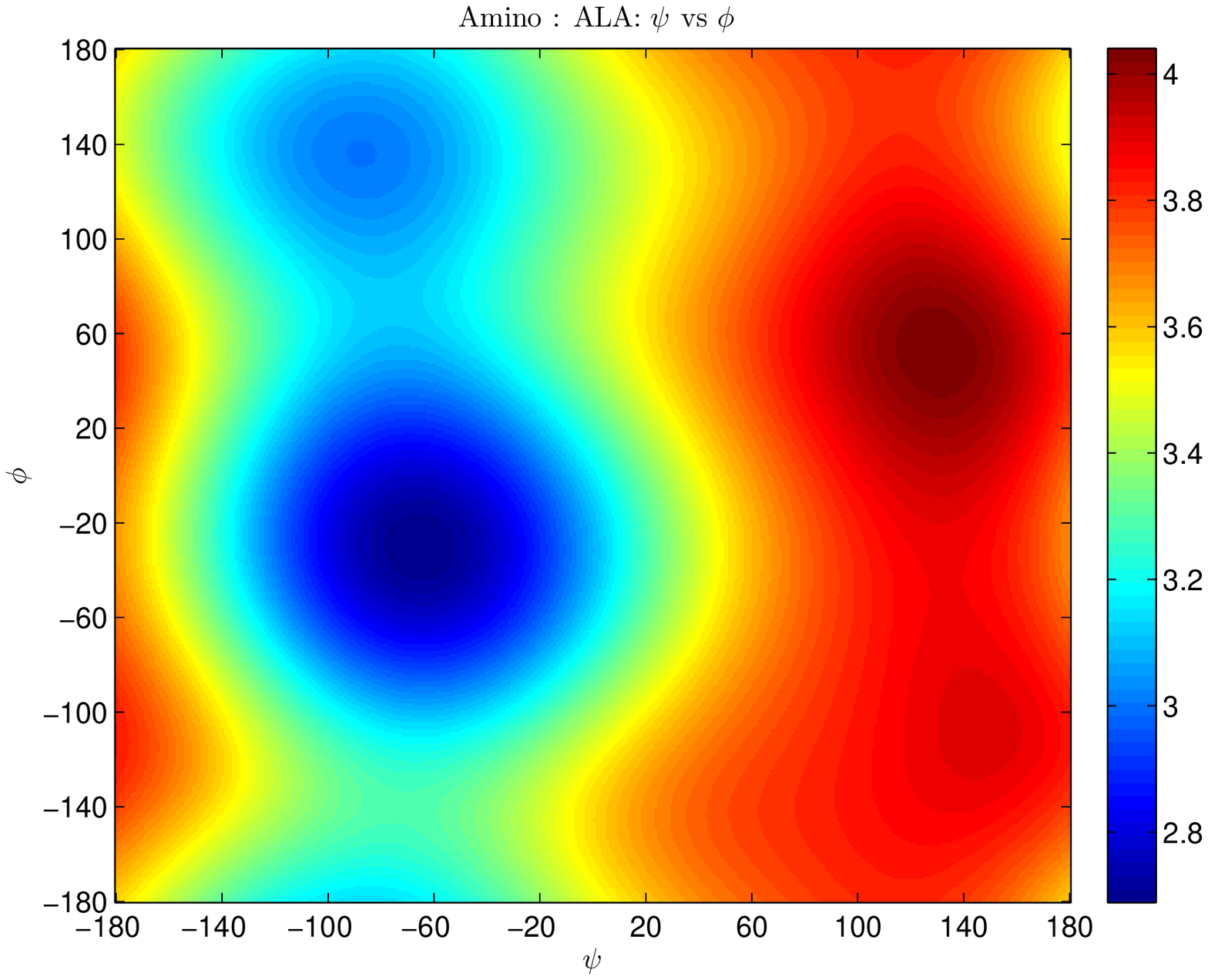} &
\includegraphics[width=200pt]{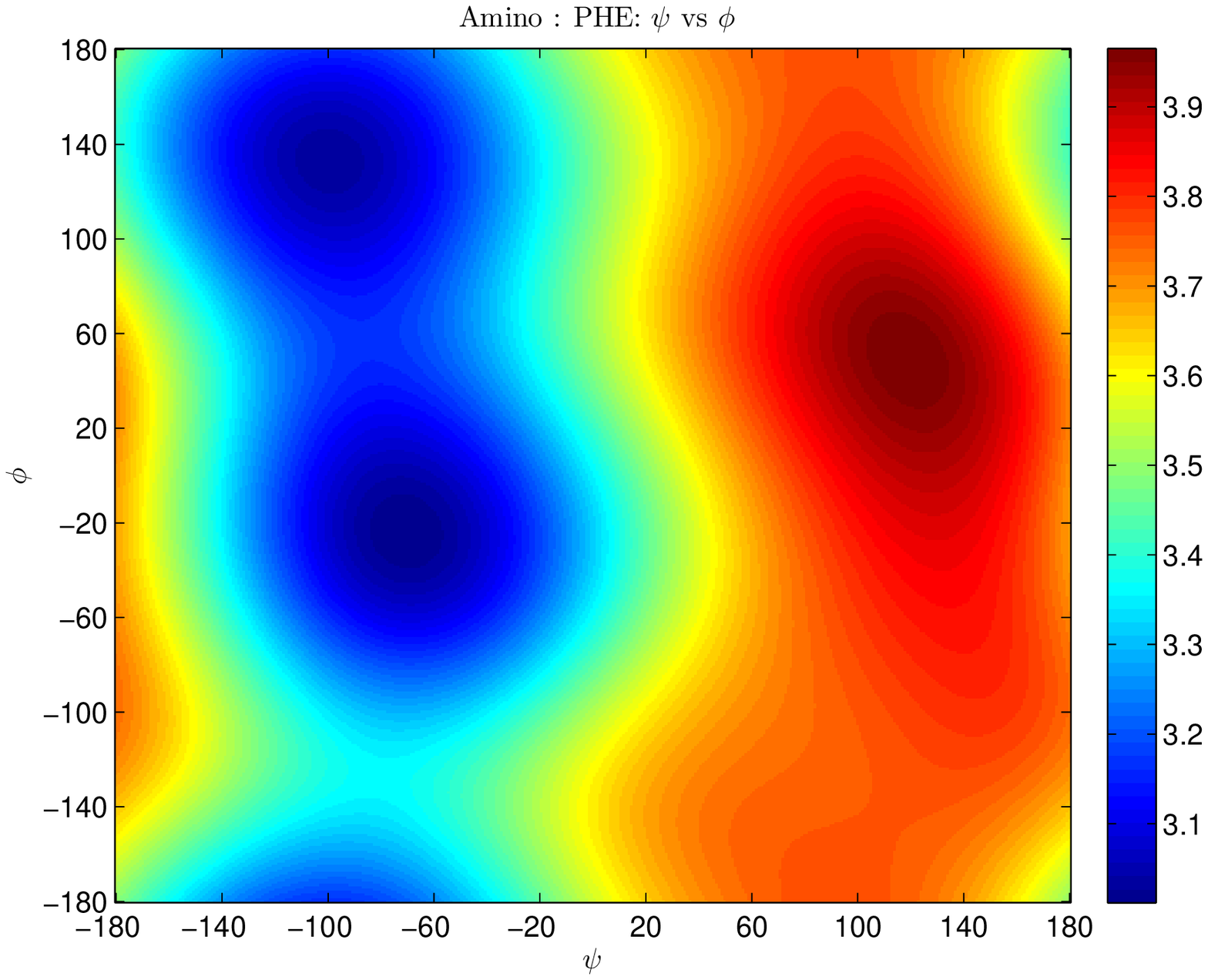}\\
\end{tabular}
\caption{ Amino acid Mises propensity distributions a)  ALA  b) PHE }\label{fig10}
\end{figure}

\noindent Ramachandran maps play a central role in developing empirical energy functions for structure prediction and simulation \cite{amakrishann 1965}\cite{Shapovalov 2011}. Those are used as a probability density function gives the probability of finding an amino acid conformation in a specific range of $\phi$, $\psi$ values. For this work these functions are given on a $2^\circ\times 2^\circ$ grid from $-180^\circ$ to $180^\circ$ in $\phi$, $\psi$ values. Such distributions was derived for each amino acid types, but since they are very irregular we used the van Mises density estimations to have soother surfaces. The quality and quantity of the data are crucial in determining distributions to approximate the system free energy and its gradients. Here all the distribution are computed using a datasets of $254$ globular proteins with a resolution cutoffs of 1.0 $\AA$, filters to reduce the number of amino acids angles evaluated over great structural stresses. Those distribution estimations were used to build a library of conformational propensities for each amino acid, applied for drive polypeptide chain kinematics. Examples of van Mises propensity distributions for two amino acids are presented in Figure \ref{fig10}. The library was developed by combining a prior estimate of the probabilities of each $(\phi,\psi)$ bin raw counts by amino acid. This standard probability distributions are quite bumpy in their variation, a result of using raw counts in the probability estimates and calculation of simple averages. In order to produce smooth and continuous estimates of the conformation probabilities, we use kernel density estimation. A kernel is a nonnegative symmetric function that integrates to $1.0$ and is centered on each data point. Density estimates at specific query points are determined by summing the values of the kernel functions centered on the data points. The smoothness of the density estimate is determined by the form of the kernel, in particular its bandwidth. For each amino acid, $aa$, we determine a probability density estimate, $P(\phi,\psi | aa)$. For that \textit{von Mises probability density function} (PDF) as the kernel are used since this density estimates are more appropriately for angles than the usual Gaussian or other nonperiodic kernels \cite{Mardia 2000}. Because Ramachandran probability density is defined for the backbone torsion angles $\phi$ and $\psi$ as two arguments, we use a nonadaptive kernel density estimators in two dimensions written as the sum over products of $\phi$ and $\psi$ van Mises kernels for $N_r$ data points of amino acid of type, $aa$:
\begin{equation}
P(\phi,\psi | aa)=\frac{1}{4\pi^2N_r}\sum_{i=1}^{N_r}\frac{1}{I_0(\kappa)^2}\exp(\kappa \cos(\phi-\phi_i)+\kappa \cos(\psi-\psi_i))
\end{equation}
In this case, $\sqrt{\frac{1}{\kappa}}$ defines a radius of the two dimensional hump covering $60\%$ of the kernel density \cite{Shapovalov 2011}, $I_0$ is the Bessel function of the first kind of order $0$, normalizing the kernel to $1$. 

To reduce the amount of computation involved on the protein kinematics analysis, the library of conformational propensities was completed with directional gradients, for each distribution. For that the central difference method was used.

Amino acid dihedral angles cannot take any arbitrary values due to atomic clashes and orientations \cite{Shapovalov 2011}. It is also consensual \cite{Bosco 2003} on the basis of the conformational enumeration of polypeptide chains and molecular dynamic simulations that the Ramachandran basin populations are affected by their nearest neighbours. The populations are affected, in particular, by the neighbour's conformation and their identity. There are a strong correlations between a residue's conformation  and that of its neighbours are responsible for cooperative effects. This works however uses only correlation defined by dihedral angles in the same amino acid and between two consecutive amino acid. On the presented extended mechanical model, the coupling band of springs, uses these potentials to find optimal conformations, by the generation of body torsions along each associated covalent bond.

The dependence of an amino acid conformation on the conformations of its adjacent residues involves too many variables to be captured in a single probability density function for the available data. Instead, we divided the probability densities in individual terms involving pairs of angles. In particular, for dihedral angles $\phi_i$, $\psi_i$, $\phi_{i+1}$, and $\psi_{i+1}$ in consecutive amino acids
we looked at the density plots involving $(\phi_i,\psi_{i+1})$, $(\psi_i,\phi_{i+1})$ and $(\phi_i,\phi_{i+1})$.  For those von Mises distribution evaluate for each pair of amino acids, and the correspondent directional gradient, where added to our propensity library.

\subsection{Minimum energy path in an amino acid}

\noindent Potentials for torsions in a single amino acid of type $aa$ can be obtained from a statistic distribution using Boltzmann \cite{Henkelman2 2000} inversion:
\[
E_{aa}(\phi,\psi)=-k_BT\log(P(\phi,\psi|aa))
\]
where $k_BT$ is the thermal energy and $\phi,\psi$ are assume to be its two dihedral angles. If we sample the amino acid $aa$ configuration space with a curve, from a conformation with  dihedral angles $\bm{A}=(\phi^{(A)},\psi^{(A)})$ to a conformation $\bm{B}=(\phi^{(B)},\psi^{(B)})$. The curve satisfies a differential equation which by construction guarantees that it evolves to the most probable transition path connecting $\bm{A}$ and $\bm{B}$ in the potential $E_{aa}$. For that, consider the system modeled with a Brownian random force $W$ acting in each conformation, by:
\begin{equation}\label{D1sysmodel}
\rho \dot{q}=-\nabla E_{aa}(q)+W(t)
\end{equation} 
where $\rho$ is the friction coefficient modeling the influence of the system surrounding in its dynamics, and $q=(\phi,\psi)$ is a conformation in the selected path. The Fluctuation-Dissipation Theorem fixes that thermal random forces $W$ obey Gaussian statistics \cite{Henkelman2 2000}. The stable states are localized around the minima of the potential $E_{aa}$ \cite{Buchenberg 2015}. Assuming $E_{aa}$ has at least two minima $\bm{A}$ and $\bm{B}$, the MEP connecting these states in the amino acid potential is a smooth curve $\vartheta^\ast$ connecting $\bm{A}$ and $\bm{B}$, and defined by conformations $q$, such that its path integral satisfies \cite{Henkelman2 2000}
\begin{equation}\label{stabeq}
(\nabla E_{aa})^\perp(\vartheta^\ast)=0,
\end{equation}
where $(\nabla E_{aa})^\perp$ is the component of $\nabla E_{aa}$ normal to the path $\vartheta^\ast$ (see Figure \ref{fig10}). If $\vartheta(\alpha)$, with $\alpha\in [0,1]$, is an arbitrary curve parametrization on the amino acid conformation space connecting $\bm{A}$ and $\bm{B}$, them
\begin{equation}
\vartheta^\ast=\min_\vartheta\int_\vartheta E_{aa} ds.
\end{equation}

\begin{figure}[h]
\begin{tabular}{cc}
\includegraphics[width=220pt]{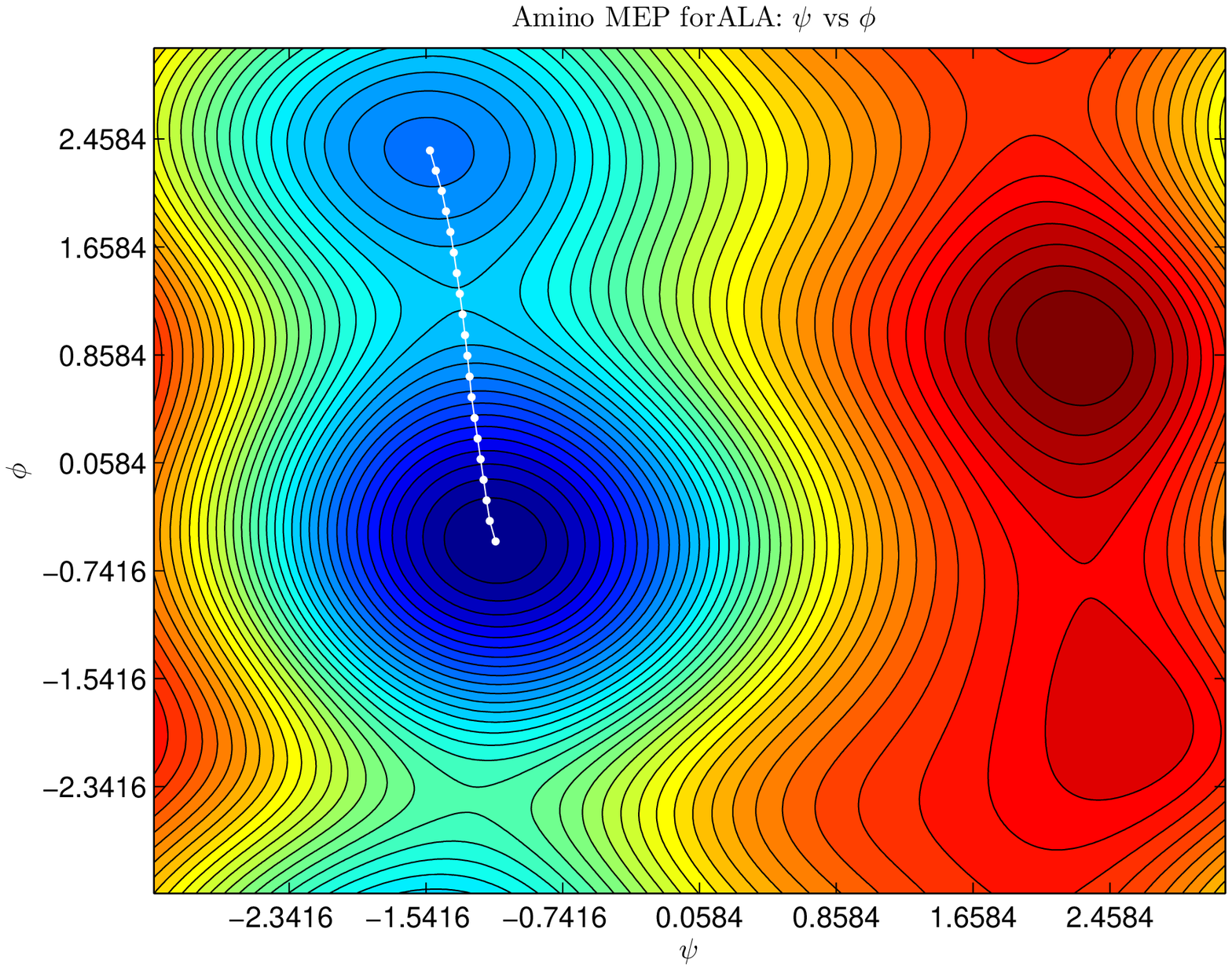} &
\includegraphics[width=220pt]{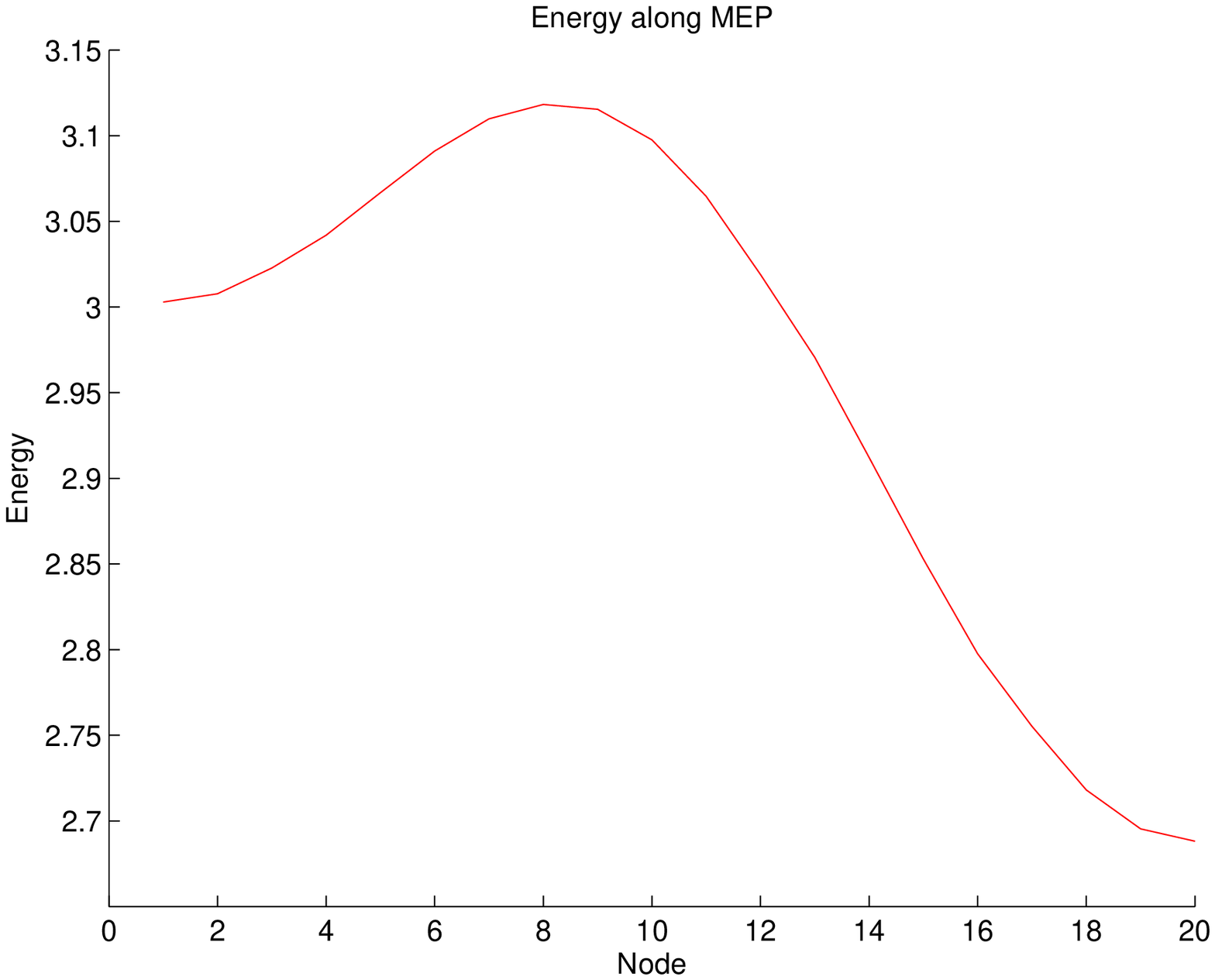}\\
\end{tabular}
\caption{ A MEP, defined using a string with 20 nodes,in the amino acid ALA van Mises propensity distributions, as shown in Fig. \ref{fig10}, between two saddle conformations.}\label{fig10-1}
\end{figure}

\subsection{Minimum energy path in a sequence with two amino acid}

\begin{figure}[h]
\begin{tabular}{ccc}
\includegraphics[width=145pt]{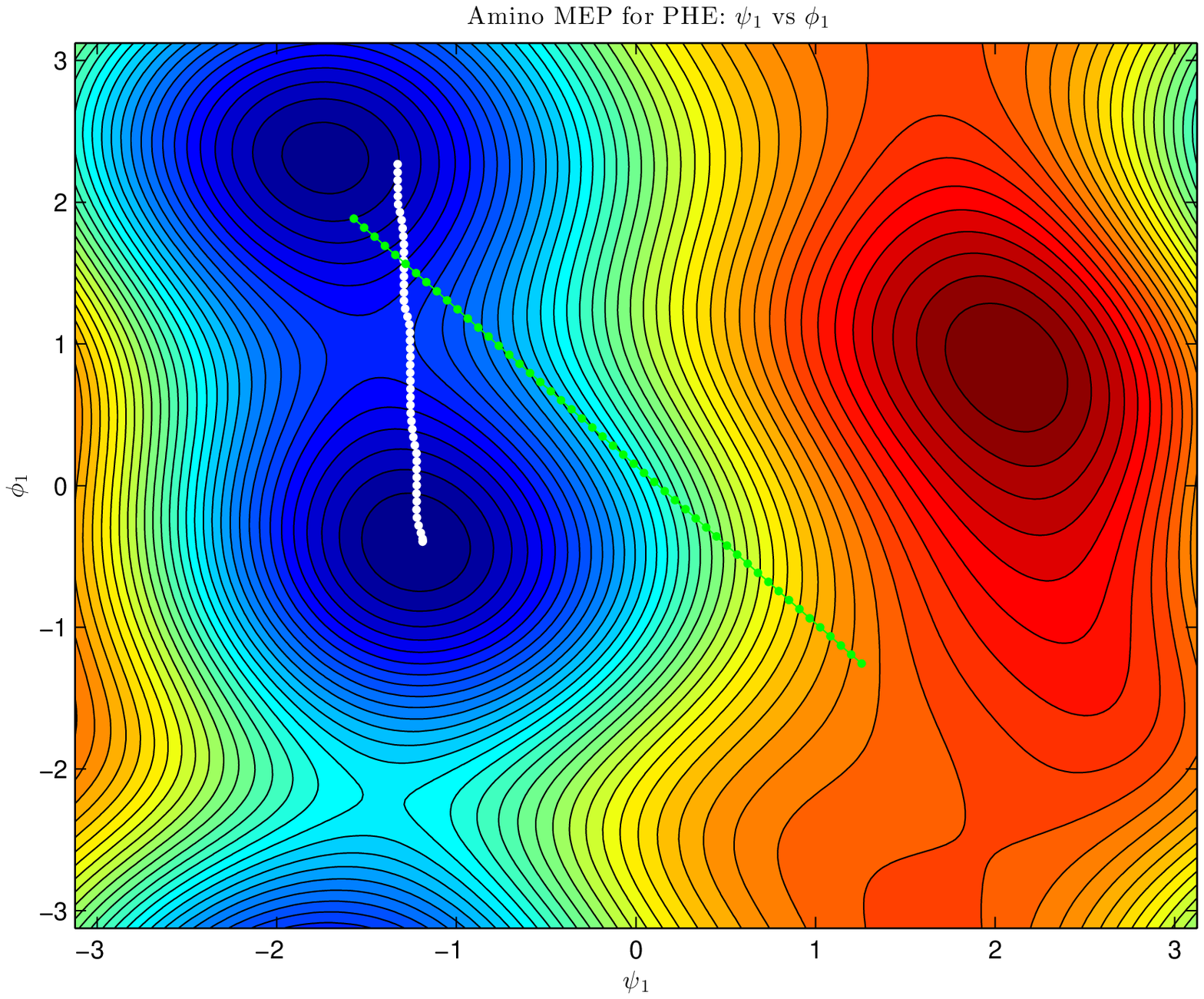} &
\includegraphics[width=145pt]{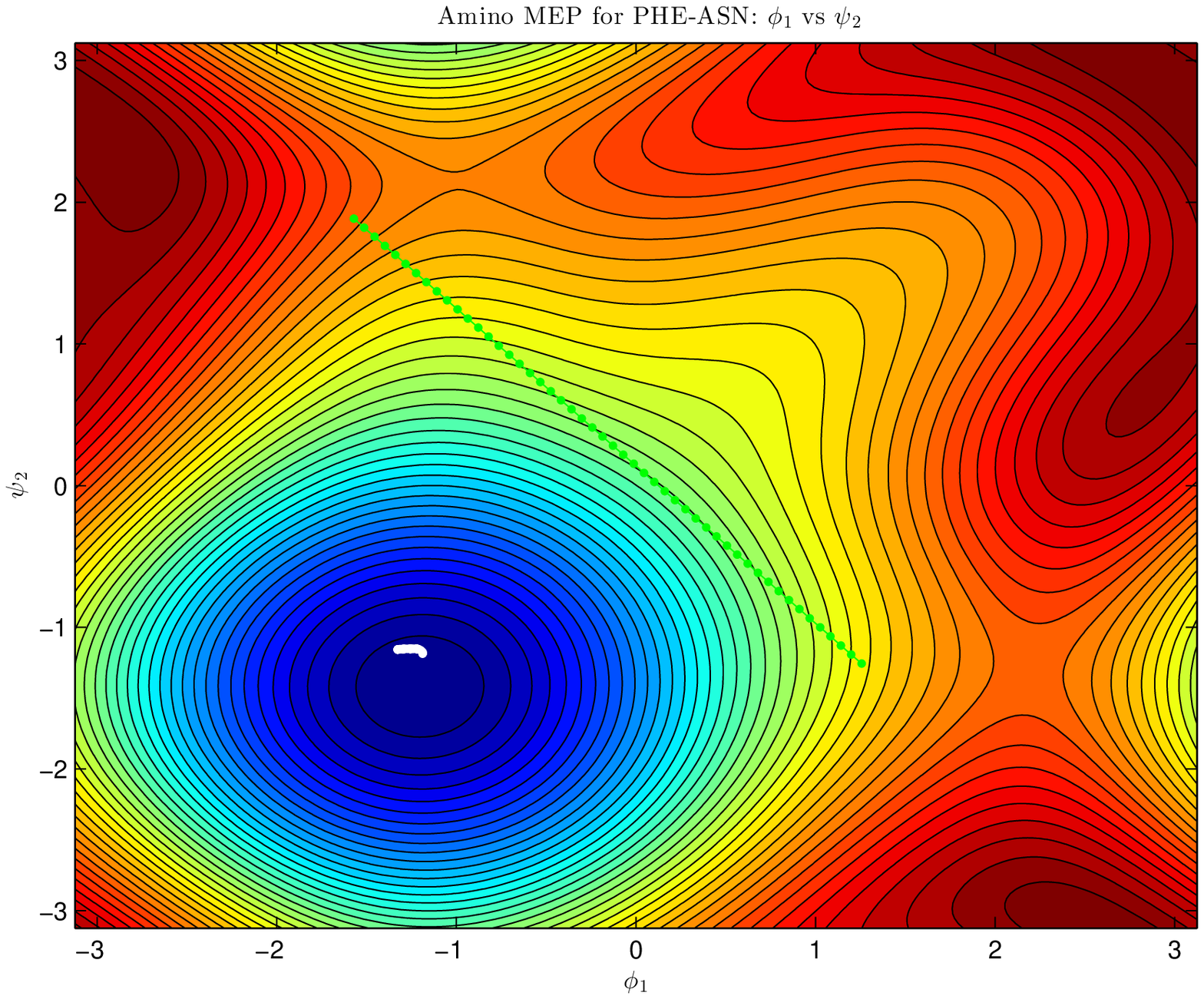} &
\includegraphics[width=145pt]{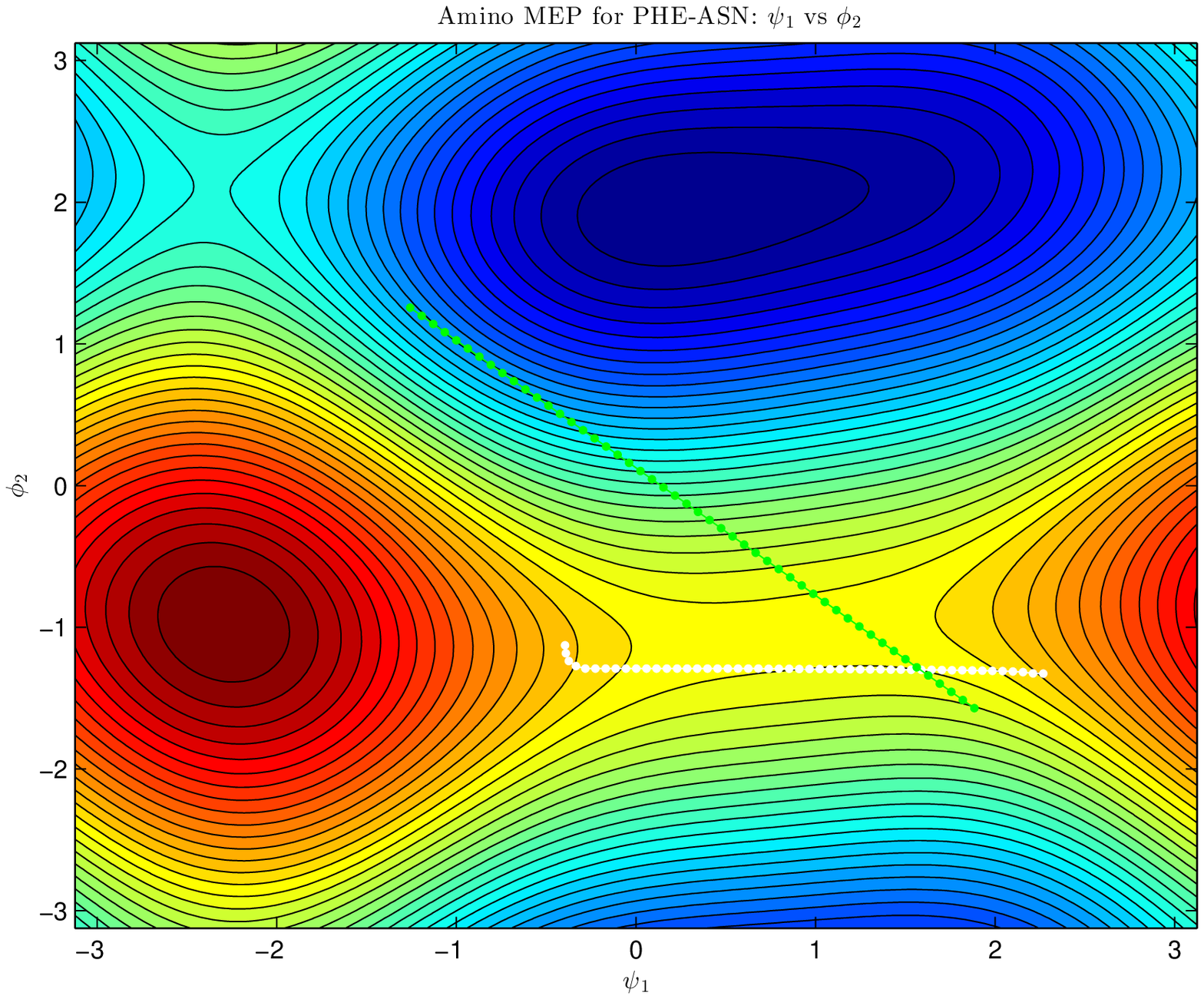} \\
\includegraphics[width=145pt]{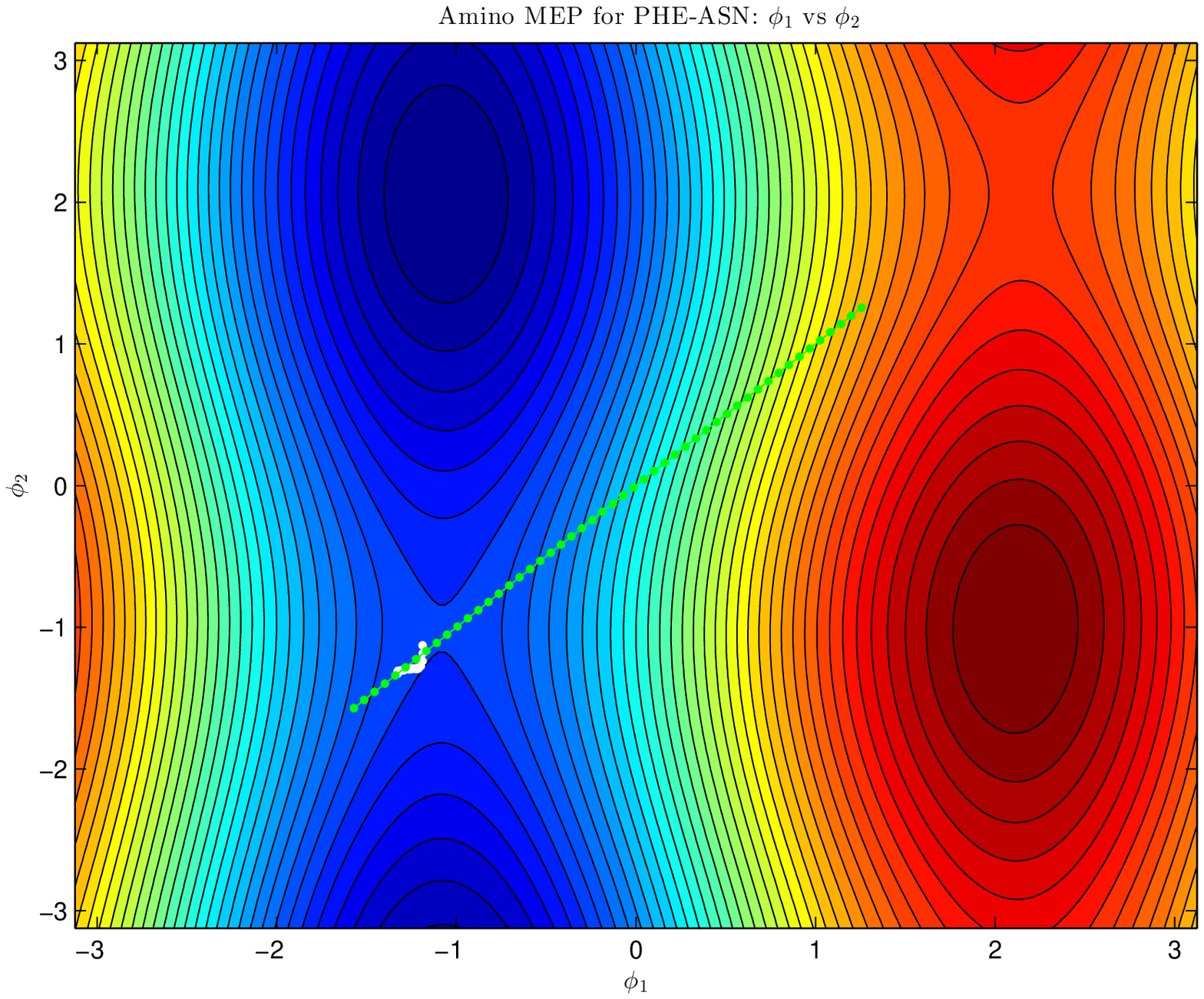} &
\includegraphics[width=145pt]{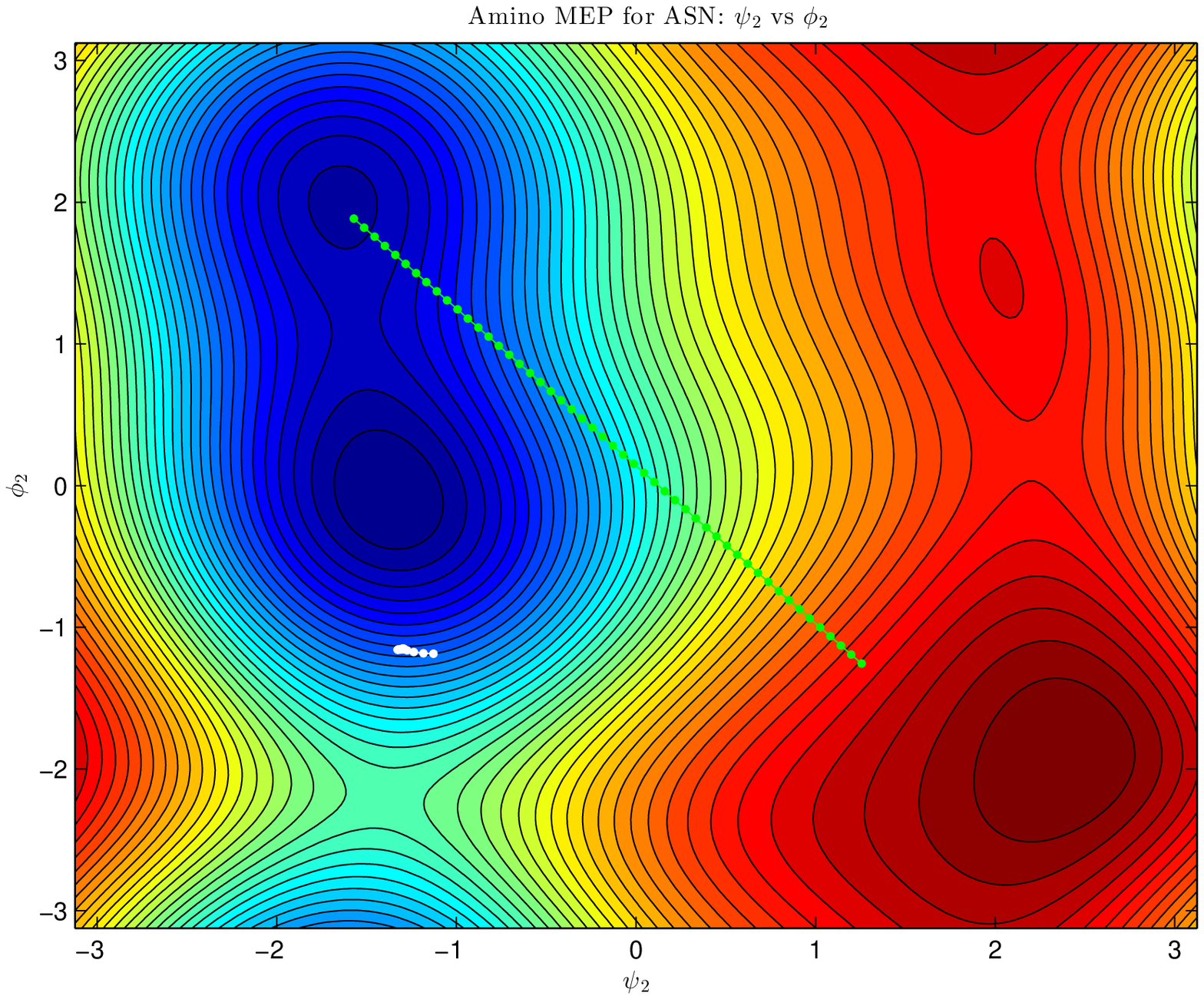}
\end{tabular}
\caption{ Views of the five projections for initial approximation (green) and final MAP (white), for a sequence of two amino acids defined by ILE and ALA. From Top to Bottom projection $(\phi_1,\psi_1)$, $(\phi_1,\psi_2)$, $(\psi_1,\phi_2)$, $(\phi_1,\phi_2)$ and $(\phi_2,\psi_2)$.}\label{fig11}
\end{figure}

\noindent Given a polypeptide chain, defined by a sequence of residues $aa_1aa_2\cdots aa_n$, a dihedral angle statistical potential for the coarse-grain model, can be defined as:
\[
E=-k_BT\sum_{i=1}^{n-1}(\log(P(\phi_i,\psi_i|aa_i))+1/2(\log(P(\phi_i,\psi_{i+1}|aa_iaa_{i+1}))+
\]
\[
+\log(P(\psi_i,\phi_{i+1}|aa_iaa_{i+1}))+\log(P(\phi_i,\phi_{i+1}|aa_iaa_{i+1})))+\log(P(\phi_n,\psi_n|aa_n))
\]
When system is defined by a sequence with only two residues $aa_1aa_2$, the above free energy description takes the form
\[
E=-k_BT(\log(P(\phi_1,\psi_1|aa_1))+\log(\phi_1,\psi_2|aa_1aa_2)+
\]
\begin{equation}\label{potential2}
+\log(\psi_1,\phi_2|aa_1aa_2)+\log(\phi_1,\phi_2|aa_1aa_2))+\log(P(\phi_2,\psi_2|aa_2)))
\end{equation}
One of the most demanding task is to find two metastables for a generic system of this type. The usual approach is by applying a Monte Carlo methodology, followed by systems like FOLDING@home.  Here to test the library of conformation propensities, we represented a trajectory discritization by a sequence of dihedral angles and use the Spring Method \cite{Henkelman 2000-2} to characterise the involved ``hierarchical'' free energy landscape. For that, given initial approximations to MEPs, the statistical potential were used to generating external forces for the motion equation, simulating the path shape change in the landscape. The shape of such paths are changed in each integration step, by applying those forces on each of its nodes, driving each nodes using the gradient method to local optimal points on the energy surface. To avoid the accumulation of points near local optima, in each integration step, the MEP approximation is reparametrized, in each iteration, using a cubic spline interpolation, imposing a regular distribution of nodes in the curve. With this, after each integration, the coordenates of nodes are moved along the curve, imposing a regular discretization. The generated path have its starting and final nodes near optimal conformation, and the tangents on each internal node is perpendicular to the force field generated by the landscape gradient. Note that, in this approach the path initial state and its final state are not fixed. This method was used to compute the MEP presented on Figures \ref{fig10-1} and \ref{fig11}, the first to analyse the energy landscape for a single amino acid, having two degrees of freedom, and the least for a system defined by two amino acids, having by degrees of freedom the four involved dihedral angles, two in each amino acid.

Let $\vartheta$ be a string define by a sequence of $N$ nodes $(\phi_1,\psi_1),(\phi_1,\psi_1),\cdots,(\phi_N,\psi_N)$, denoted by $\vartheta_1,\vartheta_2,\cdot,\vartheta_N$ in the system phase space $\Omega$, connecting an initial conformation $A$ to a final conformation $B$. Since stationary solutions of (\ref{stabeq}) satisfy (\ref{D1sysmodel}), a simple method to find MEP is to evaluate the conformation at node $\vartheta_i$, $i=1,\cdots,N$, according to
\begin{equation}\label{eqeachnode}
\dot{\vartheta_i}^\bot=-(\nabla E)^\bot(\vartheta_i),
\end{equation}
here $\dot{\vartheta_i}^\bot$ denotes the normal velocity of node $\vartheta_i$, and $(\nabla E)^\bot(\vartheta_i)=\nabla E(\vartheta_i)-(\nabla E\cdot \hat{g}_i)\hat{g}_i$, where $\hat{g}_i$ is the unit tangent vector along $\vartheta$ on $\vartheta_i$.
Since both (\ref{stabeq}) and (\ref{D1sysmodel}) are intrinsic, the path discretization can be arbitrarily chosen. For parametrizing a path $\vartheta$ it is sufficient  to define a map $\alpha$ from $[0,1]$ to the conformation space, by arc length, such that $\alpha(0)=A$ and $\alpha(1)=B$. In practice (\ref{eqeachnode}) is solved by a time-splitting scheme. A reparametrization step is applied to enforce the proper parametrization of the string. 

\begin{figure}[h]
\begin{tabular}{cc}
\includegraphics[width=210pt]{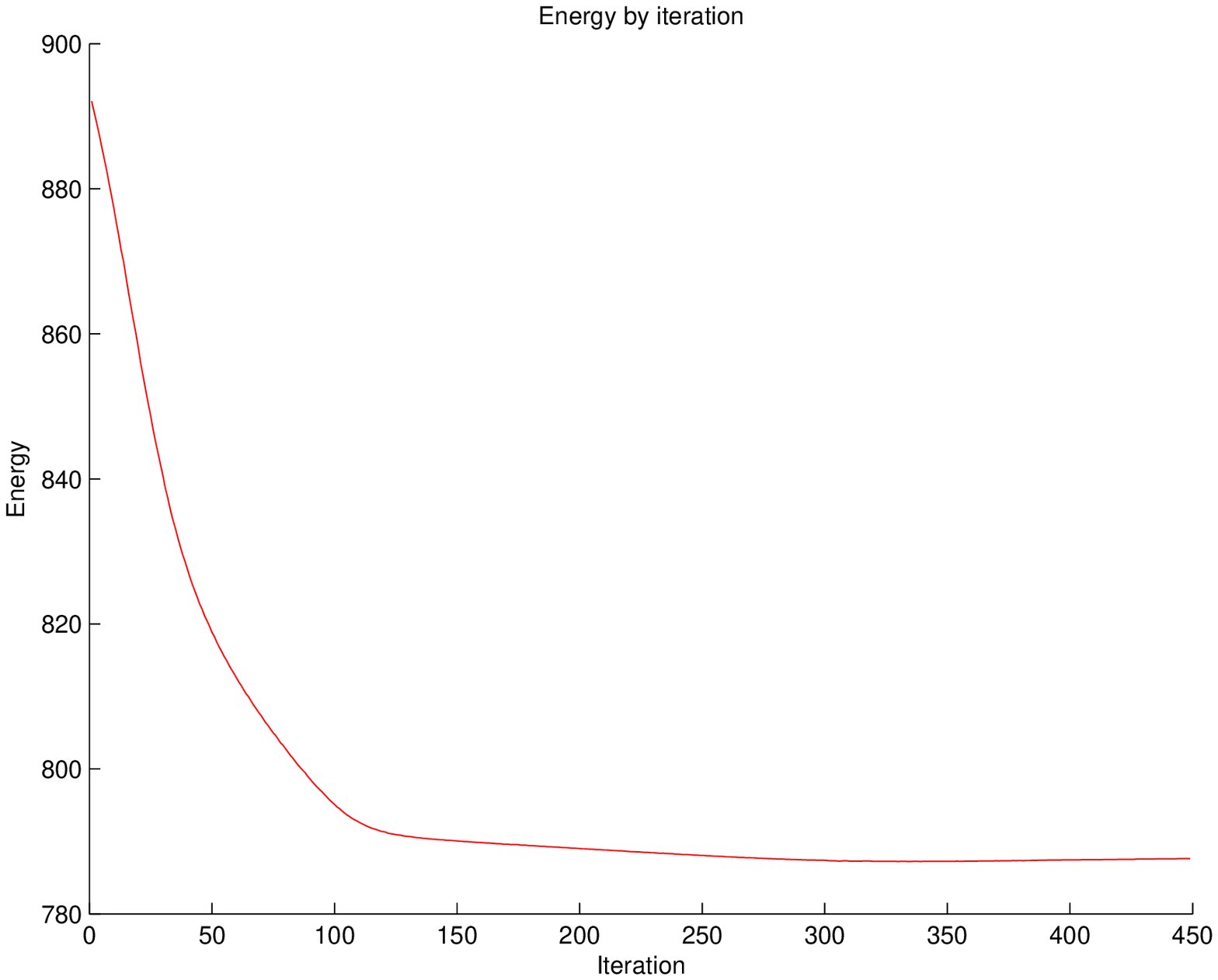} &
\includegraphics[width=210pt]{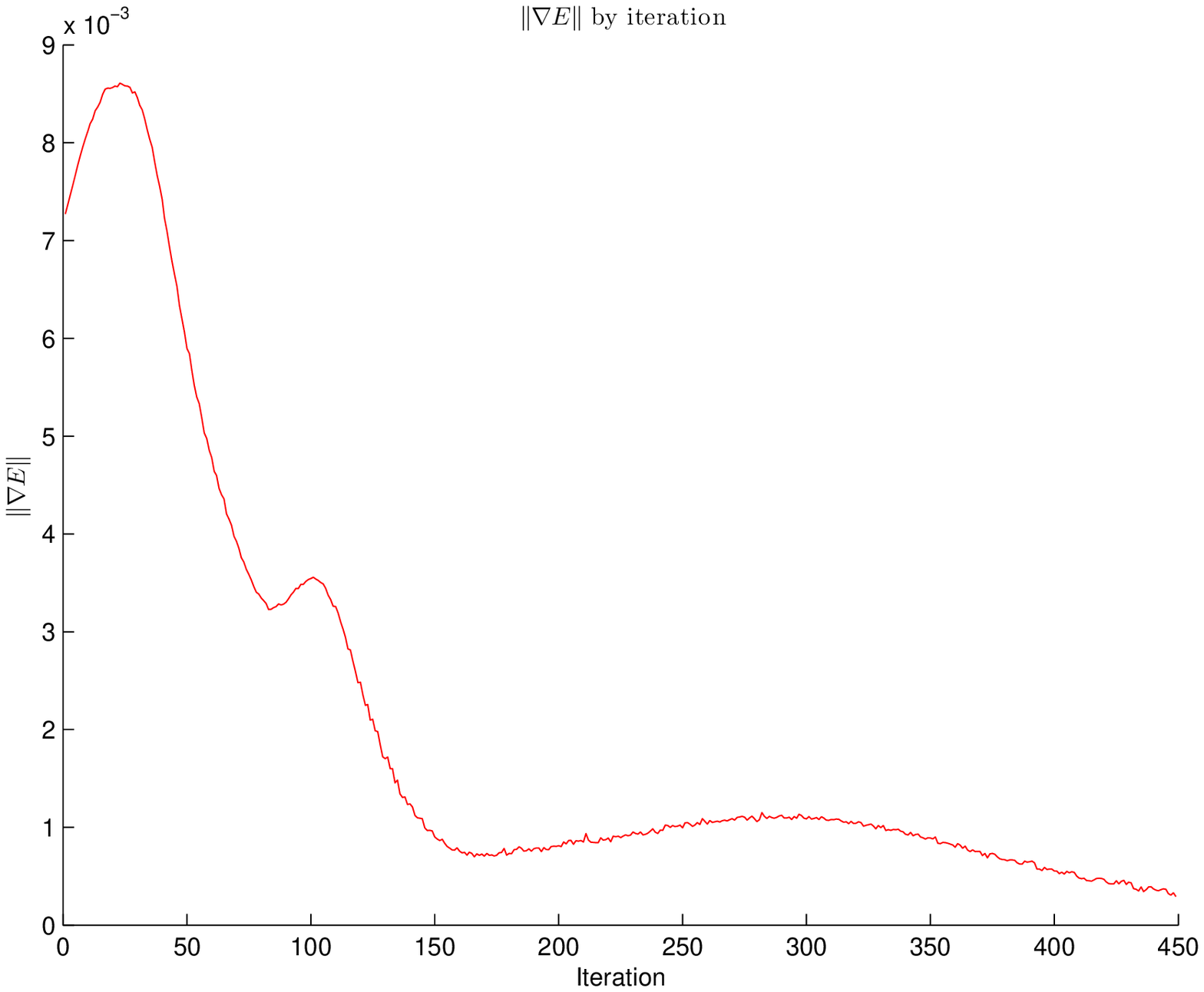}\\
\includegraphics[width=210pt]{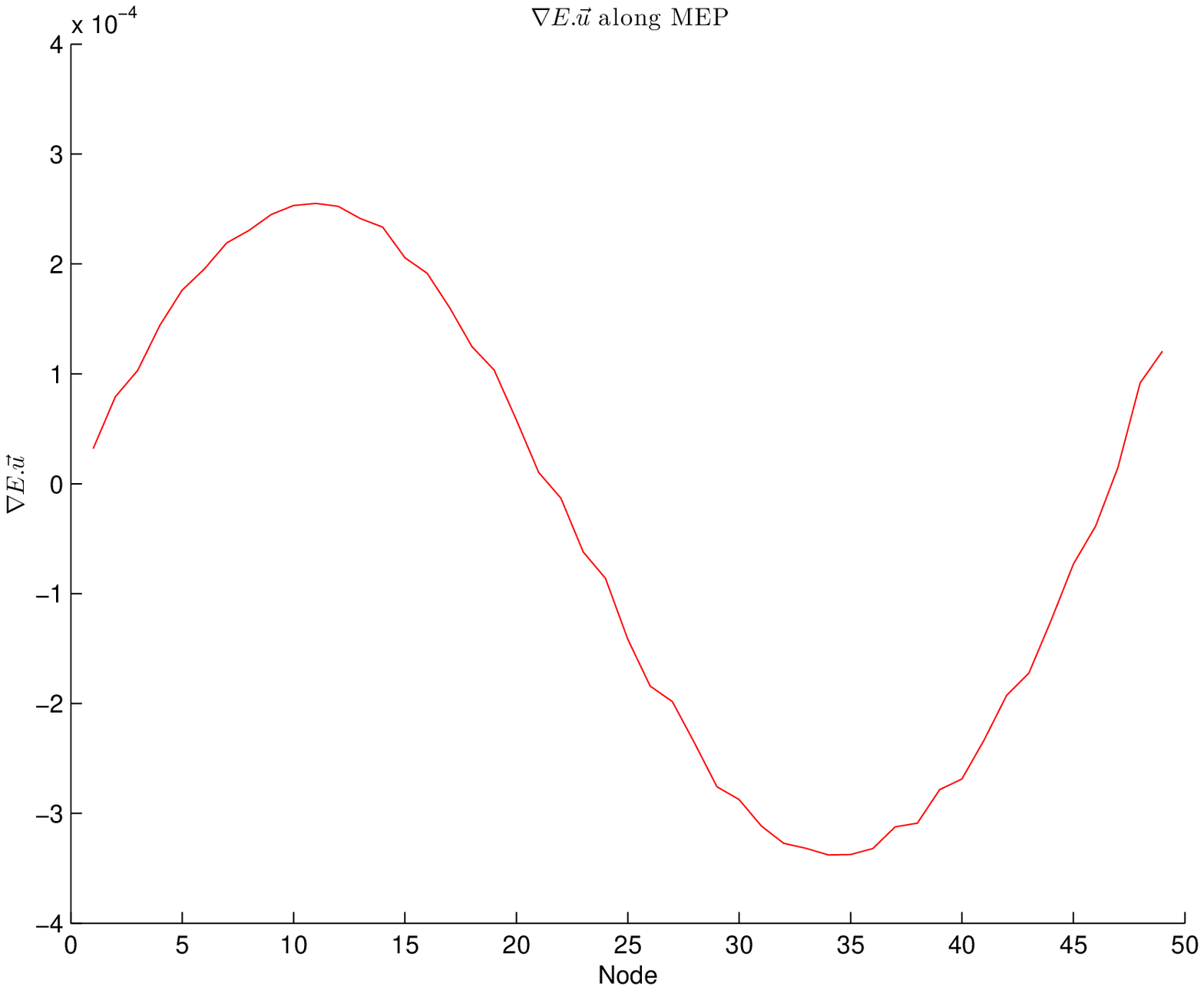} &
\includegraphics[width=210pt]{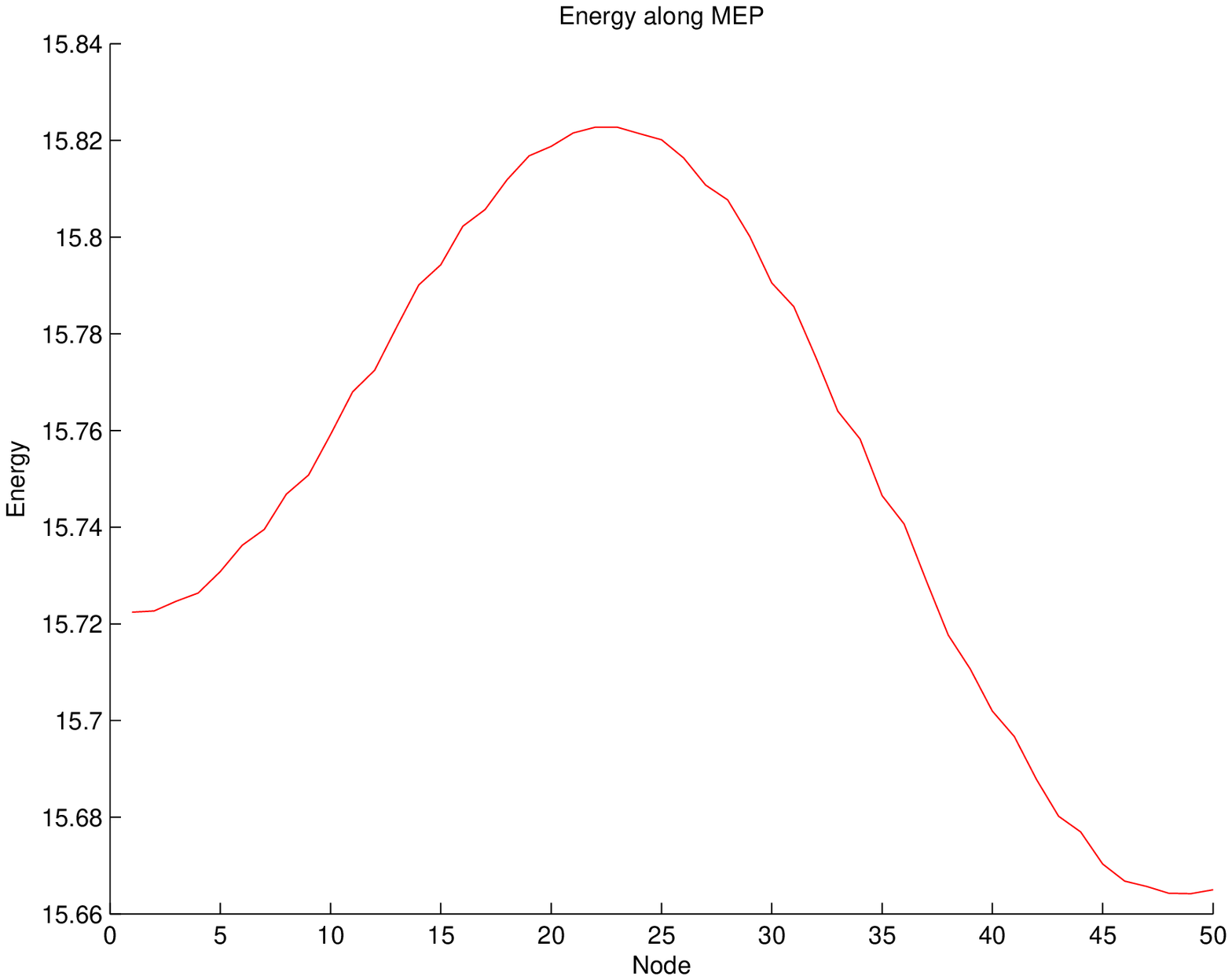}\\
\end{tabular}
\caption{ Top: Evolution of the string energy and gradient norm to the final solution (with stopping criteria $\|\nabla E\|<3\times 10^{-4}$). Bottom: Internal product between the energy gradient and curve tangential vector (for a string with 50 nodes). Final energy distribution for each node on the path.}\label{fig12-2}
\end{figure}

As an example we look at the dynamics of two amino acids iteration, in a peptide chain defined by PHE and ASN, via the potential (\ref{potential2}). Having by principal interest test the associated distribution for conformational iteration. Our numerical results are presented on Figures \ref{fig11} and \ref{fig12-2}.  For that we selected on the conformation space two arbitrary system states $A$ and $B$, and using convex combinations a string of conformation between $A$ and $B$, was generated, with equidistant 50 nodes. This path was used as initial approximation to the MEP. Equation (\ref{eqeachnode}) was solved using the Euler method for each node, and after each iteration the path update was reparametrized, using a quadratic spline, to enforce that each node have the same arc length to its neighbours. For this process a stopping criteria $\max_i\|\nabla E(\vartheta_i)\|<3\times 10^{-4}$ was used. This system as four degrees of freedom, defined by the dihedral angles $\phi_1,\psi_1,\phi_2,\psi_2$ and on Figure \ref{fig11} the projection of the initial (green) and final path are projected on five planes (white), receptively, $(\phi_1,\psi_1)$, $(\phi_1,\psi_2)$, $(\psi_1,\phi_2)$, $(\phi_1,\phi_2)$ and $(\phi_2,\psi_2)$, using as background the van Mises distributions of each of this pairs. Evolution of the string energy in each iteration and  norm of the gradient for the final solution, are presented on Figure  \ref{fig12-2}. Since the final path must be perpendicular to the energy gradient, also in Figure \ref{fig12-2}, the quality of the path is verified by the value of the internal product between the energy gradient and curve tangential vector, at each node final is present.

In the following sections we describe a generic formalization to extend MEP determination to multibody systems. However, for that we must review the multibody dynamics and discuss the type of kinematic constrains with strong conformation propensity used in this models model to perform rare event analysis.

\section{Reference multibody dynamics} \label{III}

\noindent The  motion of a system can be approximated by a path of system admissible conformations, usually  specified by time dependent driving elements that govern the system motion and internal forces. The system admissible conformations are here characterized by system elements position, velocity and acceleration, and defined by the kinematic constrain equations that describe the system topology. These constraints are geometric constraints defined using generalized coordinates, expressed as functions of the body-fixed frame displacement, having the form of an algebraic equation that reduces the system number of degrees of freedom. 

The fundamental assumption is that translational motion of a rigid body $i$ can be defined by a vector $\bm{r}_i$ that describes the position of the origin of the body reference coordinate system with respect to the global $XYZ$-coordinate system, while the orientation of the body $\xi_i\eta_i\zeta_i$-coordinate system, with respect to global coordinate system defined using a vector of angles $\bm{w}_i$, and can be described by a set of Euler parameter $\bm{p}_i$. These rotational coordinates provide a $3\times 3$ rotational transformation matrix, denoted by $\bm{A}(\bm{p}_i)$. Therefore, the vector $\bm{q}_i$ of the generalized absolute cartesian coordinates of body $i$ in a multi-body system can be represented as, $\bm{q}_i=[\bm{r}_i,\bm{w}_i]^T$.

\begin{figure}[h]
\begin{center}
\includegraphics[width=150pt]{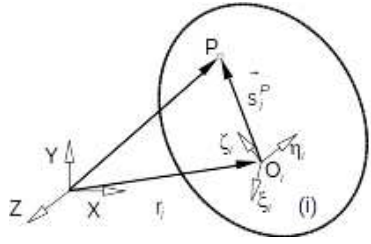}
\end{center}
\caption{ Global $XYZ$-coordinates for a point $P$ with local coordinates $\bm{s}_i^{\prime P}$. }\label{fig11-3}
\end{figure}

In this sense, a vector $\bm{s}$ with $\bm{s}_i^\prime$ components in the body $\xi_i\eta_i\zeta_i$-coordinate system as global components $\bm{s}_i=\bm{A}(\bm{p}_i)\bm{s}_i^\prime$. The global components of the vector can be transformed in terms of the body coordinate system as $\bm{s}_i^\prime=\bm{A}^T(\bm{p}_i)\bm{s}_i$. Every point $\bm{P}$ in the body $i$ with local coordinates $\bm{s}_i^{\prime P}$, is describe in the global $XYZ$-coordinate system by $\bm{s}_i^P=\bm{r}_i+\bm{A}(\bm{p}_i)\bm{s}_i^{\prime P}$ (see Figure \ref{fig11-3}).

The state of a multibody system consisting of $N$ rigid bodies, can be described by a vector $\bm{q=}[\bm{q}_1,\cdots,\bm{q}_N]^T$, where $\bm{q}_i$ is the vector of generalized coordinates of the body $i$. Points and vectors in this bodies can be used to defined sets of kinematic constrains, to impose restrictions on the relative motion between this bodies.

\subsection{Cylindrical and revolute joints}

\noindent Here we center our description in two types of kinematic constrains, named cylindrical and revolute joints. There are two relative degrees of freedom between two bodies connected by a cylindrical joint, while only one for the revolute joint. These are used in our coarse-grain models to describe the rotations allowed on atomic bounds and its possible length fluctuations.

\begin{figure}[h]
\begin{tabular}{cc}
\includegraphics[width=230pt]{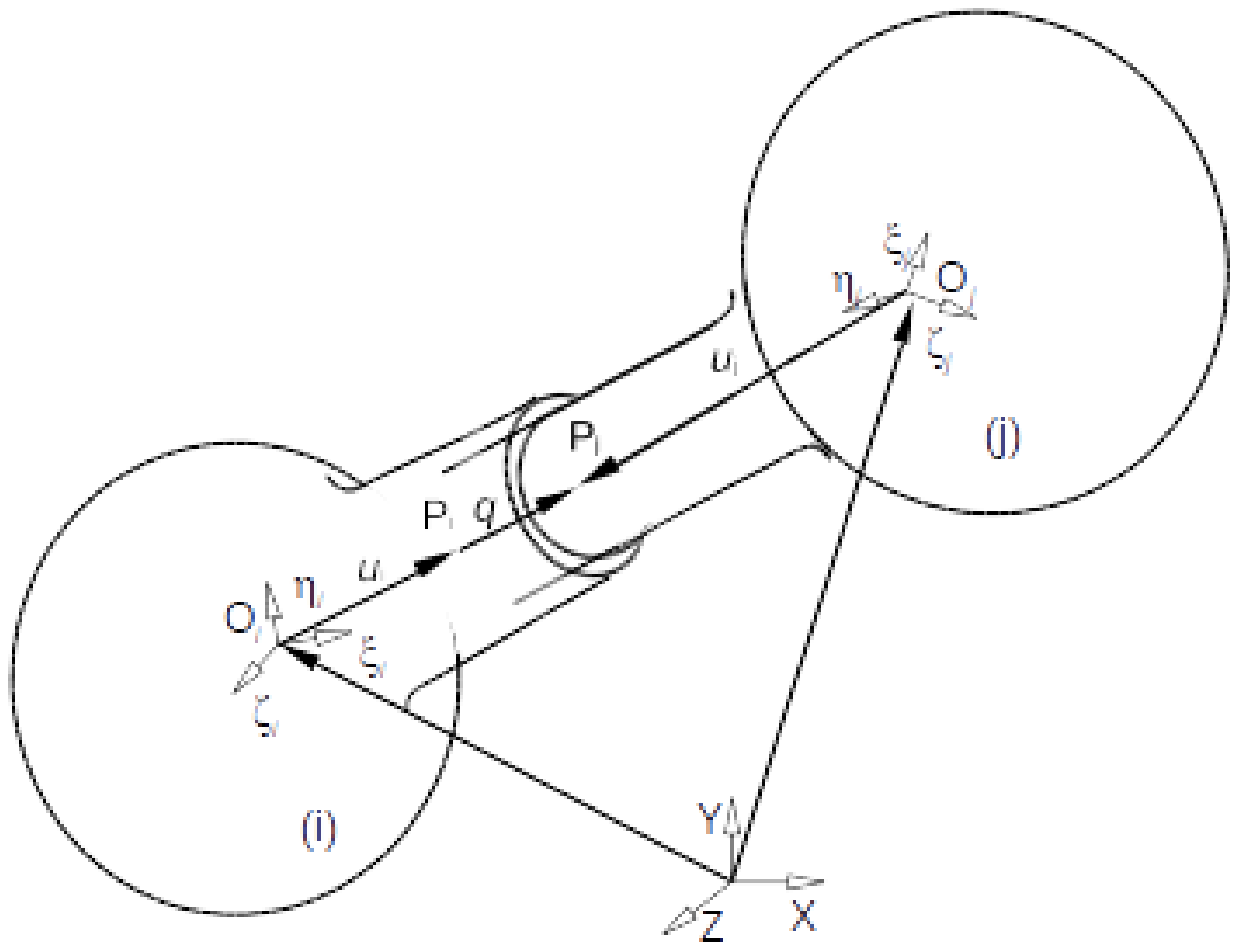} &
\includegraphics[width=230pt]{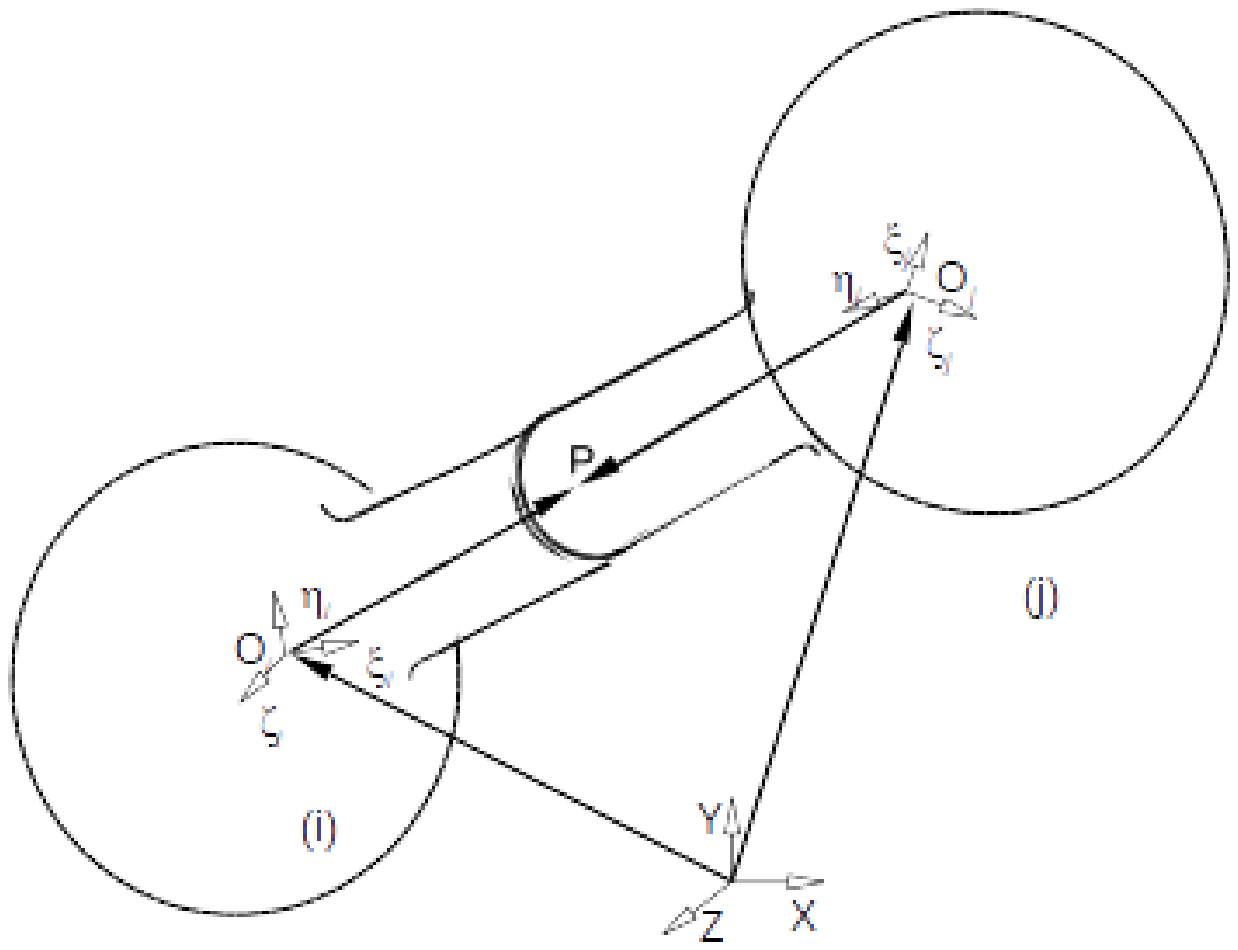}\\
\end{tabular}
\caption{a)  A cylindrical joint on a covalent bond. b) A revolute joint. }\label{fig5}
\end{figure}

A cylindrical joint constrains two bodies $i$ and $j$ to move along a common axis, but allows relative rotation about this axis. To derive equations of constraint for this joint, four points, $\bm{P}_i$ and $\bm{Q}_i$ on body $i$ and $\bm{P}_j$ and $\bm{Q}_j$ on body $j$, are arbitrary chosen on the joint axis, as shown in Figure \ref{fig5}. It is required that the vectors $\bm{u}_i$ and $\bm{v}_j$ of constant magnitude and $\bm{d}$ of variable magnitude remain collinear. Therefore, four constraint equations are needed to define a cylindrical joint; they can be found from two external product conditions: $\bm{\Phi}^{(p1,2)}\equiv \tilde{\bm{u}}_i\bm{u}_j=0$ and $\bm{\Phi}^{(p2,2)}\equiv \tilde{\bm{u}}_i\bm{d}=0$. 

A revolute join can be seen as a particular case of a cylindrical joint, only allowing relative rotation about its axis. As shown in Figure \ref{fig5}, it is required that the vectors $\bm{u}_i$, $\bm{v}_j$ of remain collinear, and the point $\bm{P}$ must have constant coordinates with respect to the $\xi_i\eta_i\zeta_i$ and $\xi_j\eta_j\zeta_j$ coordinate system. Therefore, the six constraint equations needed to define a revolute   joint are: $\bm{\Phi}^{(p1,2)}\equiv \tilde{\bm{u}}_i\bm{u}_j=0$ and $\bm{\Phi}^{(s,3)}\equiv \bm{r}_i+\bm{A}(\bm{p}_i)\bm{s}_i^{\prime P}-\bm{r}_j-\bm{A}(\bm{p}_j)\bm{s}_j^{\prime \bm{P}}=0$. And we name the vector $\bm{l}_i=\bm{u}_i/|\bm{u}_i|$ the joint axis.

\subsection{Propensity}\label{propensity}

\noindent Here we are interested in analysing the kinematic of open-chain multibody systems with strong local conformation propensities. For that the conformation propensity is coupled to the system geometric constrains as a energy constrain. The idea is to add to the system dynamic constraint a set of correction forces, such that the integrations moves the joint angles in the direction that most rapidly reduces the system energy. This process, based on the steepest descendent algorithm, impose a local propensity to the system by reducing its potential energy on its joints, i.e. the propensities are achieved by reaction forces acting on the linked bodies, adjusting the joint conformation to a  minimum for the system entropy.  This is done here using pure moments applied on the bodies.

In the literature of bimolecular dynamics and solid state physics, a great diversity of analytic defined potentials can be found for complex systems modelled as open chains mechanisms. The system energy is defined by the system degrees of freedom and the forces acting on its bodies are gradients extracted from the system energy landscape, usually named force fields. The approach proposed in this work can also be used in this context. However here the goal is assumes a given partial description for the system energy using geometrical projections, without an explicit analytic definition. Each projection defines a view of the system potential field, for a restricted number of its degrees of freedom. These views are pre-compiled and stored on a local conformation libraries, for a fixed parameter space discritization. All these different views of the system energy landscape can be integrated at runtime to define a system energy gradient, in each iteration, and used on the system kinematic analysis, to a minimize the system entropy. 

Let $\bm{E}$ be the implicit system energy, assumed to be a unknown real function form the system parametrization space. In this sense if the system has $n$ degrees of freedom given by a sequence of parameters $\bm{\Theta}=(\theta_1,\theta_2,\cdots,\theta_n)\in \Omega\subset \mathbb{R}^n$, $\bm{E}(\bm{\Theta})\in \mathbb{R}$, and a view for the system potential field $\nabla \bm{E}:\mathbb{R}^n\rightarrow \mathbb{R}^n$ is defined by the image of $(\bm{T}\circ \nabla \bm{E})(\Omega)$, with $\bm{T}$ a linear transformation $\bm{T}:\mathbb{R}^n\rightarrow \mathbb{R}^n$. 

A set of transformations $(\bm{T}_i)_{i=1,\cdots N}$ is regular if the sum of all involved linear transformations are regular, i.e. if $\sum_{i=1}^{N}\bm{T}_i$ is invertible. Here we must note that, if $(\bm{T}_i)_{i=1,\cdots, N}$ is a regular set of views defined by $N$ endomorphism in the vectorial space $\mathbb{R}^n$, then for every vector $\bm{v}\in \mathbb{R}^n$,
\begin{equation}
\bm{v}=(\sum_{i=1}^{N}\bm{T}_i)^{-1}(\sum_{i=1}^{N}\bm{T}_i(\bm{v})).
\end{equation}
To proof this result it is sufficient to fix a based in the involved vectorial field, and use the matricidal presentation of each transformation to show that $(\sum_{i=1}^{N}\bm{T}_i)(\bm{v})=
\sum_{i=1}^{N}\bm{T}_i(\bm{v})$.

When each view $\bm{T}_i$ in a regular set $(\bm{T}_i)_{i=1,\cdots, N}$ is a orthogonal projection represented, in the global coordinate system, by a diagonal matrix $\bm{A}_i$, then also $\sum_{i=1}^{N}\bm{T}_i$ is a diagonal  matrix $\bm{M}$. The potential field $\nabla \bm{E}$ in this case is defined by its projections $\bm{A}_i\nabla \bm{E}$, with $i=1,\cdots, N$. And can be recover from this projection, since $\bm{M}^{-1}\sum_{i=1}^{N}(\bm{A}_i\nabla \bm{E})(\bm{\Theta})=\nabla \bm{E}(\bm{\Theta})$, of every parametrization $\bm{\Theta}\in \Omega$. Here each $\bm{A}_i\nabla \bm{E}$ define a view and the set of this views constitute a conformation library, denoted by $(\bm{f}_i)_{i=1,\cdots, N}$ assumed to be a set of known non-linear functions.

For an open-chain system defined by revolute joints, having its relative degrees of freedom given by $n$ joint angles $\bm{\Theta}=(\theta_1,\theta_2,\cdots,\theta_n)\in \bm{\Omega}\subset \mathbb{R}^n$. Its potential field can be recover from a regular set of projections $(\bm{T}_{i,j})_{i,j=1,\cdots, n}$, and the associated conformation library $(\bm{f}_{i,j}(\Omega))_{i,j=1,\cdots, n}$, since the energy gradient at $\bm{\Theta}\in \Omega$, is given by $\nabla \bm{E}(\bm{\Theta})=(\sum_{i,j=1}^{N}\bm{T}_{i,j})^{-1}(\sum_{i,j=1}^{N}\bm{f}_{i,j}(\bm{\Theta}))$. Used to determine the contribution to the moment applied to bodies to minimize the system entropy. Assuming that joint of order $i$ links the two bodies of order $i$ and $i+1$, and that its conformation is described by the angle $\theta_i$. Denote as above by $\bm{w}_i$ and $\bm{w}_{i+1}$ the bodies rotational coordinates vectors, and $\bm{l}_{i,i+1}$ the axis of joint $i$. When
\begin{equation}
\nabla \bm{E}(\bm{\Theta})=(\sum_{i,j=1}^{N}\bm{T}_{i,j})^{-1}(\sum_{i,j=1}^{N}\bm{f}_{i,j}(\bm{\Theta}))=[\nu_1,\cdots,\nu_n]^T,
\end{equation}
its contribution to the moment applied to bodies $i$ and $i+1$, to impose the propensity in the joint $i$, describe by potential energy $\bm{E}$, is in the bodies $i$ given by
\begin{equation}
\bm{n}_i=-\nu_{i}\tilde{\bm{w}}_i\bm{l}_{i,i+1}/\|\bm{l}_{i,i+1}\|,
\end{equation}
 and in the body $i+1$ by
\begin{equation}
\bm{n}_{i+1}=\nu_{i}\tilde{\bm{w}}_{i+1}\bm{l}_{i,i+1}/\|\bm{l}_{i,i+1}\|.
 \end{equation}
Here the scalar $\nu_{i}$ is the component of order $i$ extracted from the gradient $\nabla \bm{E}(\bm{\Theta})$. The sum moment of all contributions applied to a body $i$ will be denotes by $\bm{n}_i^\ast$, and is the sum of all contribution to this body moments defined by the angles of adjacent joints. This directional gradient changes the system conformation in the direction of the greatest rate of decrease in the system energy $\bm{E}$.

\begin{figure}[h]
\begin{center}
\includegraphics[width=330pt]{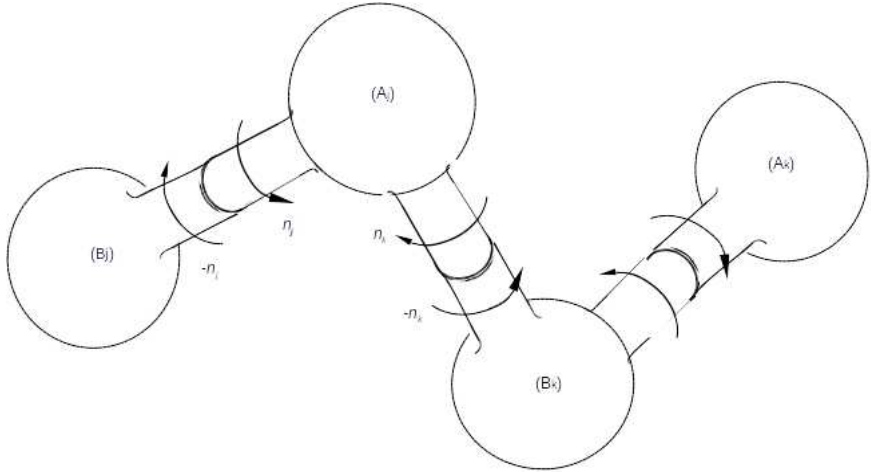}
\end{center}
\caption{Moments  between three bodies linked by two revolute joints.}\label{fig6}
\end{figure}

\subsection{Motion with kinematic constrains}
\noindent The set of all kinematic joints in the system can be represented as $m$ independent constrains, normally nonlinear equations in terms of $\bm{q}$, as \cite{Nikravesh 1988}
\begin{equation}\label{constrains}
\bm{\Phi}(\bm{q})=\bm{0}.
\end{equation}
If the system vector of coordinates $\bm{q}$ satisfies these equations, $\bm{q}$ is named an \textit{admissible conformation}.

Each kinematic joint introduces reaction forces,  $\bm{g}^{(c)}=[\bm{g}_1^{(c)},\cdots,\bm{g}_N^{(c)}]^T$, where $\bm{g}_i^{(c)}$, $i=1,\cdots,N$ is the vector of joint reaction forces acting on body $i$. Denoting the vector of propensity forces by $\bm{g}^{(p)}=[\bm{g}_1^{(p)},\cdots,\bm{g}_N^{(p)}]^T$, where $\bm{g}_i^{(p)}=[\bm{0},\bm{n}_i^\ast-\tilde{\bm{w}_i}\bm{J}_i\bm{w}_i]$, $i=1,\cdots,N$ is the vector of moments used to minimize the system entropy by acting on body $i$, defined by the kinematic joint propensities, and $\bm{J}_i$ is the body $3\times 3$ constant rotational inertia matrix. The equation of the motion for this type of system can be described by
\begin{equation}\label{motion1}
\bm{M}\ddot{\bm{q}}+\bm{b}=\bm{g}^{(c)}+\bm{g}^{(p)},
\end{equation}
where $\bm{M}=diag[\bm{M}_1,\cdots,\bm{M}_N]$ is the system mass matrix, and $\bm{b}=[\bm{b}_1,\cdots,\bm{b}_N]^T$ contains the quadratic velocity terms. Equations \ref{constrains} and \ref{motion1} together present the equation of motion for a system of constrained bodies. However the constraint force vector $\bm{g}^{(c)}$ can be expressed in terms of the constraint equations by $\bm{g}^{(c)}=\bm{\Phi}_q^T\bm{\lambda}$, where $\bm{\Phi}_q$ is the Jacobian matrix of the constrain equation.

The first and second time derivatives of the position constraint equations yield the kinematic velocity and acceleration equations, respectively,
\begin{equation}\label{velocity}
\dot{\bm{\Phi}}(\dot{\bm{q}},\bm{q})=0 \Leftrightarrow \bm{\Phi}_q\dot{\bm{q}}=-\bm{\Phi}_t,
\end{equation}
where  $\bm{\Phi}_t$  is the partial derivative of $\bm{\Phi}$ with respect to time. 
\begin{equation}\label{EqLambda}
\ddot{\bm{\Phi}}(\ddot{\bm{q}},\dot{\bm{q}},\bm{q})=0 \Leftrightarrow \bm{\Phi}_q\ddot{\bm{q}}=-(\bm{\Phi}_q\dot{\bm{q}})_q\dot{\bm{q}}
-2\bm{\Phi}_{qt}\dot{\bm{q}}-\bm{\Phi}_{tt}
\end{equation}
Therefore the Equation (\ref{motion1}) can be written as:
\begin{equation}\label{ConstEqMove}
\bm{M}\ddot{\bm{q}}+\bm{\Phi}_q^T\bm{\lambda} = \bm{g}^{(p)}. 
\end{equation}

Equations (\ref{EqLambda}) and (\ref{ConstEqMove}) form a system of differential-algebraic equations that must be solved together to obtain the accelerations $\ddot{\bm{q}}$ and Lagrange multipliers $\bm{\lambda}$:
\begin{equation}
\left[
  \begin{array}{cc}
    \bm{M}    & \bm{\Phi}^T_q\\
    \bm{\Phi}_q& \bm{O} \\
  \end{array}
\right] 
\left[
  \begin{array}{c}
    \ddot{\bm{q}}\\
    \bm{\lambda} \\
  \end{array}
\right] 
=
\left[
  \begin{array}{c}
    \bm{g}^{(p)}\\
    \bm{\gamma} \\
  \end{array}
\right]\label{EqMatrMoveExt}
\end{equation}
where $\bm{\gamma}=-(\bm{\Phi}_q\dot{\bm{q}})_q\dot{\bm{q}}-2\bm{\Phi}_{qt}
\dot{\bm{q}}-\bm{\Phi}_{tt}$,
for the numerical solution of these equations we used here the Direct Integration Algorithm (DIA) described in \cite{Nikravesh 1988} and \cite{Flores 2004}. The DIA can be summarized by the following steps:
\begin{enumerate}
\item Give initial conditions for position and velocity $\bm{q}_0$.
\item Evaluate the Jacobian matrix $\bm{\Phi}_{\bm{q}}$, construct vector $\bm{\Phi}$, determine the right-hand-side of acceleration $\bm{\gamma}$, and calculate the forces $\bm{g}^{(p)}$.
\item Solve the linear system of motion (\ref{EqMatrMoveExt}) in order to obtain the accelerations $\ddot{q}$ at time $t$ and the Lagrange multipliers $\lambda$.
\item Assemble an auxiliary vector $\dot{y}_t$ containing velocities and accelerations for instant $t$.
\item Integrate numerically $\dot{\bm{q}}$ and $\ddot{\bm{q}}$ for time $t+\Delta t$ and obtain the new position and velocities.
\item Update the time and go to step 2) and proceed with the process for the new time step. Perform these until time of analysis is reached.
\end{enumerate}
The DIM is highly prone to integration errors, therefore the method requires constrain stabilization. It should be note that this process is quite sensitive to initial conditions, which can be other source of errors. 

Although the Euler parameter are ideal for representing angular transformations of a body in space, and avoid singularities on the integration, they yield to many equations when their time derivatives are used explicitly in the equation of motion. For the integration each body is described by its global coordinates, moment using Euler parameter, velocity and angular velocity, such that $\bm{q}_i=[\bm{r}_i,\bm{p}_i,\dot{\bm{r}}_i,\dot{\bm{w}}_i]$. However, to accelerate the resolution of the linear system (\ref{EqMatrMoveExt}), the Euler parameter are transformed on its correspondent rotation angles, defining $\bm{q}_i=[\bm{r}_i,\bm{w}_i,\dot{\bm{r}}_i,\dot{\bm{w}}_i]$, and using  $\dot{\bm{p}}_i=\frac{1}{2}\bm{G}^T\bm{w}_i$ \cite{Nikravesh 1988}.

Using this angular representation, the first time derivative of a position constrain yield the corresponding velocity constraint. For the two fundamental constraints, used on our joints, the velocity constraints are:
\begin{equation}\label{E1}
\dot{\bm{\Phi}}^{(p1,1)}\equiv -\bm{s}^T_j\tilde{\bm{s}}_i\bm{w}_i-\bm{s}^T_i\tilde{\bm{s}}_j\bm{w}_j=0
\end{equation}
\begin{equation}\label{E2}
\dot{\bm{\Phi}}^{(p1,2)}\equiv -\bm{d}^T\tilde{\bm{s}}_i\bm{w}_i+\bm{s}^T_i(\dot{\bm{r}}_j
-\tilde{\bm{s}}_j^B\bm{w}_j- \dot{\bm{r}}_i+\tilde{\bm{s}}_i^B\bm{w}_i)=0
\end{equation}
\begin{equation}\label{E3}
\dot{\bm{\Phi}}^{(s,3)}\equiv \dot{\bm{r}}_i-\tilde{\bm{s}}_i^B\bm{w}_i-
 \dot{\bm{r}}_j+\tilde{\bm{s}}_j^B\bm{w}_j=0
\end{equation}
We consider a matrix of velocities for the two bodies constraining the translations and rotational velocities for bodies $i$ and $j$ in this order. Each velocity equation, whether it contains one or several algebraic equations, can be described as $\bm{D}\bm{q}=\bm{D}_i\bm{q}_i+\bm{D}_j\bm{q}_j=0$, where $\bm{q}_i$ and $\bm{q}_j$ each contain six velocity components, and $\bm{D}_i$ and $\bm{D}_j$ are the corresponding sub-Jacobian. The time derivative of a velocity constrain yields the corresponding acceleration constraint. The acceleration constrain for a joint can be expressed as $\bm{D}\dot{\bm{q}}+\dot{\bm{D}}\bm{q}=0$. Note that, acceleration constraints contain quadratic terms that are moved to the right-hand-side of the equation $\bm{\gamma}=-\dot{\bm{D}}\bm{q}$. For the constrains defined by (\ref{E1}), (\ref{E2}) and (\ref{E3}), they yield respectively\cite{Nikravesh 1988}: 
\begin{equation}
\bm{s}_i^T\tilde{\bm{w}}_j\dot{\bm{s}}_j+\bm{s}_j^T\tilde{\bm{w}}_i
\dot{\bm{s}}_i
+2\dot{\bm{s}}^T_j\dot{\bm{s}}_i=-\bm{s}_j^T\tilde{\bm{s}}_i\bm{w}_i-
\bm{s}_i^T\tilde{\bm{s}}_i\bm{w}_j
\end{equation}
\begin{equation}
\bm{d}^T\tilde{\bm{w}}_i\dot{\bm{s}}_i+\bm{s}_i^T(\tilde{\bm{w}}_j
\dot{\bm{s}}^B_j-
\tilde{\bm{w}}_i\dot{\bm{s}}^B_i)
+2\dot{\bm{d}}^T\dot{\bm{s}}_i=-\bm{s}_i^T\dot{\bm{r}}_i-
\bm{s}_j^T\tilde{\bm{s}}_j\bm{w}_i+\bm{s}_i^T\dot{\bm{r}}_j-
\bm{s}_i^T\tilde{\bm{s}}_j\bm{w}_j
\end{equation}
\begin{equation}
-\tilde{\dot{\bm{s}}}_i^B\bm{w}_i+\tilde{\dot{\bm{s}}}_j^B\bm{w}_j=
\dot{\bm{r}}-\tilde{\bm{s}}_i^B\bm{w}_i-\dot{\bm{r}}
+\tilde{\bm{s}}_j^B\bm{w}_j
\end{equation}

\subsection{Motion with kinematic propensity}
\noindent We must note however that if the system is defined by an open chains of rigid bodies linked together using revolute joints, every system configuration can be described by the system joint angles. A system  with $N$ bodies, have in this case, $N-1$ relative degrees of freedom defined by number of dihedral angles in each joint. When the system is describe using cylindrical joints, instead of revolute joints, the number of relative degree of freedom becomes $2(N-1)$. Having its conformation described by a vector
\[\bm{\Theta}=(d_{1,2},\theta_{1,2},d_{2,3},\theta_{2,3},\cdots,d_{N-1,N},\theta_{N-1,N})
\]
where $d_{i,i+1}$ is the joint length between bodies $i$ and $i+1$, and $\theta_{i,i+1}$ is the correspondent dihedral angle. For that we assumed the joint length fluctuations are undergo Gaussian-distributed fluctuations about their mean position, modelled here, as usual, by a spring defined between two point, on the joint axis, one in the body $i$ and the other on the body $i+1$. For points $\bm{B}_i$ and $\bm{A}_j$ (see Figure \ref{fig6}) the spring force magnitude is given by $f^{(s)}_{ij}=k_{ij}(l_{ij}-l_{ij}^{(0)})$, where $l_{ij}^{(0)}$ is the spring undeformed length and $k_{ij}$ its stiffness. For our case of study this values can be found on Table \ref{Stiffness}. The force of the element on body $i$ and body $j$ is determined as $\bm{f}_{ij}=f^{(s)}_{ij}(\bm{r}^A-\bm{r}^B)/|\bm{r}^A-\bm{r}^B|$ and $\bm{f}_{ji}=-f^{(s)}_{ij}(\bm{r}^A-\bm{r}^B)/|\bm{r}^A-\bm{r}^B|$. Since this force have moments arms, when we used them, their moments must be include in the motion equation as well, $\bm{n}_i=\tilde{\bm{s}}_i^{B}\bm{f}_{ij}$ and $\bm{n}_j=\tilde{\bm{s}}_j^{A}\bm{f}_{ji}$. The linear forces and moments generated the set of cylindrical joints, applies to the system bodies are denoted by the time dependent vector $\bm{g}^{(s)}$.

For an open-chains of bodies linked together using cylindrical joints, with conformation propensity, and axial linear fluctuation and Brownian random forces, (\ref{EqMatrMoveExt}) is rewriting as:
\begin{equation}\label{EqMatrMoveExt2}
\left[
  \begin{array}{cc}
    \bm{M}    & \bm{\Phi}^T_q\\
    \bm{\Phi}_q& O \\
  \end{array}
\right] 
\left[
  \begin{array}{c}
    \ddot{\bm{q}}\\
    \bm{\lambda} \\
  \end{array}
\right] 
=
\left[
  \begin{array}{c}
    \bm{g}^{(p)}+\bm{g}^{(s)}+\bm{W}-\rho\dot{\bm{q}}\\
    \bm{\gamma} \\
  \end{array}
\right]
\end{equation}
where $\bm{g}^{(p)}$ is the vector of moments generated by joint propensity and $\bm{g}^{(s)}$ is the vector of linear forces and moments generated by axial fluctuations on the joints.
The friction coefficient $\rho$ is here used to characterize the contra reaction to the motion imposed to the system surrounding, to maintain the system in constant thermal fluctuation a random external force $\bm{W}=\sqrt{2\rho \beta^{-1}}\eta(t)$ that is Gaussian and white was added. Here $\beta=1/k_bT$, the inverse temperature, the vector $\eta(t)$ is a white-noise.

We are interested in characterizing the mechanism of transition between conformations of this type of systems, particularly when the joint propensity library is defined by probabilistic distribution with several saddle points. Example of those are the libraries generated from data of amino acid propensities described above. In the simplest situations, when the energy landscape is smooth and the system conformations are separated by a few isolated barriers, this is usually done by identifying conformation, which are saddle points on the free energy landscape. The most probable path for the transition is the so-called minimum energy path. The dynamic defined by (\ref{EqMatrMoveExt2}) is characterized by long waits periods around this saddle conformations followed by sudden jumps, produced here by the added random external force. Usually this changes occur with lower probability and on a time scale larger than the time scale used to simulate the system dynamics. Because of the wide separation of time scales, instead of considering this trajectory as a dynamic problem parametrized by the physical time, in the following sections the transition paths is viewed as a kinematic problem in the in configuration space. Figure \ref{fig11-4} presents the variation on the energy and position for a system with two stable states, having its free energy defined, by the amino acid PHE van Mises propensity potential, as shown in Figure \ref{fig10} and described by a system with three bodies, linked by two revolute joints. Most of the time the reaction coordinate fluctuate near the stable states. Rarely, on the time scale of the stable fluctuations, the system switches between metastates.
\subsection{Open chain conformation sampling}
\noindent The solution to the motion equation of a given system can be interpreted at a point moving in state-variable space. The necessary conditions for an admissible solution is that the point moves along the commune intersection of all the constrain surfaces in the phase space. The energy constrains can be used to impose a bias to this motion. However this bias and  the numerical errors produce disturbances that send the point away of the commune intersection. 

Here $\bm{q}=[\bm{q}_1,\bm{q}_2,\cdots \bm{q}_N]^T$ is the vector of coordinates for the system, with $N$ bodies, and $$\bm{\Phi}(q)=[\bm{\Phi}_1(\bm{q}),\bm{\Phi}_2(\bm{q}),\cdots,\bm{\Phi}_m(\bm{q})]^T$$ is a vector of constraint functions ($m<N$), used to describe the system kinematic constraints. This constrains restrict possible motions of the system to a $N-m$ dimensional configuration manifold
\[
Q=\{\bm{q}:\;\bm{\Phi}(\bm{q})=0\},
\]
describing the commune intersection of all the constrain surfaces in the phase space.
The configuration manifold coincides with the number of degrees of freedom of the system. The constraints give rise to constraint forces of the form
\[
R(\bm{q},\bm{\lambda})=\sum_{j=1}^m \lambda_j\nabla \bm{\Phi}_j(\bm{q}) = \bm{\Phi}_{\bm{q}}\bm{\lambda},
\]
here $\lambda_j$ is the Lagrange multiplier for constraint equation $j=1,\cdots,m$ and evaluated from  (\ref{EqMatrMoveExt2}).
For an open-chain multibody system, for the right selection of constraint equations, the constraint gradients $\nabla \bm{\Phi}_i(\bm{q})$ are linear independent, i.e. $\bm{\Phi}_{\bm{q}}$ is full rank $m$, for every $\bm{\Theta}\in \Omega$. 

Each multibody trajectory in $Q$ can be represented by sequence of conformation states in $\Omega$, defined by all the possible system parametrization for each degrees of freedom,
\[
\bm{\Theta}_0,\bm{\Theta}_{\Delta t},\bm{\Theta}_{2\Delta t},\cdots,\bm{\Theta}_{T},
\]
here each $\bm{\Theta}_t$, $t=1,\cdots,T$ must be seen as a system micro-sate, describe the complete state of the system or conformation at time $t$ in the phase space $Q$, describe by a vector  \[\bm{\Theta}_t=(d^t_{1,2},\theta^t_{1,2},d^t_{2,3},\theta^t_{2,3},\cdots,d^t_{N-1,N},\theta^t_{N-1,N})\in \Omega.
\] 

Since the underlying dynamics in the commune intersection of all the constrain surfaces in the phase space is assumed to be Markovian, the probability density for observing a particular trajectory can be written as the product of the distribution $\rho(\bm{\Theta}_0)$ of the initial micro-state $\bm{\Theta}_0$ with the product of all single time step transitions probabilities
\begin{equation}
P[\bm{\Theta}_0,T]=\rho(\bm{\Theta}_0)\prod_{i=0}^{T/(\Delta t -1)}\rho(\bm{\Theta}_{i\Delta t}\rightarrow \bm{\Theta}_{(i+1)\Delta t}). \label{ProbDist}
\end{equation}
Naturally, the form of the single time step transition probabilities
\[
\rho(\bm{\Theta}_{i\Delta t}\rightarrow \bm{\Theta}_{(i+1)\Delta t}),
\]
depends on the system dynamics. The dynamic in (\ref{EqMatrMoveExt2}) is known to be ergodic with respect to the Boltzmann-Gibbs probability density function
\begin{equation}\label{Boltzmanndensity}
\rho(\bm{\Theta})=\frac{1}{\int_\Omega e^{-\beta E(\bm{\theta})}d\bm{\theta}}e^{-\beta E(\bm{\Theta})}.
\end{equation}

If we assume that (\ref{EqMatrMoveExt2}) is metastable over a family of sets $(A_i)_{i\in J}$ defined by micro-states in the conformation space $\Omega$, or section on the the commune intersection of all the constrain surfaces. The volume of each set o micro-sates $A_i\subset \Omega$ is relatively small, and yet the probability to find the system inside one of these sets is close to one:
\begin{equation}
1\approx\frac{1}{\int_\Omega e^{-\beta E(\bm{\theta})}d\bm{\theta}}\sum_{\bm{\Theta}\in\bigcup_iA_i}e^{-\beta E(\bm{\Theta})}
\end{equation}
By ergodicity, transitions between these sets must occur. In this context our main goal is to understand how they occur without running any dynamical trajectory. For that in this context is usual to call \textit{reaction coordinate} to part of a trajectory connecting conformation belonging to two of this sets, in the configuration manifold $Q$.
 
\section{Minimum energy path}\label{IV}
\noindent The idea behind the current techniques for finding minimum energy paths for complex systems is to introduce some bias on the dynamics to enhance the probability to observe desired reactive trajectories. 

As is well known, the best reaction coordinate to describe this bias, is the \textit{committor function} which, at any point in the configuration manifold $Q$, defined by probability that a trajectory initiated at this state will reach first the goal state rather than the initial state, under the system dynamics in the commune intersection of all the constrain surfaces \cite{Weinan 2005}. For two meta-states $A$ and $B$, the committor $\Upsilon(\Theta)$ of a micro-state $\Theta$ can be computed by initializing a large number of trajectories from $\Theta$ and recording how many arrive in $B$, without use states in $A$. However can be shown that the committor can be identified directly, without sampling actual reactive trajectories. For that we must note that the committor function $\Upsilon$, satisfies the backward constrained Kolmogorov equation
\begin{equation}
-\nabla E.\nabla \Upsilon + \beta^{-1}\Delta \Upsilon = 0,,\;\Upsilon|_{\Theta\in A}=0,\;\Upsilon|_{\Theta\in B}=1
\end{equation}
However also this partial differential equation is way too complex to be solved, in the configuration manifold $Q$, by traditional numerical methods \cite{Weinan 2005}.

Let $\alpha\in [0,1]$, the isosurface $S_\alpha$ of level $\alpha$ is the set of micro-states with have a uniform probability $\alpha$, i.e.
  $$S_\alpha=\{\Theta\in \Omega\;:\;\Upsilon(\Theta)=\alpha\}.$$   
These surfaces can be approximated locally by a family of planes, and represented by a parameterized curve in the configuration space $\vartheta(\alpha)\subset\Omega$, with $\alpha\in [0,1]$, such that:
\begin{enumerate}
\item the plane $P_\alpha$ indexed by $\alpha$, contains a micro-state $\Theta=\vartheta(\alpha)$,
\item the unit normal vector to $P_\alpha$ is $\hat{g}(\alpha)$, and
\item the micro-state $\Theta=\vartheta(\alpha)$ is the mean position in the plane $P_\alpha$ with respect to the equilibrium density (\ref{Boltzmanndensity}) restricted to $P_\alpha$, i.e.
\begin{equation}
\vartheta(\alpha)=\frac{1}{\int_{P_\alpha} e^{-\beta E(\bm{\theta})}d\bm{\theta}}\int_{P_\alpha} \bm{\theta} e^{-\beta E(\bm{\theta})}d\bm{\theta}.
\end{equation}
\end{enumerate}

The sequence of planes $P_\alpha$ is such that the mean position within these planes form a curve $\vartheta$ which is everywhere perpendicular to the planes. In fact is the MEP meta-states $A$ and $B$. The definition of the MEP is quite simple and natural, but it is more complicated to give a full justification of why the MEP is relevant to understand the mechanism of a reaction and, in particular, to determine the isosurface $S_\alpha$. 

A MEP is a path which connects two minima of $E$ via a saddle point and corresponds to the steepest descent path on $E$ from this saddle point. In other words, a MEP is composed of two orbits
connecting a saddle point to two minima by the steepest descent dynamics $\dot{\bm{\Theta}}=-\nabla E(\bm{\Theta})$, but impossible of full fill using the system dynamic. It is useful to change perspective, to look at the MEP as a curve in configuration
space and represent it using a different parametrization than
time. This is done by transform a time dependent dynamic problem in a steady kinematic problem. To understand why it is advantageous to do so, consider a distinguished reaction coordinate as a parametrization for  $\vartheta$ defining conformations for an open chain system, like the one presented on Figure \ref{fig3}, with $N$ bodies $b_1,\cdots,b_N$. The smooth curve parametrization should be seen as a map $$\vartheta:[0,1]\rightarrow \Omega,$$ where if the system is defined using cylindrical joints, the parametrization space $\Omega\subset\mathbb{R}^{2(N-1)}$  and $\vartheta$ assigning to each value in $\alpha\in[0,1]$ a list of joint axis lengths and joint angles, on the system parametrization space $$\bm{\Theta}_{\alpha}=(d_{1,2}(\alpha),\theta_{1,2}(\alpha),d_{2,3}(\alpha),\theta_{2,3}(\alpha),\cdots,d_{N-1,N}(\alpha),\theta_{N-1,N}(\alpha))\in \Omega.$$ Note that, $\vartheta(0)$ and $\vartheta(1)$ can be seen as the initial and final conformation on the system trajectory $\vartheta$, respectively, the meta-states $A$ and $B$. Each value of $\vartheta(\alpha)$ defines a system conformation, given by setting a value to each of its joint angles. The constraint equation represented by Equation (\ref{constrains}) are, in general, non-linear in terms of $q(\vartheta)$ and in general can be solved by employing the Newton-Raphson method. Here however, when the only degrees of freedom are defined only by joint angles, its conformation can be determined using only geometric transformations. In this case each parametrization $\bm{\Theta}_\alpha=\vartheta(\alpha)$ defines a system state given, for each joint angle $\beta_i$, with $i=1,\cdots,N-1$, by rotating the bodies $b_{i+1}$ to $b_{N}$ having by axis the joint axis between bodies $b_{i}$ and $b_{i+1}$. This procedure is particularly useful in the MEP initial approximation, as described bellow.

Since by definition the force $-\nabla E$ must be everywhere tangent to the MEP, if $\vartheta$ is a MEP we must have
\begin{equation}\label{MEPCond}
\frac{d\alpha(\lambda)}{d\lambda}\text{ parallel to } \frac{\partial E(\vartheta(\lambda))}{\partial \alpha}
\end{equation}
If the MEP connects the two minima of $E(\bm{\Theta})$ located at $\bm{\Theta}_A$
and $\bm{\Theta}_B$ , (\ref{MEPCond}) must be supplemented by the boundary conditions $\vartheta(0)=\bm{\Theta}_A$ and $\vartheta(1)=\bm{\Theta}_B$. Notice that this implies that
the component of the force perpendicular to the curve $\vartheta$ is
zero everywhere along the curve.

\subsection{Nudged-Elastic Band method}
\noindent For determining the MEP in this work we used the NEB method. 
In the NEB method a string of replicas of the initial system are created and connected together with springs in such a way as to form a discrete representation of the path conformation from a reactant conformation $\bm{q}(\bm{\Theta}_1)$ to a product conformation $\bm{q}(\bm{\Theta}_r)$.
Here this springs couple together only angles of the same joint in different replicas. This defined a band of springs between each replica. In this sense, since an open-chain system with $N$ bodies are linked by $N-1$ joints, for a MEP discretization defined by a string of $r$ replicas, each replica is coupled to the next  using $N-1$ springs. The new system defined by all this replicas, have $r\times N$ bodies and $(N-1)\times (N-1)$ spring. Each conformation of this new systems describe a node approximation to the MEP for the the original system in different time steps, in this sense it is usually called a \textit{frame}. Those conformations are relaxed to the MEP, by the NEB method, through a force projection scheme in which potential forces act perpendicular to the band, and spring forces act along the band \cite{Henkelman 2000}. 

A NEB calculation is started from an initial conformation for the extended system, for that a pathway $\vartheta_0$ connecting initial state $\bm{\Theta}_1$ and final states $\bm{\Theta}_r$ are generate using a convex combination.  Note that, for every $\alpha_j\in [0,1]$, each convex combination $\vartheta_0(\alpha_j)=\alpha_j \bm{\Theta}_1 +(1-\alpha_j)\bm{\Theta}_r$ describes a conformation in the commune intersection of all the constrained surfaces, i.e. such that $\bm{\Phi}(\bm{q}(\vartheta_0(\alpha_j)))=0$. A trajectory, discretized by $r$ replicas, is defined by a list of parametrizations $$\bm{\Theta}=[\bm{\Theta}_1,\dots,\bm{\Theta}_j,\dots,\bm{\Theta}_r],$$ where the end-points are fixed, but the internal conformations $\bm{\Theta}_j$ are adjusted. Initially the list $\bm{\Theta}$ is initialized as $\bm{\Theta}_j=\vartheta_0(\alpha_j)$, with $\alpha_j=(j-1)/(r-1)$, for all $j=1,\cdots,r$.  For the finding of the system state in the phase space, needed to initialize the multibody kinematic analysis, we use here geometric transformation along the path.

To optimize the extended system conformation the objective function is defined as in \cite{Henkelman 2000} by
\begin{equation}
S(\bm{\Theta})=\sum_{j=2}^{r-1}E(\bm{\Theta}_j)+\sum_{j=1}^{r}
\sum_{i=1}^{N-1}\frac{k}{2}(\|\bm{\Theta}_{j+1,i}-\bm{\Theta}_{j,i}\|^2-\|\bm{\Theta}_{j,i}-\bm{\Theta}_{j-1,i}\|^2),
\end{equation}
where $E(\bm{\Theta}_j)$ is original system free energy at $\bm{q}(\bm{\Theta}_j)$ and $\bm{\Theta}_{j,i}$ in the $i$-component in the parametrization vector $\bm{\Theta}_{j}$, i.e. the angle in the joint of order $i$ in the replica of order $j$.  
The function $S(\bm{\Theta})$ is minimize with respect to the set of conformations in $\Omega$. This models an elastic band made up to $r$ beads and $r-1$ sets of $N-1$ springs for each body in the system.

To make use of the projection method, the tangent along the path, in the replica $j$, is defined by a vector $\bm{g}_j$ to the higher energy neighboring conformation. The tangent to conformation $\bm{\Theta}_j$, is given in \cite{Henkelman 2000} and \cite{Henkelman2 2000} by
\begin{equation}
\bm{g}_j=\left\{
  \begin{array}{rcl}
    \bm{g}^+_j & \text{ if } & E(\bm{\Theta}_{j+1})>E(\bm{\Theta}_{j})>E(\bm{\Theta}_{j-1}) \\
    \bm{g}^-_j & \text{ if } & E(\bm{\Theta}_{j+1})<E(\bm{\Theta}_{j})<E(\bm{\Theta}_{j-1}) \\
  \end{array}
\right.
\end{equation}
where
\[
\bm{g}^+_j=\bm{\Theta}_{j+1}-\bm{\Theta}_{j} \text{ and } \bm{g}^-_j=\bm{\Theta}_{j}-\bm{\Theta}_{j-1}.
\]
If adjacent configurations are either lower in energy, or both are higher in energy than configurations $j$, the tangent is taken to be a weighted average of the vectors to the two neighboring configurations. The weight is determined from the energy. The weighted average is only used at extrema along the MEP, allowing a smoothly switch between the two possible tangents $\bm{g}^+_j$ and $\bm{g}^-_j$. Otherwise, there is a discontinuity in the tangent as $j$ one conformation becomes higher in energy than another. If conformation $\bm{\Theta}_{j}$ is a minimum $E(\bm{\Theta}_{j+1})>E(\bm{\Theta}_{j})<E(\bm{\Theta}_{j-1})$ or at maximum $E(\bm{\Theta}_{j+1})<E(\bm{\Theta}_{j})>E(\bm{\Theta}_{j-1})$ the tangent estimate becomes
\begin{equation}
\bm{g}_j=\left\{
  \begin{array}{rcl}
    \bm{g}^+_j\nabla E_j^{\max}+\bm{g}^-_j\nabla E_j^{\min}\ & \text{ if } & E(\bm{\Theta}_{j+1})>E(\bm{\Theta}_{j-1}) \\
    \bm{g}^+_j\nabla E_j^{\min}+\bm{g}^-_j\nabla E_j^{\max} & \text{ if } & E(\bm{\Theta}_{j+1})<E(\bm{\Theta}_{j-1}) \\
  \end{array}
\right.
\end{equation}
where $$\nabla E_j^{\max}=\max(\|E(\bm{\Theta}_{j+1})-E(\bm{\Theta}_{j})\|,\|E(\bm{\Theta}_{j})-E(\bm{\Theta}_{j-1})\|),$$ and
$$\nabla E_j^{\min}=\min(\|E(\bm{\Theta}_{j+1})-E(\bm{\Theta}_{j})\|,\|E(\bm{\Theta}_{j})-E(\bm{\Theta}_{j-1})\|).$$
The tangent normalization is denoted bellow as $\hat{\bm{g}}_j=\frac{\bm{g}_j}{\|\bm{g}_j\|}$. With this approximation the elastic band is well behaved and its relaxation converges to the MEP if sufficient number of replicas are used in the band \cite{Henkelman2 2000} and if the energy landscape is sufficiently smooth.

Let $\nabla E(\bm{\Theta}_j)$ be the gradient of the energy with respect to conformation $\bm{\Theta}_j$ and denoting by $\bm{F}^s_j$ the spring force  acting in it. The force on each conformation should only contain the parallel components of the spring forces, and perpendicular components to the true force. Therefore, since the perpendicular component of the gradient is
\begin{equation}
\nabla E(\bm{\Theta}_j)_\bot=\nabla E(\bm{\Theta}_j)-(\nabla E(\bm{\Theta}_j).\hat{\bm{g}}_j)\hat{\bm{g}}_j,
\end{equation}
the force on each acting on conformation $\bm{\Theta}_j$ is
\begin{equation}
\bm{F}_j=-\nabla E(\bm{\Theta}_j)_\bot+(\bm{F}^s_j.\hat{\bm{g}}_j)\hat{\bm{g}}_j. \label{MEPequation}
\end{equation}
These decouple the dynamics of the path itself from the particular distribution of conformations chosen in the initial approximation to the path. And when $\nabla E(\bm{\Theta}_j)_\bot=0$ the spring force don't change the configuration parameters relaxation, the spring forces only affects the distribution of the parameters within the path. The decoupling of the relaxation system conformation and the chosen discretization for the path is essential to ensure convergence to the MEP and its proof can be found in \cite{Henkelman 2000}, and this convergence is independent from the springs stiffness $k$.

The parallel components of the spring forces, applied to restrain the separations between adjacent parametrizations, are calculated in our implementation as
\begin{equation}
\bm{F}_j^s=k(\|\bm{\Theta}_{j+1}-\bm{\Theta}_j\|-\|\bm{\Theta}_j-\bm{\Theta}_{j-1}\|)\hat{\bm{g}}_j,
\end{equation}
where $k$ is the stiffness for the strings coupling the three replicas. The initial path  optimization can be performed using the described DIA, by solving the motion equation for the extend multibody system.

\section{Kinematic analysis as a steady problem }\label{V}
\noindent For an open-chain mechanism having $N$ bodies $b_1,b_2,\dots,b_N$, we analyse its kinematic, between two stationery conformation, as a steady problem. For that we used an unique multibody system having $r\times N$ bodies $$b_{1,1},b_{1,2},\dots,b_{1,N},\dots,b_{r,1},b_{r,2},\dots,b_{r,N},$$ here $r$ is the number of time steps used on the analysis, defining a size for the time discretization or the number of replicas defining the elastic band. In the sense used in the previous section each cluster $b_{j,1},b_{j,2},\dots,b_{j,N}$ define a replica of the original system, possibly with different floating frame orientation, but having the same kinematic constraints. Those orientation are described using a list of parametrizations $\bm{\Theta}=[\bm{\Theta}_1,\dots,\bm{\Theta}_j,\dots,\bm{\Theta}_r]^T$, such that $\bm{\Theta}_j$  defining the conformations on  cluster $j$, by setting angles for each of its joints, $\bm{\Theta}_j=[\beta_{j,1},\cdots,\beta_{j,N-1}]^T$. For that each angle $\beta_{j,i}$, in the cluster $j$, is coupled to the angles $\beta_{j+1,i}$ and $\beta_{j-1,i}$, in the cluster $j+1$ and $j-1$, respectively, using a springs. The set of all the springs can be seen as an elastic band coupling cluster angular degrees of freedom. 

The motion of such band is given extending the linear system (\ref{EqMatrMoveExt2}) to $r$ replicas, coupling those replicas using the forces generated by NEB method:
\begin{equation}
\left[
  \begin{array}{cc}
    \bm{M}^\ast    & \bm{\Phi}^{\ast T}_q\\
    \bm{\Phi}^\ast_q& \bm{O} \\
  \end{array}
\right] 
\left[
  \begin{array}{c}
    \ddot{\bm{q}}\\
    \bm{\lambda} \\
  \end{array}
\right] 
=
\left[
  \begin{array}{c}
    \bm{g}_\ast^{(s)}+\bm{g}_\ast^{(b)}\\
    \bm{\gamma}_\ast \\
  \end{array}
\right]\label{EqMatrMoveExt3}
\end{equation}
In the system $\bm{M}^\ast=diag(\bm{M},\cdots,\bm{M})$ is the constant mass matrix for the extended system defined for each replicas, defined as a block diagonal matrix formed by $r$ copies of $\bm{M}$, the mass matrix of the original system. The dynamic constrains are iteration dependent and coded in $\bm{\Phi}^\ast_q=diag(\bm{\Phi}^1_q,\cdots,\bm{\Phi}^j_q,\cdots,\bm{\Phi}^r_q)$, where $\bm{\Phi}^j_q$ describes the constrains for replica of order $j$, changing during the optimization process, for all $1\leq j\leq r$. Vectors $\bm{g}_\ast^{(b)}$ describe the net of elastic band forces applied to each body. When the system is defined by cylindrical joints, the vector $\bm{g}_\ast^{(s)}$ describe the net of spring forces in each joints applied to each body in each replica, otherwise it is set to zero. 

It is quit simple the application of multibody dynamic simulation, based for instance in the DIM, to solve this kinematic type of problem. Whoever this strategy increases dramatically the system size, and its energy and the gradients need to be evaluated for each conformation in the elastic band using some description of the the energetics of the system. Moreover, for each two adjacent conformations we must estimate the local tangent to the path, project out the perpendicular component to the gradient and add the parallel component of the spring force. At each iteration the coordinates and velocities are updated from the coupled first order equation of motion based on the force evaluated at the current coordinates. If the angular velocities are zeroed at each step, the algorithm gives a steepest descent minimization. 

\subsection{Numeric Aspect in Elastic band dynamics}

\noindent The integration of the velocities and accelerations of the multibody system introduce numerical errors in the new positions and velocities obtained. These errors are due to truncation and because the system of motion equations do not use explicitly the position and velocity associates with the kinematic constrains. Consequently the original constrain equations are increasingly violated due to the the instability of the integration process. To mitigate this problems we tested two stabilization methods: the Baumgarte Stabilization Method and the Augmented Lagrangian Formulation. 

Baumgarte's method is based on feedback control and is known to be very hard to parametrize, without control of the error allowed to the simulation. It damps out the acceleration constraint violations by feeding back the violation of the position and velocity constraints. When the Baumgarte stabilization technique is used (\ref{EqMatrMoveExt4}) is modified to,
\begin{equation}
\left[
  \begin{array}{cc}
    \bm{M}^\ast    & \bm{\Phi}^{\ast T}_q\\
    \bm{\Phi}^\ast_q& \bm{O} \\
  \end{array}
\right] 
\left[
  \begin{array}{c}
    \ddot{\bm{q}}\\
    \bm{\lambda} \\
  \end{array}
\right] 
=
\left[
  \begin{array}{c}
    \bm{g}_\ast^{(s)}+\bm{g}_\ast^{(b)}\\
    \bm{\gamma}_\ast -2\alpha\dot{\bm{\Phi}}^\ast-\beta^2\bm{\Phi}\\
  \end{array}
\right]\label{EqMatrMoveExt4}
\end{equation}
where $\alpha$ and $\beta$ are positive constants that represent the feedback control parameters for the velocity and position constraint violation. Note however that is well known the importance of an adequate selection of values of $\alpha$ and $\beta$ to keep under control the constraint violations. There is a lack of a criteria in the choice of this parameter. Here we only used the stabilizations values $\alpha=\beta=5$, due to the size of the involved systems it is not practical perform numerical experiments to test this parameters.

The Augmented Lagrangian Formulation is a methodology that penalizes the constraint violations, much in the same form as the Baumgarte's \cite{Neto 2003}.
It is based on Hamilton's principle and the constraint equation are taken into account using a penalty approach, allowing a control over the maximum error allowed during simulation. The method consists in solving the system (\ref{EqMatrMoveExt3}), using an iterative process.
The evaluation of the system accelerations in a given time step starts as 
\[
\bm{M}^\ast\ddot{\bm{q}}^\ast_0=\bm{g}_\ast
\]
here $\bm{g}_\ast=\bm{g}_\ast^{(s)}+\bm{g}_\ast^{(b)}$.
The iterative process to evaluate the system accelerations proceeds with the evaluation of
\[
\bar{M}^\ast\ddot{\bm{q}}^\ast_{i+1}=\bar{\bm{g}}_\ast
\]
where the generalized mass matrix $\bar{\bm{M}}^\ast$ and the generalized force vector $\bar{\bm{g}}_\ast$ are given by
\[
\bar{\bm{M}}^\ast=\bm{M}^\ast+\alpha(\bm{\Phi}^\ast_q)^T \bm{\Phi}^\ast_q
\]
\[
\bar{\bm{g}}_\ast=\bar{\bm{M}}^\ast\ddot{\bm{q}}^\ast_{i}+(\bm{\Phi}^\ast_q)^T\alpha
(\bm{\gamma}_\ast-2\omega\beta\bm{\Phi}^\ast_q\ddot{\bm{q}}^\ast_{i}
-\omega^2\bm{\Phi}^\ast).
\]
The penalty terms $\alpha$, $\beta$ and $\omega$ ensure that the constraint violations feedback are accounted for during the solution of the system equations. The iterative process continues until
\[
\|\ddot{\bm{q}}^\ast_{i+1}-\ddot{\bm{q}}^\ast_{i}\|<\varepsilon.
\]
To compute the solution of the involved linear system we tested two methods: LU-factorization and the Moore-Penrose generalized inverse \cite{Neto 2003}. 

It is critical for the simulation performance to have an adequate methodology to describe the system propensity. As was described on Section \ref{propensity} the system propensity can be locally described using a regular set of projections $(\bm{T}_{i,j})_{i,j=1,\cdots, n}$, and the associated conformation library $(\bm{f}_{i,j}(\Omega))_{i,j=1,\cdots, n}$. The instances in this libraries are evaluated on a fixed discretization of the state-variable space $\Omega$. Assuming that the system degrees of freedom are joint angles, the space $\Omega=]-\pi,\pi]^n$ can be discretize using a fixed number of bins in each axis. For  an approximation defined by $m$ bins, a grid on $m\times n$ states are generates in $\Omega$, 
$$\hat{\Omega}=\{(\omega_1,\cdots,\omega_n):\omega_i=
2\pi\times j/m-\pi, j=1,\cdots,m,\;i=1,\cdots,n\},$$
each view $\bm{f}_{i,j}(\Omega)$ are approximate in the nodes of this $n$-dimensional grid. However to simplified the representation and reduce the size of required storage memory, here we only used orthogonal projection on the system referential defined by pair of joint angles. Each projection $\bm{T}_{i,j}$ is defined as projecting on the axis of the state-variable space defined by the angles on joints of order $i$ and $j$. In this cases $\bm{T}_{i,j}$ is represented by a diagonal matrix $A_{i,j}$, with only two ones in the diagonal, at positions $(i,i)$ and $(j,j)$. Making the computation of 
$(\sum_{i,j=1}^{N}\bm{T}_{i,j})^{-1}$ a trivial task, since in this case it is also diagonal and the storage of $\bm{f}_{i,j}(\hat{\Omega})$ can be done using a single matrix of type $m\times m$. Note that $M=\sum_{i,j=1}^{N}\bm{T}_{i,j}$, is a matrix of type $N\times N$ with the value $N$ in the diagonal. However it is not practical to define a family of projects $(\bm{T}_{i,j})_{i,j=1,\cdots, n}$, with one transformation for each pair of joint angles. For our case of study, nano mechanical systems, modelling polypeptide chain, those projection only are defined between pairs of angles in joints linking adjacent bodies, and are classified according to a correlation type. The potential for a model of a polypeptide chain is characterized by 20 types of views describing the correlation between the two dihedral angles in the same amino acid, and $6\times 20 \times 20$ views for pairs of dihedral angles in two adjacent amino acids.

\section{Application: Protein folding}\label{VI}

\noindent The method described in the least sections is here used to analyse the kinematic behaviour of a protein chain. For that two multibody  models are used. First the kinematic is analysed for an open chain multibody system, defined using the refinement proposed in Figure \ref{fig3}. Its results are compared against the same coarse-grained model, where the different bodies are linked using cylindrical joints with linear Gaussian propensity imposed by springs. For that a simplified representation for the polypeptide chain was used, it includes only backbone atoms $H$, $N$, $O$, $C$ and $C_\alpha$. This simplification reduces dramatically the model complexity, and the number of degrees of freedom. This choice is supported by the fact that the backbone local conformation change time scale is very higher that the side chain micro time scale motion. In each iteration we can assume that the side chain conformation is optimal and its shape are implicit described on the backbone dihedral angles local propensities. Similar simplified backbone models appear in literature, used for coarse-grained protein simulation \cite{Altis 2008}.

\subsection{ The protein coarse-grained model }
\noindent For this approach each body $b_i$, with an even index $i$ represents an atom $C_\alpha$ in the chain, while each body with an odd indexes $i$, represents a peptide plane in the peptide chain. We rigidly attached to each body $b_i$, with even index, four points $C^i$, $N_i$, $C^i_{\alpha l}$ and $C^i_{\alpha r}$, representing by $C^i_{\alpha l}$ the $C_\alpha$ atom at left-hand-side of the peptide plane, and by $C^i_{\alpha r}$ the atom $C_\alpha$ at its right-hand-side. Each body $b_i$, with even index, have associated two dihedral angles $(\phi_i,\psi_i)$, defined by the peptide planes $b_{i-1}$ and $b_{i+1}$. Bodies $b_{i-1}$, $b_{i}$ and $b_{i+1}$ are linked by two joints, defined between $N^{i-1}$ and $C^{i}_\alpha$, and between $C^{i}_\alpha$ and $C^{i+1}$. The axis of the first joint is described by a vectors, from $C^{i-1}$ to $C^i_\alpha$, rigidly attached to $C^{i-1}$ and $C^i_\alpha$. Similarly the second joint is described by a vectors, from $C^i_\alpha$ to $N^{i+1}$, rigidly attached to $C^i_\alpha$ and $N^{i+1}$. When the model is described by cylindrical joints springs are used to restrict the linear motion between $N^{i+1}$,  $C^i_\alpha$ and $N^{i+1}$. Table \ref{Stiffness} presents values for stiffness of this springs and its equilibrium values.

Denoting  by $\bm{r}_j^C$, $\bm{r}_j^N$, $\bm{r}_j^{C_\alpha^l}$ and $\bm{r}_j^{C_\alpha^r}$, the atoms global coordinates. The dihedral angles cosine $(cos(\phi_i),cos(\psi_i))$, for body $b_i$, with even index, is given by:
\[
cos(\phi_i)=\frac{\bm{R}_i^T\bm{S}_i}{\|\bm{R}_i\|\|\bm{S}_i\|}, \text{ and } cos(\psi_i)=\frac{\bm{R}_i^{\prime T}\bm{S}^{\prime}_i}{\|\bm{R}^{\prime}_i\|\|\bm{S}^{\prime}_i\|},
\]
here
$\bm{R}_i=\tilde{\bm{r}}^{C^{i-1}N^{i-1}}\bm{r}^{C^{i}_{\alpha l}C^{i+1}}
$,
$\bm{S}_i=\tilde{\bm{r}}^{C^{i}_{\alpha l}C^{i+1}}\bm{r}^{C^{i}_{\alpha l}C^{i+1}}
$,
$\bm{R}^{\prime}_i=\tilde{\bm{r}}^{N^{i-1}C^{i}_{\alpha r}}\bm{r}^{C^{i}_{\alpha l}C^{i+1}}$
and
$\bm{S}^{\prime}_i=\tilde{\bm{r}}^{C^{i}_{\alpha l}C^{i+1}}
\bm{r}^{C^{i+1}N^{i+1}}.$  
Note that, $\dot{\phi_i}=\bm{r}^{N^{i-1}C^{i}_{\alpha r}\prime}\omega^\prime_i$
and $\dot{\psi_i}=\bm{r}^{C^{i}_{\alpha l}\prime}\omega^\prime_i$.
The peptide plane remains rigid during system dynamics. As a result, the $\phi_i$ and $\psi_i$ dihedral angles on the chain are the essential degrees of freedom that dictate the position of the polypeptide backbone atoms $C_\alpha$, $C$ and $N$. However its dynamic is restricted by the chain strong local propensity and by the Brownian nature of the conformation.

\begin{figure}[h]
\begin{center}
\begin{tabular}{|c|c|}
\hline
Stiffness Constants (kcal/mole \r{A}$^2$) & Equilibrium Value \\
\hline
$k_{N-C_\alpha}=370$  & $b_{1}=1.490$\r{A}\\ 
$k_{C_\alpha-C}=320$  & $b_{2}=1.430$\r{A}\\
\hline
\end{tabular}
\end{center}
\caption{Stiffness Constants and equilibrium values used in the CHARMM potencial field.}\label{Stiffness}
\end{figure}
      
\subsection{Numeric test in the protein 1UAO}
\begin{figure}[H]
\begin{tabular}{cc}
\includegraphics[width=220pt]{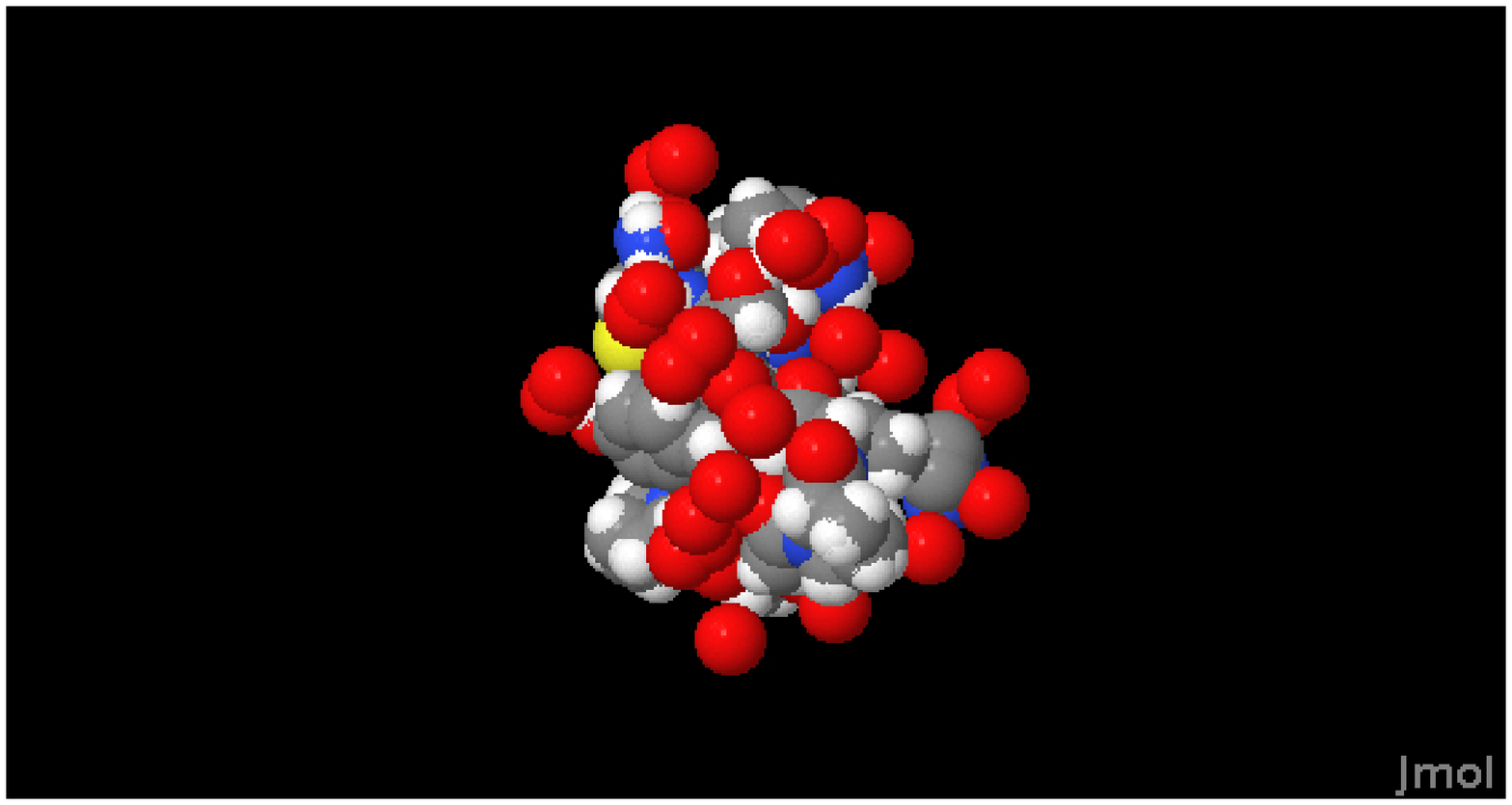} &
\includegraphics[width=220pt]{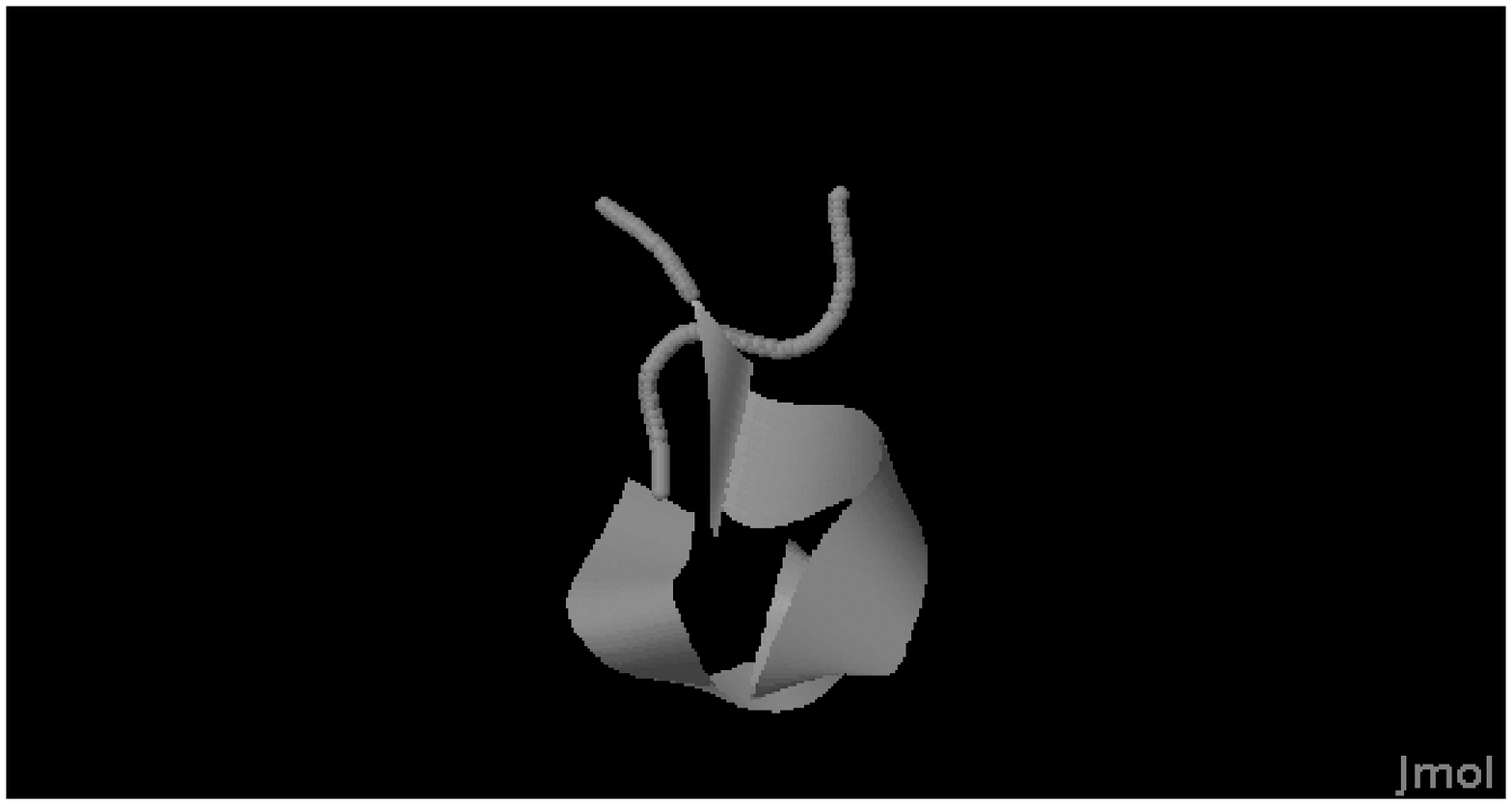}\\
\includegraphics[width=220pt]{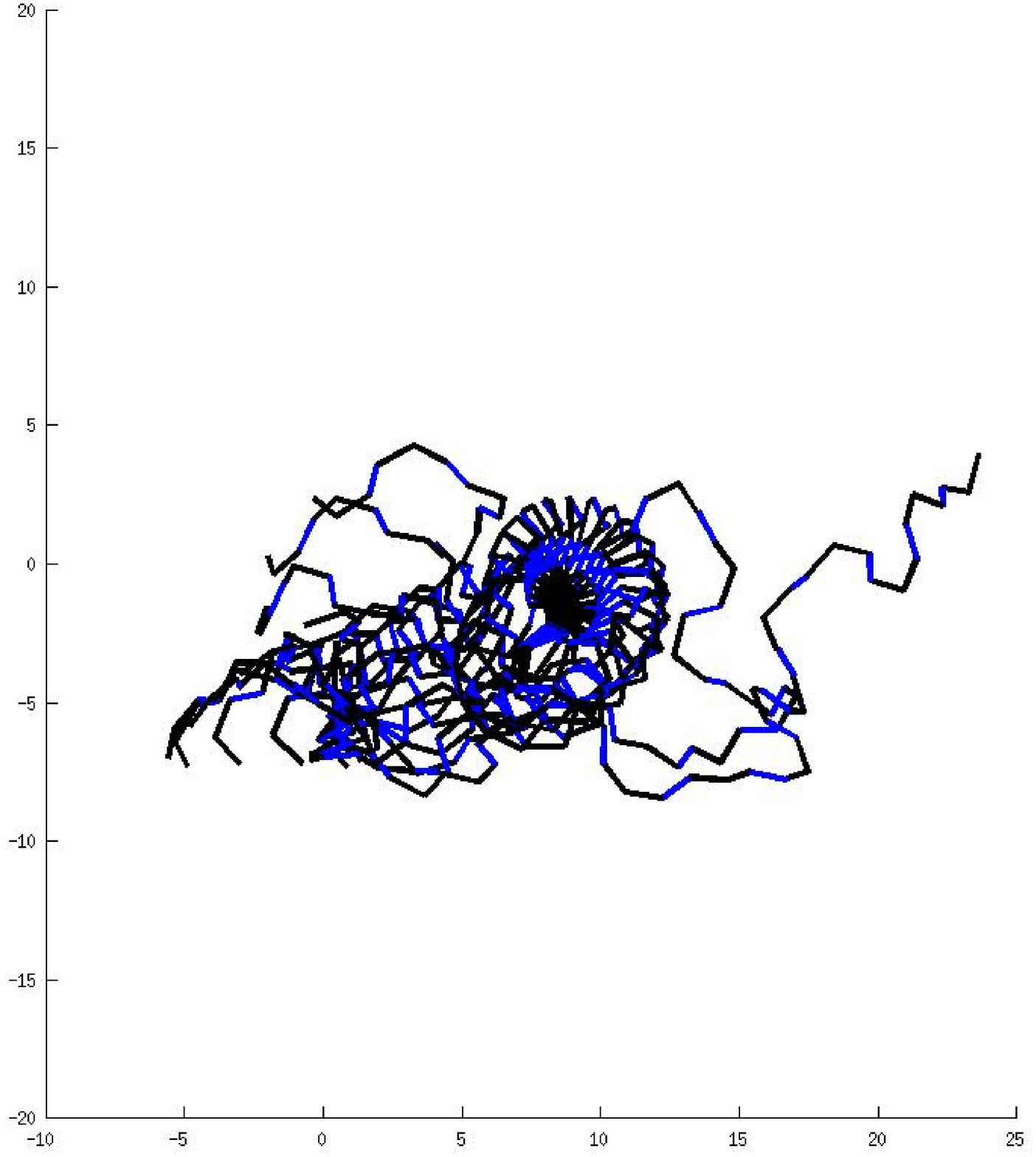}&\\
\end{tabular}
\caption{ Atomic structure and a cartoon representation for protein 1HJE. Initial band conformation.}\label{chainstable}
\end{figure}

\noindent The presented methodology is tested in this section for a simple protein chain. For that we selected the protein 1UAO Chignolin from the PDB. Chignolin is an artificial single chain protein defined by 10 amino acids, arranged in the sequence
\[
GLY-TYR-ASP-PRO-GLU-THR-GLY-THR-GLY,
\]
cooperatively fold into a $\beta$-hairpin structure in water.
Figure \ref{chainstable} presents the atomic structure and a cartoon representation to its stable conformation. Its initial coarse-grained model is defined by 50 bodies and 49 kinematic joints.

A library of van Mises propensity distribution and directional gradients was compile, for each involved amino acid and for pairs of two consecutive amino acids. Note that, the method is very sensitive to the energy landscape irregularities and on the method used to compute its directional gradients. And moreover the density estimator impose a strong bias to the conformation change dynamics. Here the propensity potential can be assumed unrealistic in the bimolecular perspective. The need of a smooth surface using the described  kernel parametrisation leaded to a loss of detail in the potential.

The protein coarse-grained was extend with 20 replicas and an elastic band. This new multibody system have 1000 bodies and 980 joints, coupled with an elastic band defined by 15200 springs. Only the gradients of singles amino acid were used to generate the conformation forces in the elastic band. The gradients for pair of adjacent amino acid are subtracted directly to the involved bodies moments. This allow to drive the conformation to a optimum in the spirit of the gradient descent method, without perturbing the forces generated by the NEB method for each amino acid.

For the DIM we selected a Runge-kutta integrator with order 2-3 (the ode23 MatLab implementation). The Baumargarte's and the Augmented  Lagrangian Formulation stabilization methods where tested here. To solve involved linear systems we tested LU-factorization (using the MatLab backslash implementation) and the Generalized inverse. As stopping criteria the DIM number of iterations was used.

The first approximation to MEP was generated by a convex combination, between the protein native and random conformation. The tests were perform for the same band initialization, presented on Figure \ref{chainstable}, the band configuration was relaxed using the DIM.

\begin{figure}[H]
\begin{tabular}{cc}
\includegraphics[width=220pt]{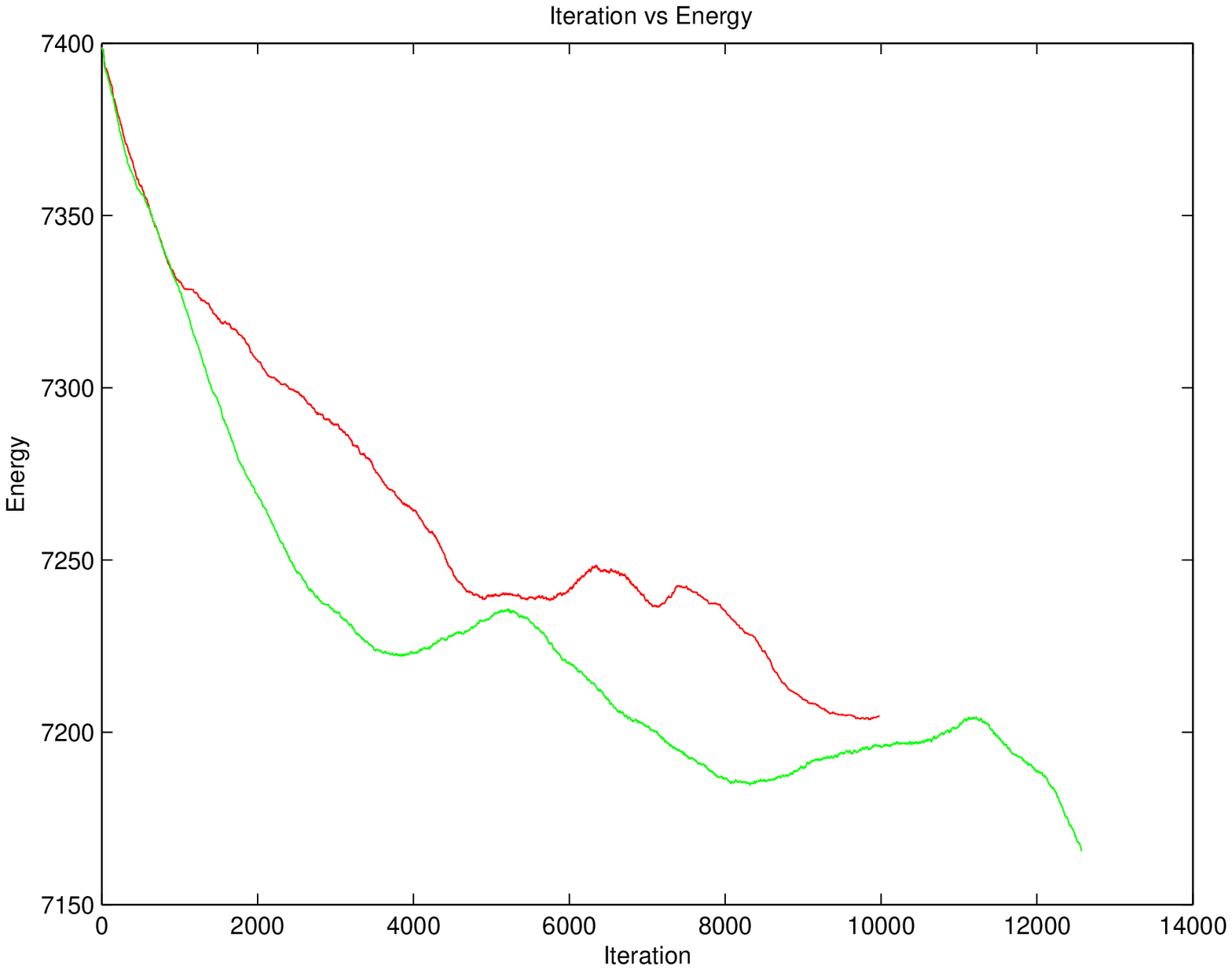} &
\includegraphics[width=220pt]{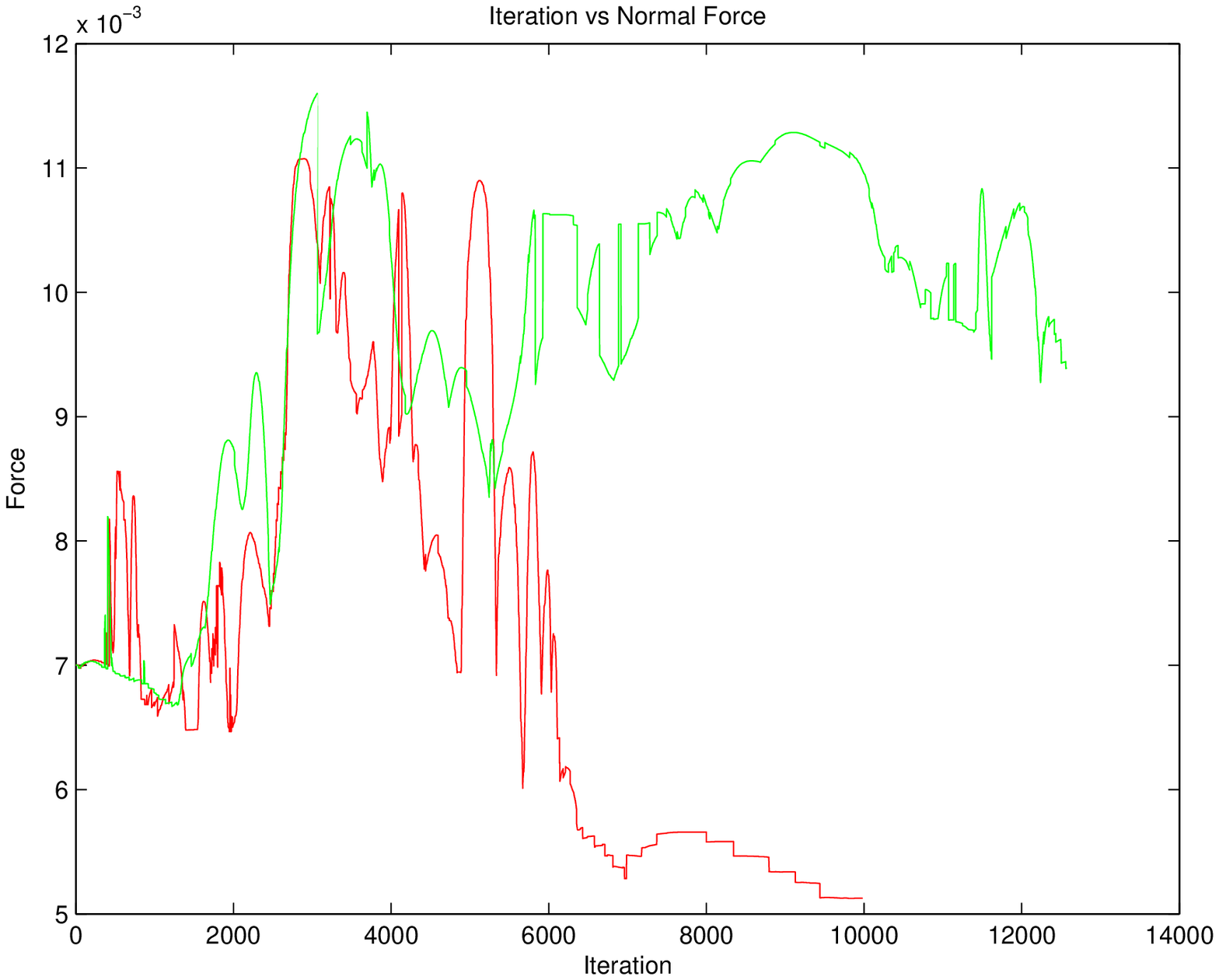}\\
\includegraphics[width=220pt]{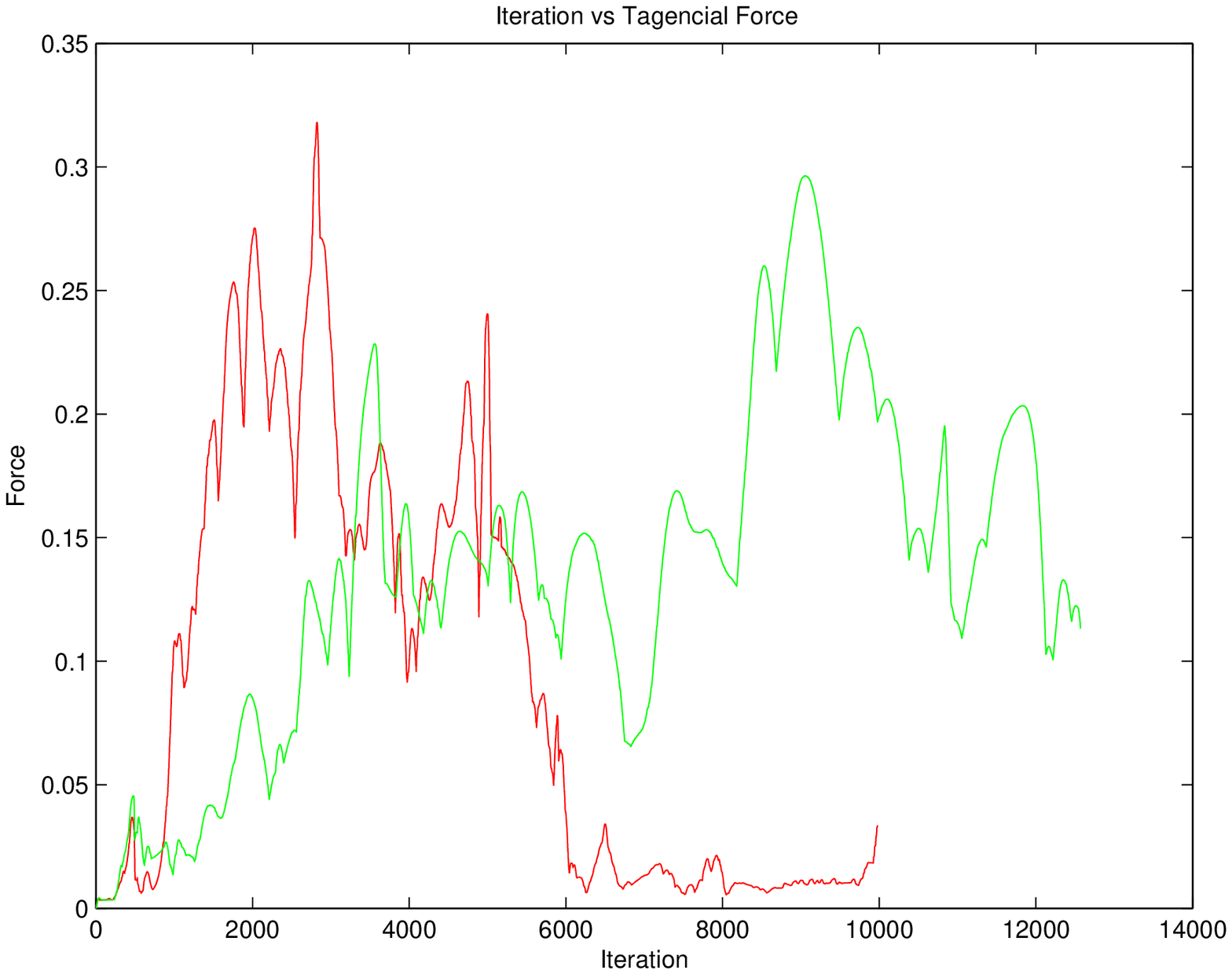} &
\includegraphics[width=220pt]{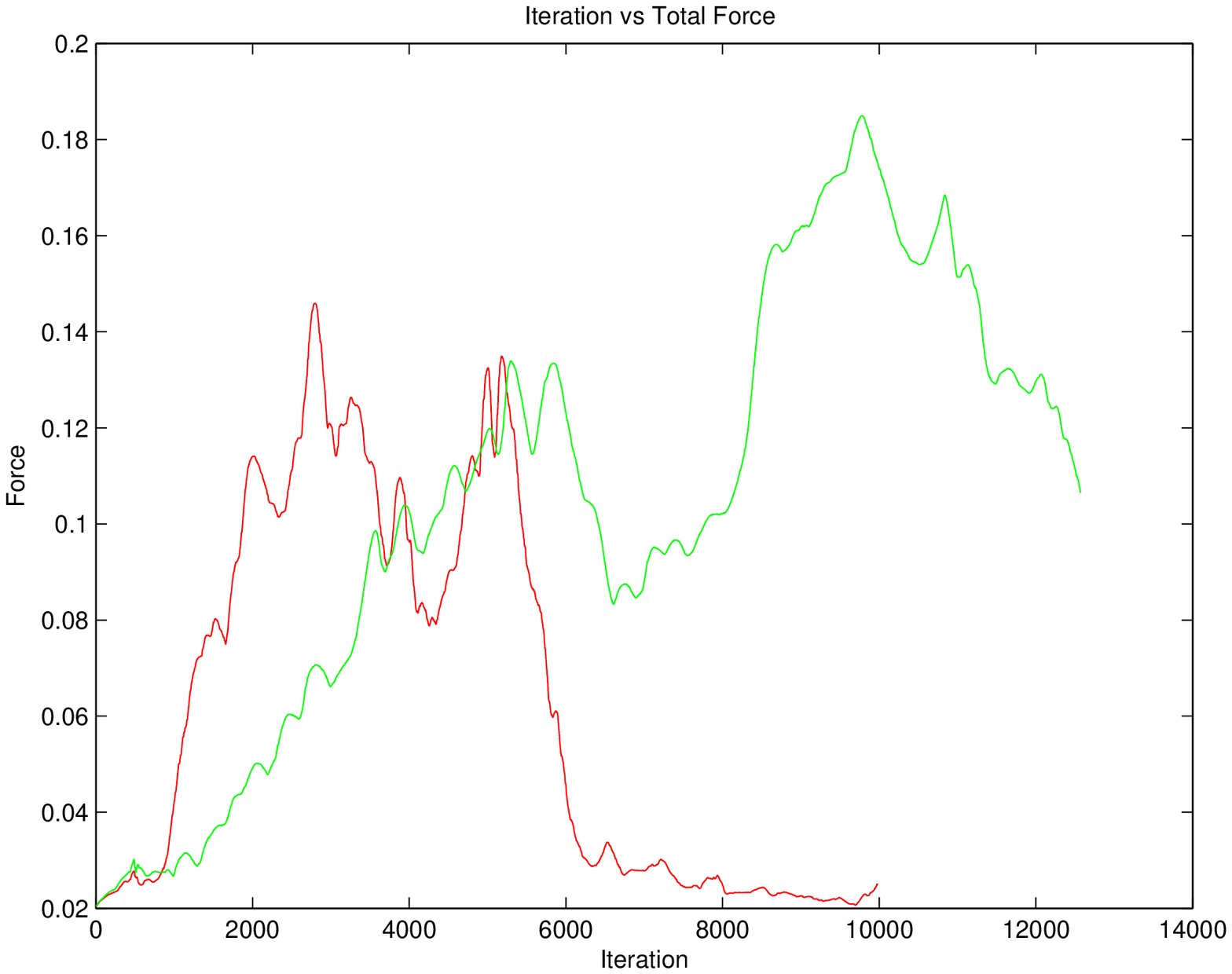}\\
\end{tabular}
\caption{ Baumargarte's  Method and Generalized inverse: Atomic structure and a cartoon representation for protein 1HJE (Green model with revolute joints, red model with cylindrical joints).}\label{MGI1}
\end{figure}

\begin{figure}[H]
\begin{tabular}{cc}
\includegraphics[width=220pt]{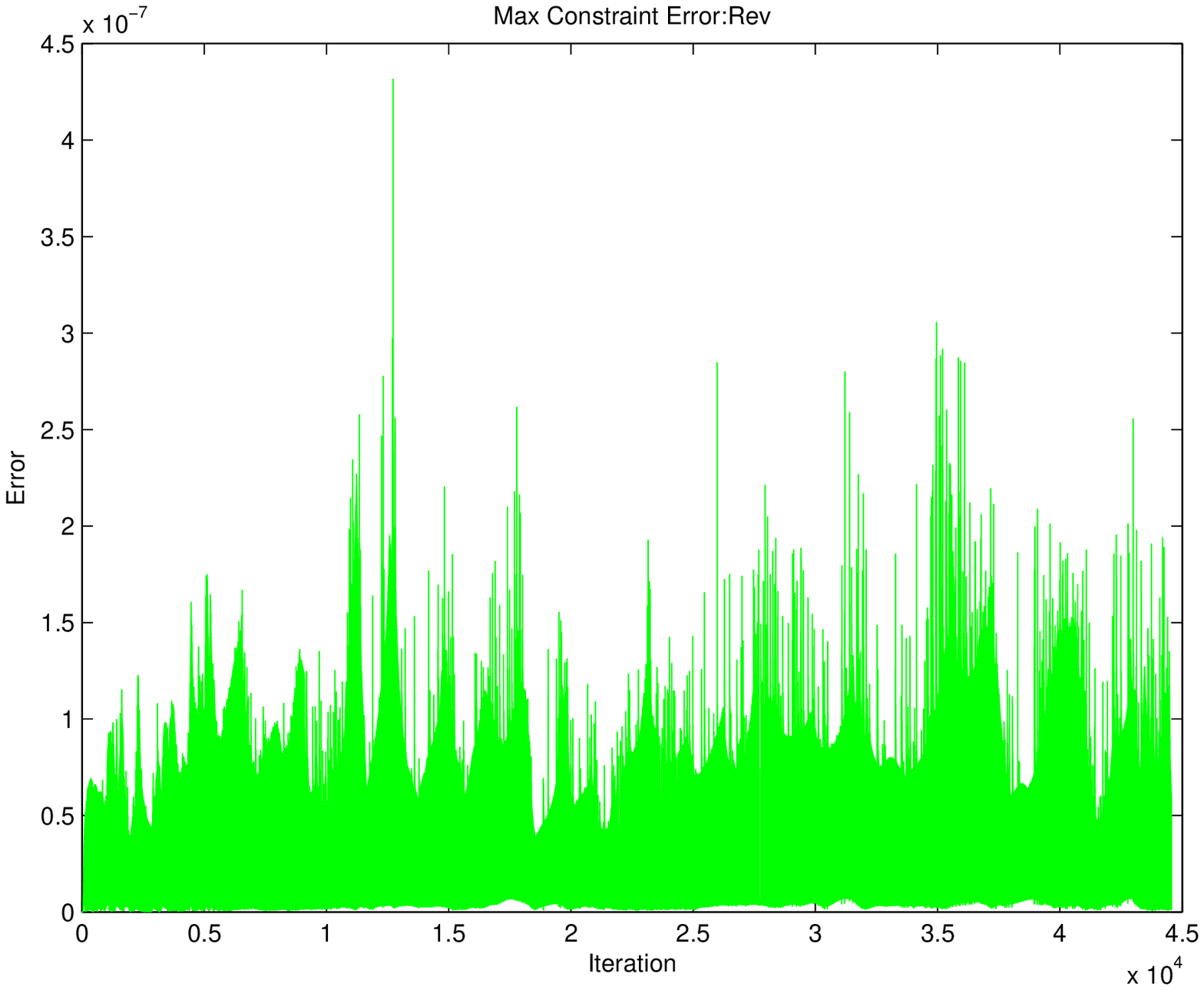} &
\includegraphics[width=220pt]{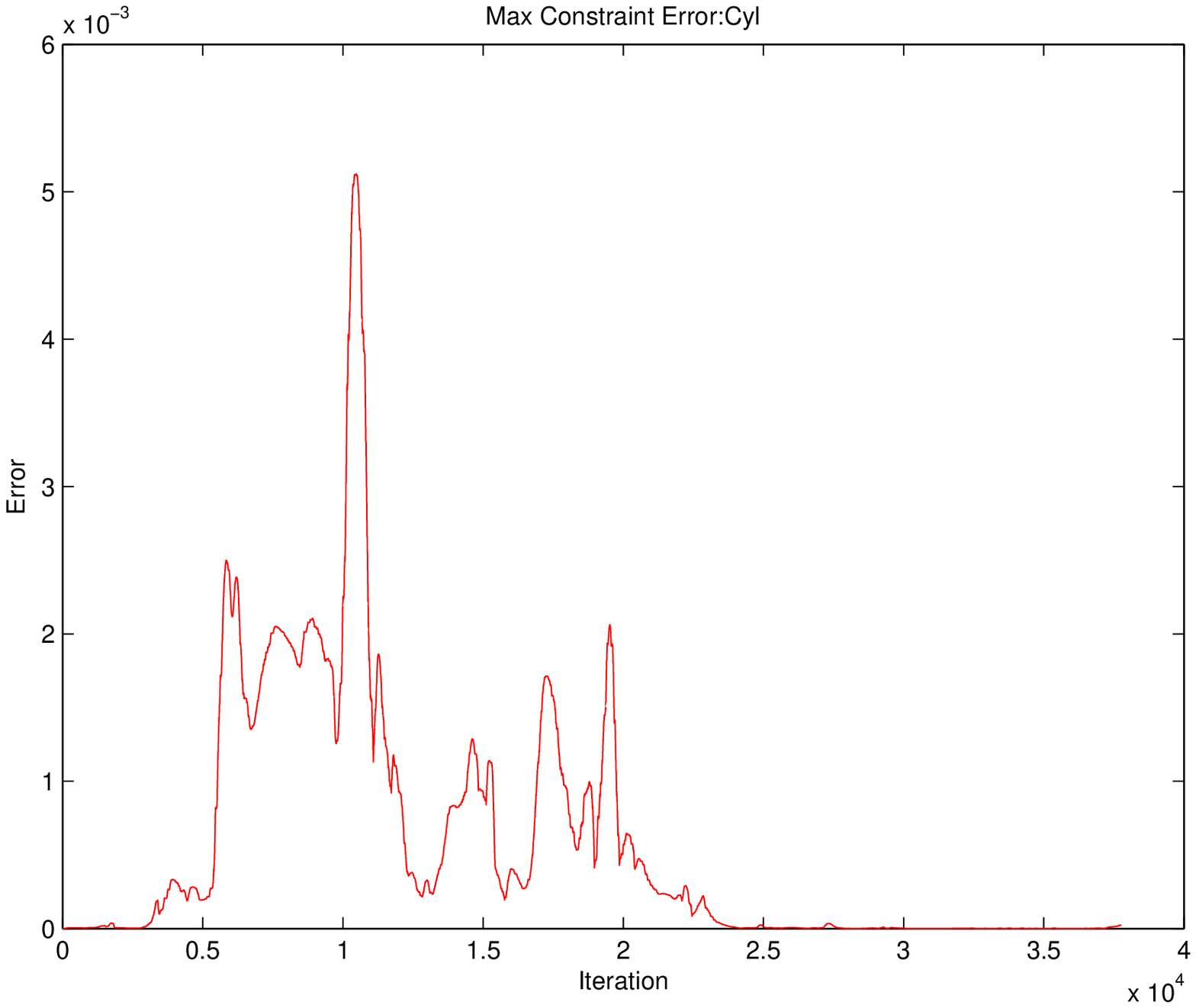}\\
\end{tabular}
\caption{ Baumargarte's  Method and Generalized inverse: The maximum error on the dynamic constraints by iteration (Green model with revolute joints, red model with cylindrical joints).}\label{MGI2}
\end{figure}

For the first test between models defined using revolute joints and cylindrical, we used the Baumargarte Stabilization Method. For that the linear systems were solved using Generalized inverse. The tests performed using Baumargarte's  Method and LU-factorization don't converged. The results are presented on Figure \ref{MGI1}, where the energy decrement between iterations can be compared. The model defined using revolute joints (green) outperform the model using cylindrical joints (red). In the same figure we can compare the maximum amplitude of the normal force, tangent force and the total force applied to change the dihedral angles in each iteration. Here we can note that the tangential and normal forces have grater intensities for revolute joints, representing a faster path reparametrization and a faster convergence to the optimal solution. The error on the dynamic constraints are presented on Figure \ref{MGI2}.  For the parameters used on the stabilization method, $\alpha=\beta=5$, have better impact on the error propagation control for revolute joints.

\begin{figure}[H]
\begin{tabular}{cc}
\includegraphics[width=220pt]{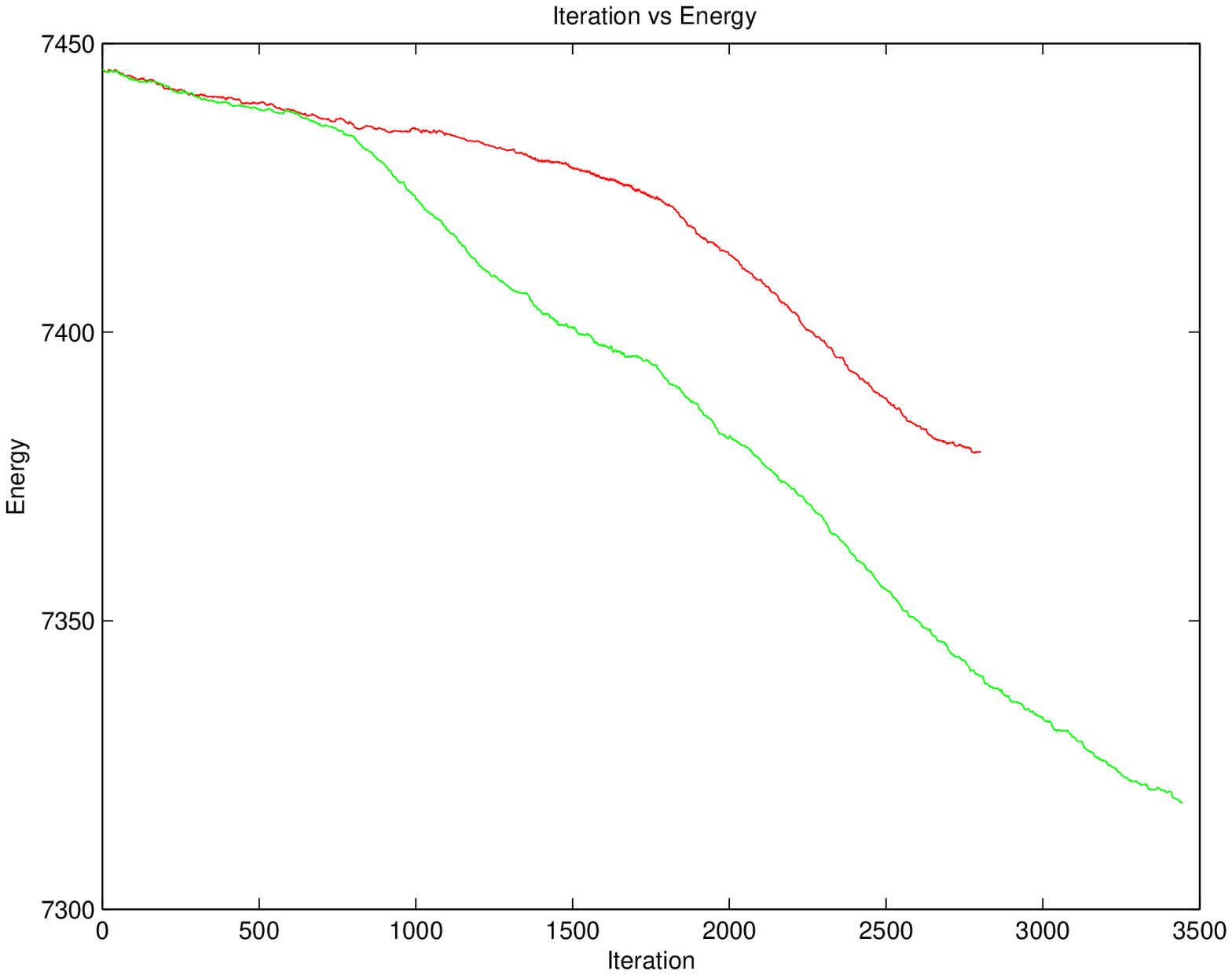} &
\includegraphics[width=220pt]{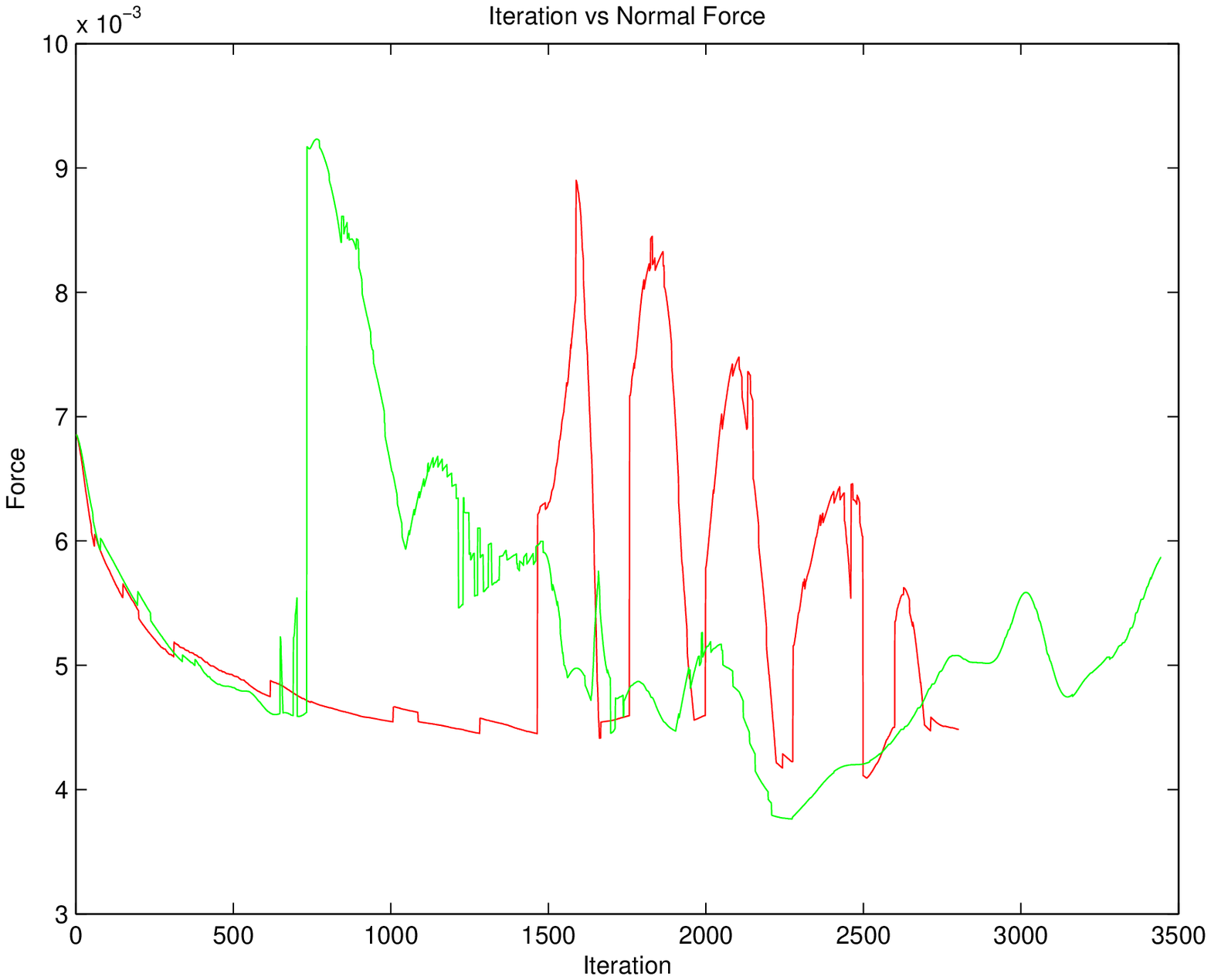}\\
\includegraphics[width=220pt]{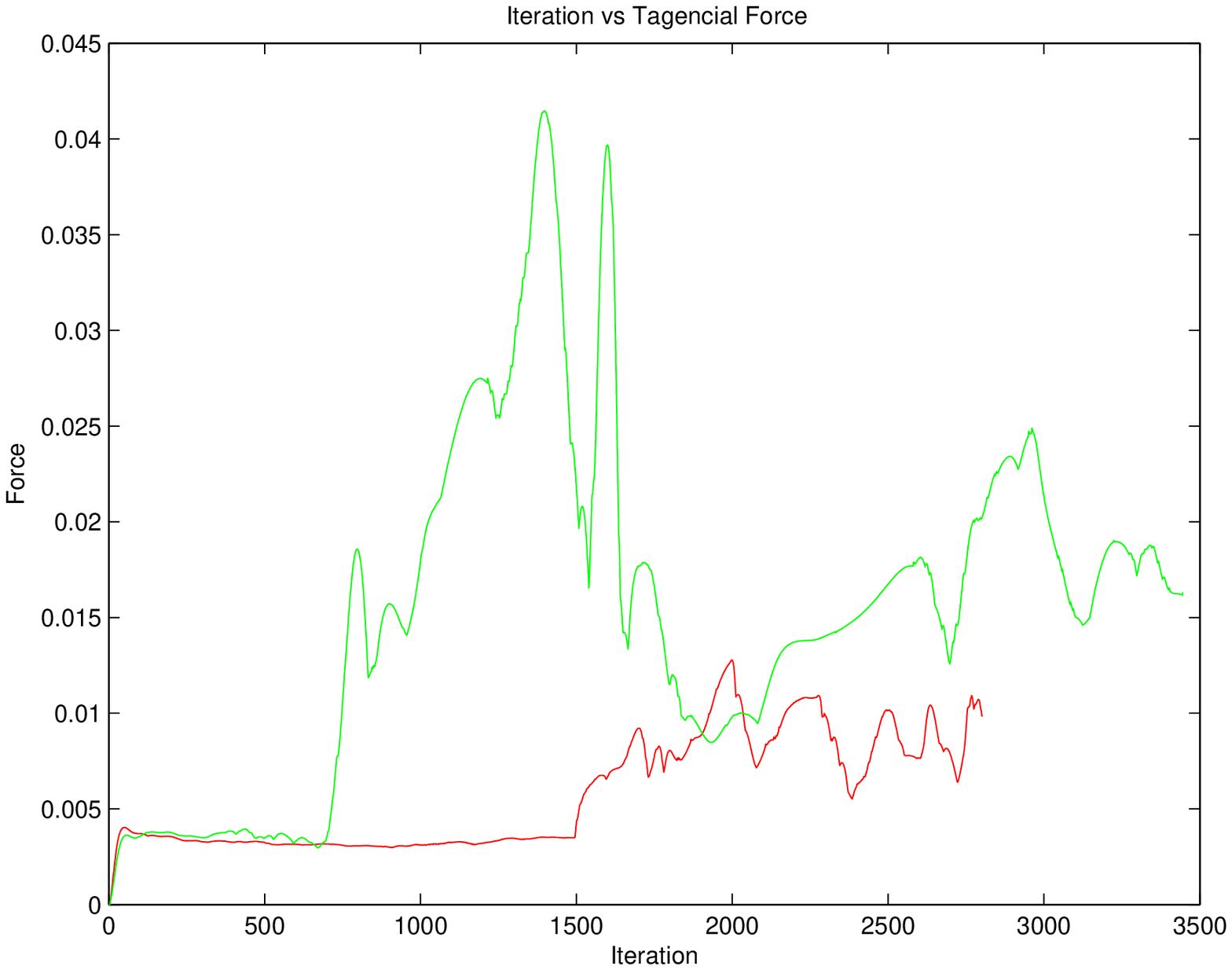} &
\includegraphics[width=220pt]{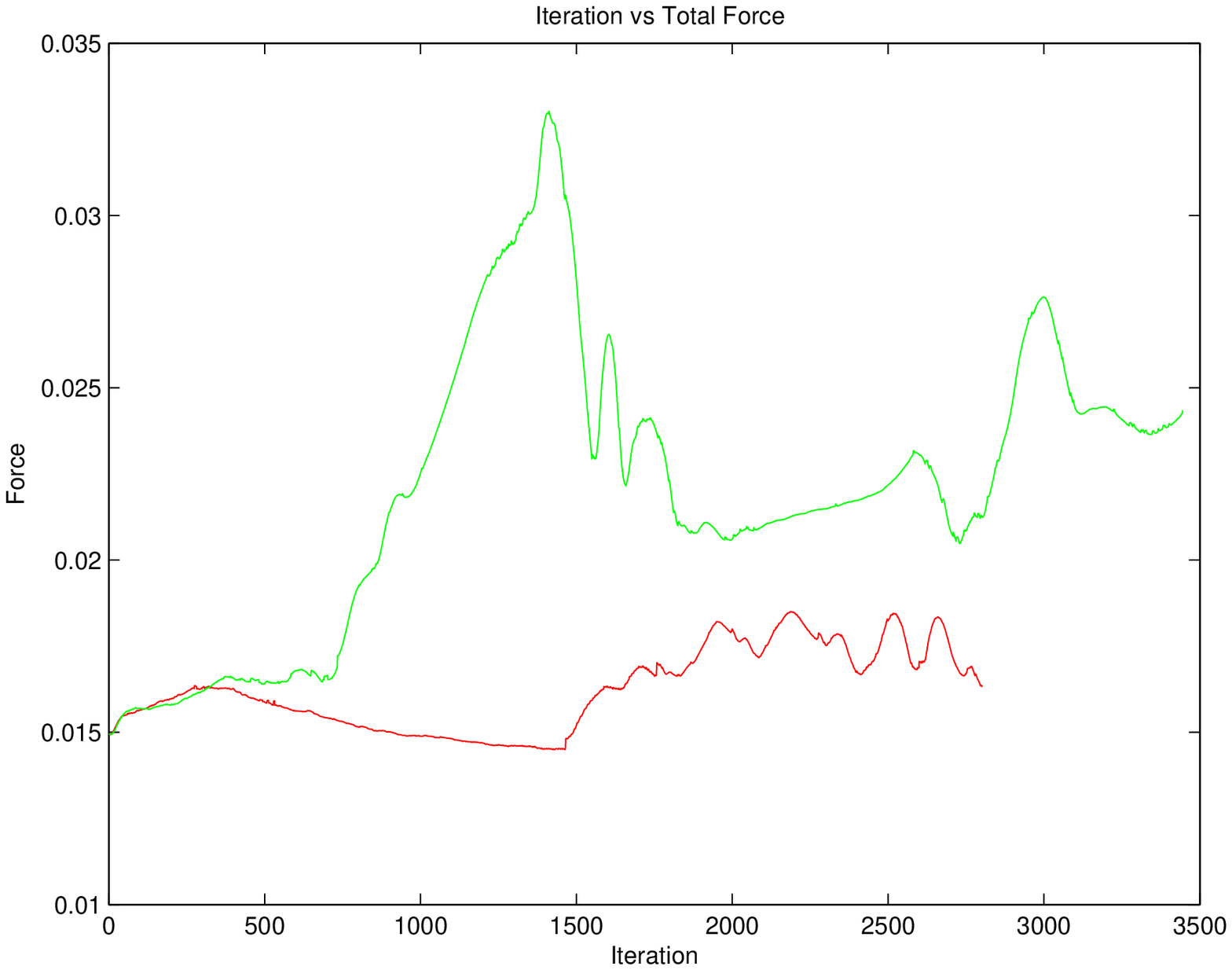}\\
\end{tabular}
\caption{ Baumargarte's  Method and Generalized inverse: Atomic structure and a cartoon representation for protein 1HJE (Green model with revolute joints, red model with cylindrical joints).}\label{ALF1}
\end{figure}

\begin{figure}[H]
\begin{tabular}{cc}
\includegraphics[width=220pt]{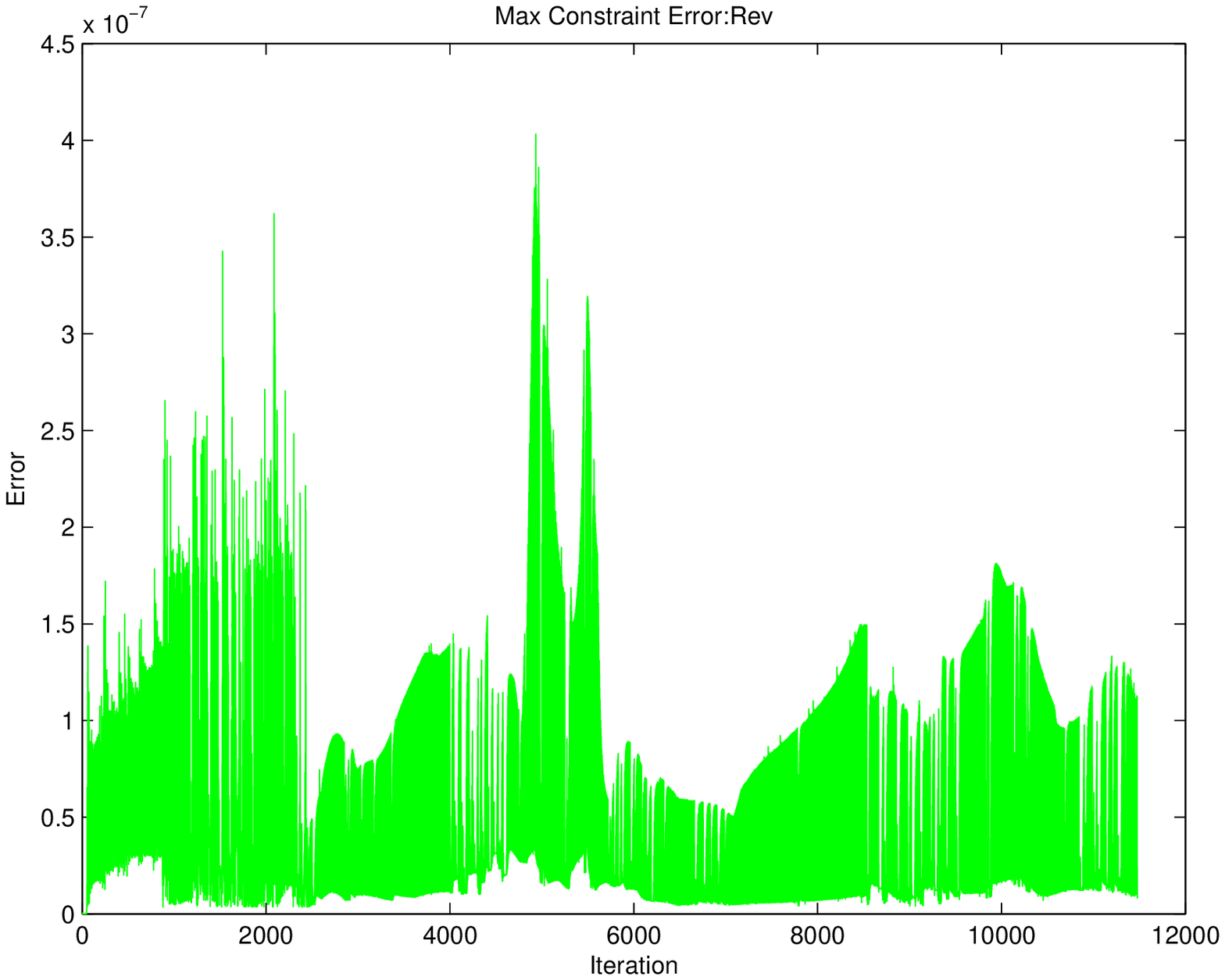} &
\includegraphics[width=220pt]{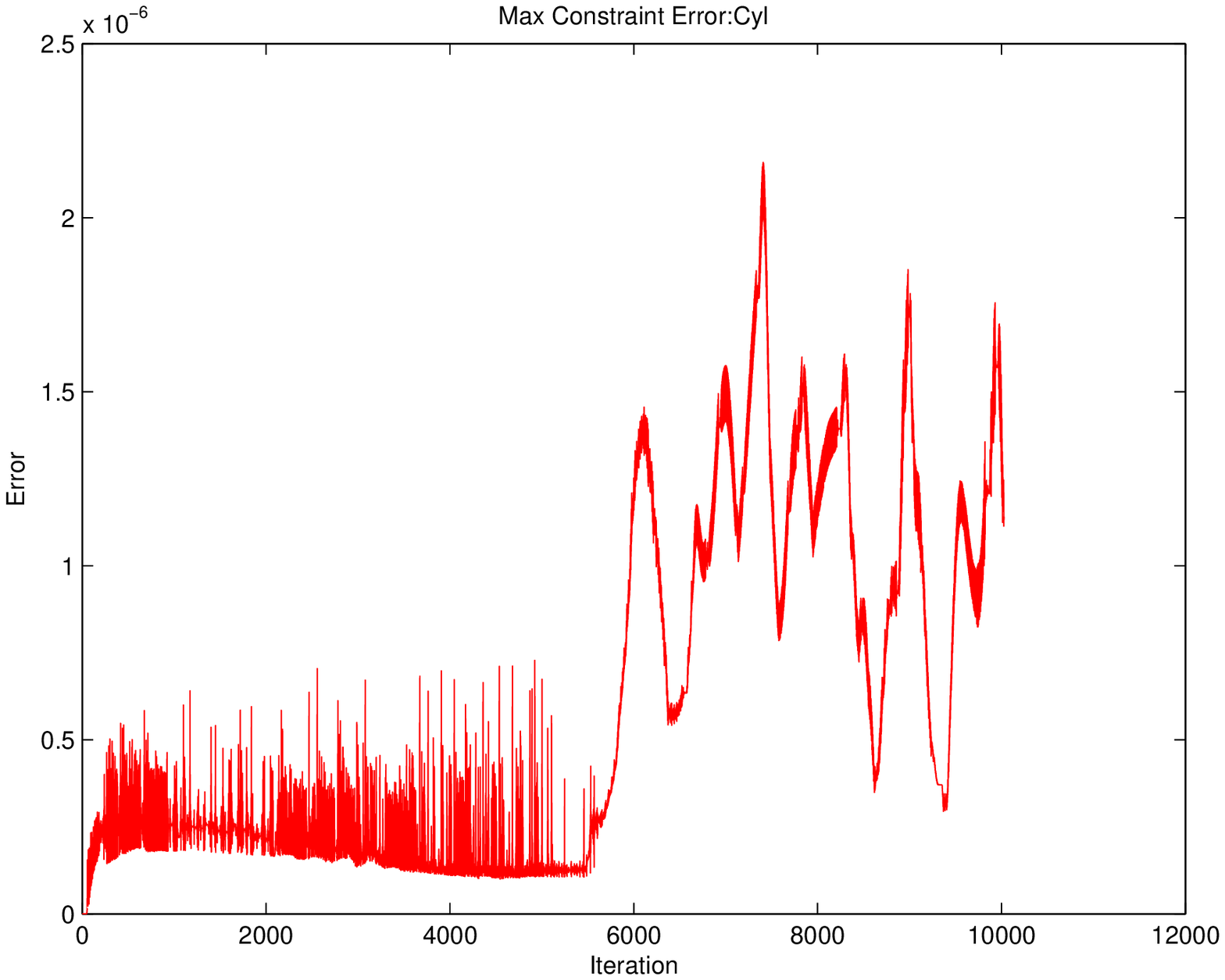} \\
\end{tabular}
\caption{Augmented  Lagrangian Formulation and LU-factorization: The maximum error on the dynamic constraints by iteration (Green model with revolute joints, red model with cylindrical joints).}\label{ALF2}
\end{figure}

\begin{figure}[H]
\begin{tabular}{cc}
\includegraphics[width=220pt]{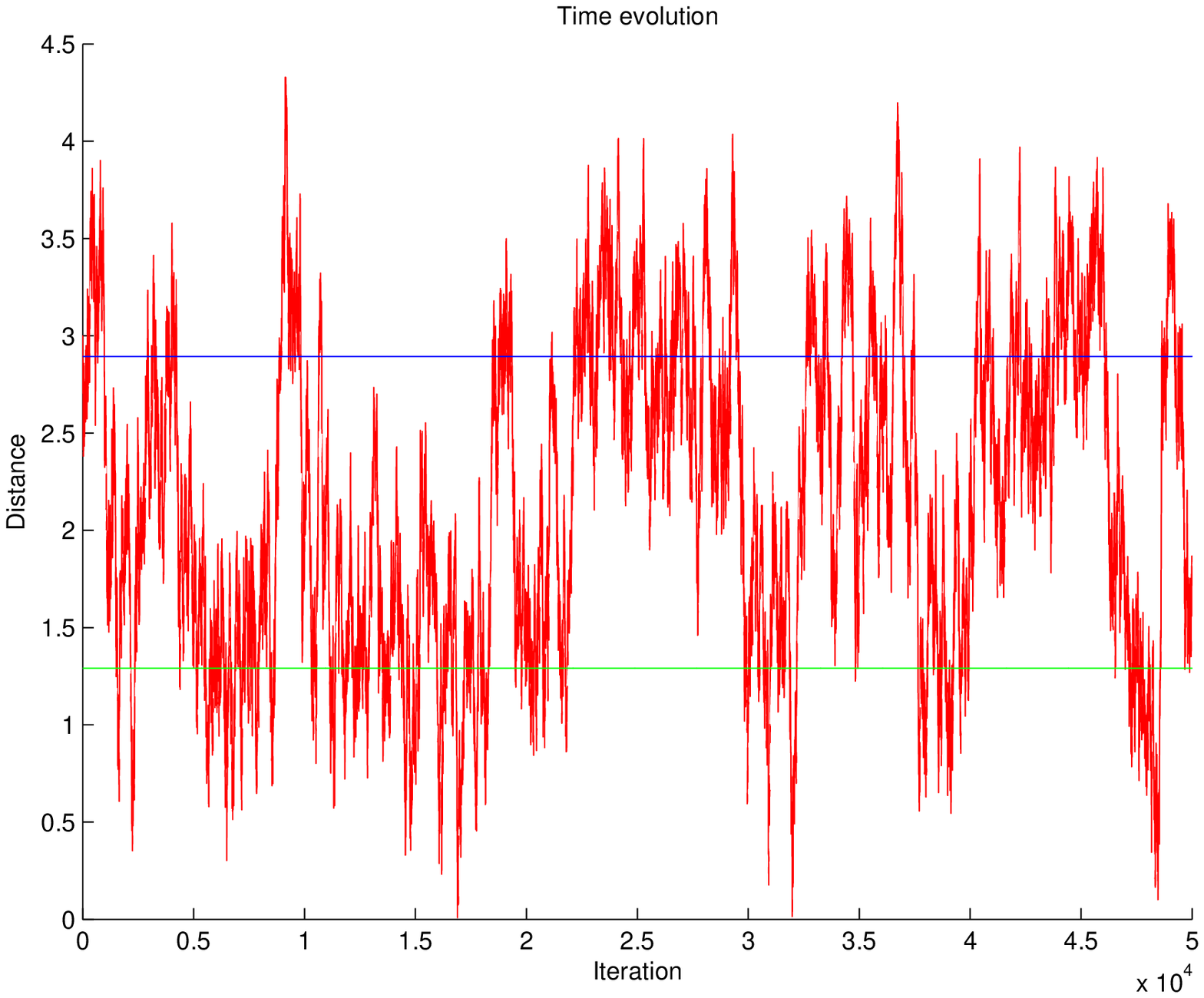} &
\includegraphics[width=220pt]{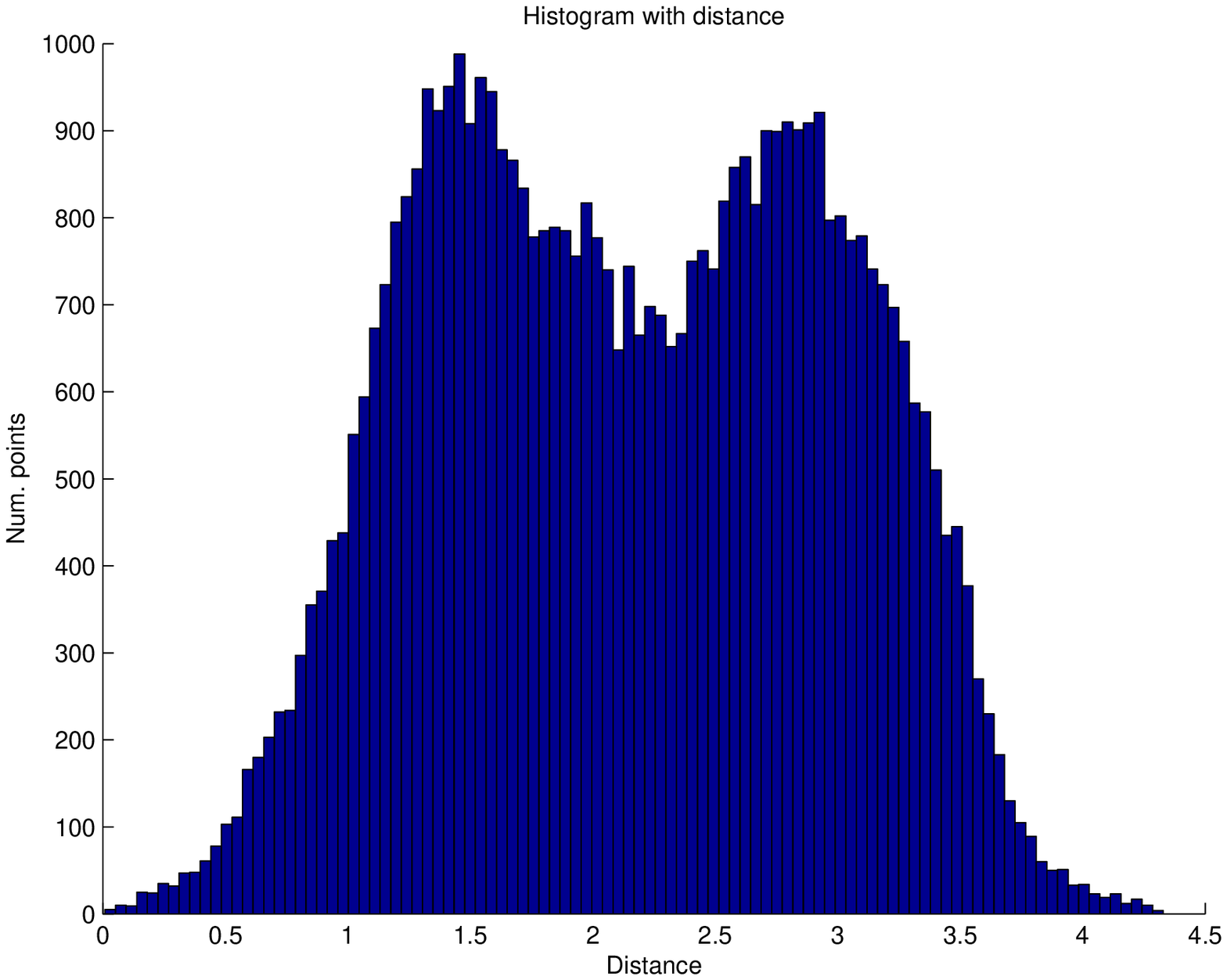}\\
\includegraphics[width=220pt]{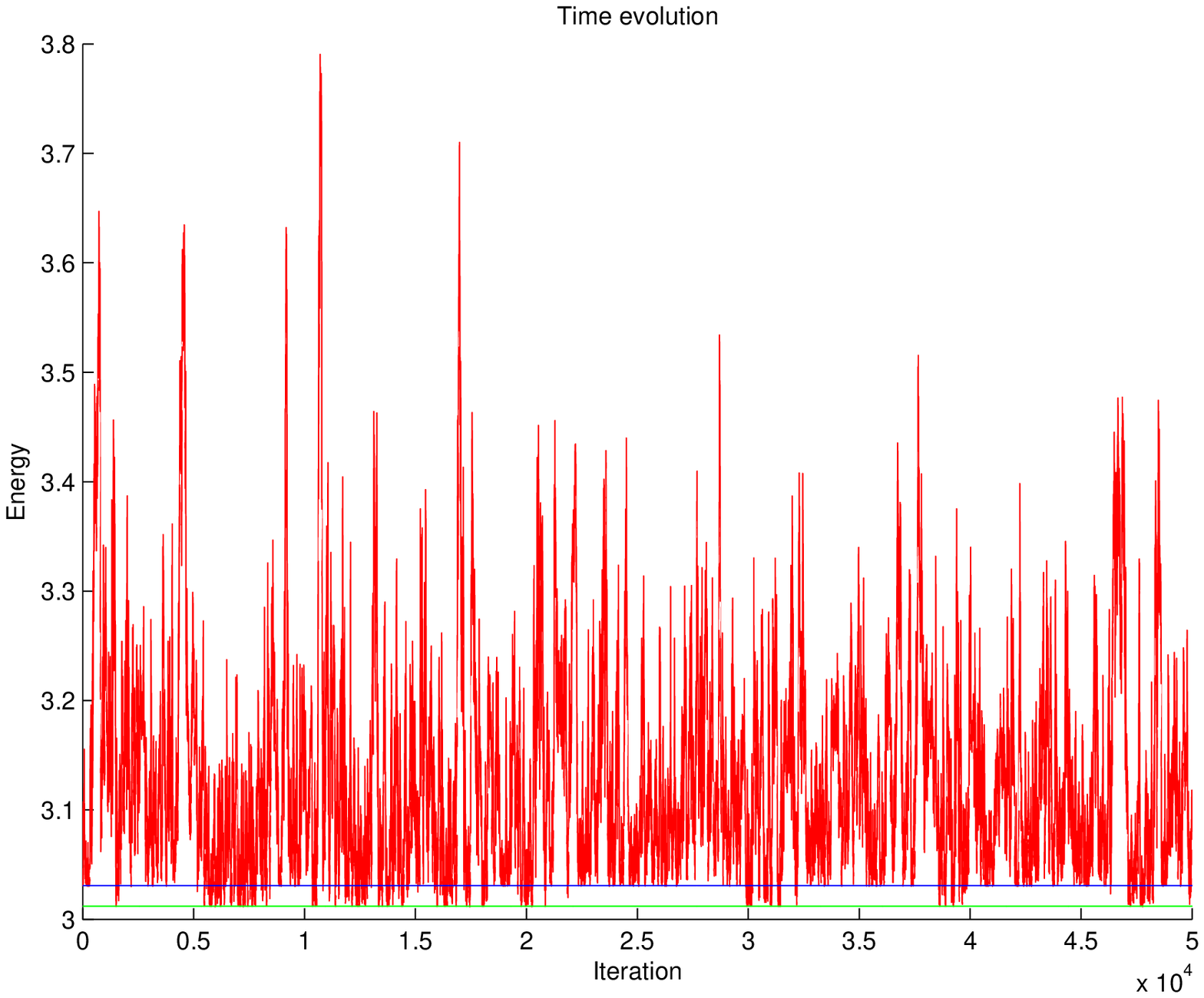} &
\includegraphics[width=220pt]{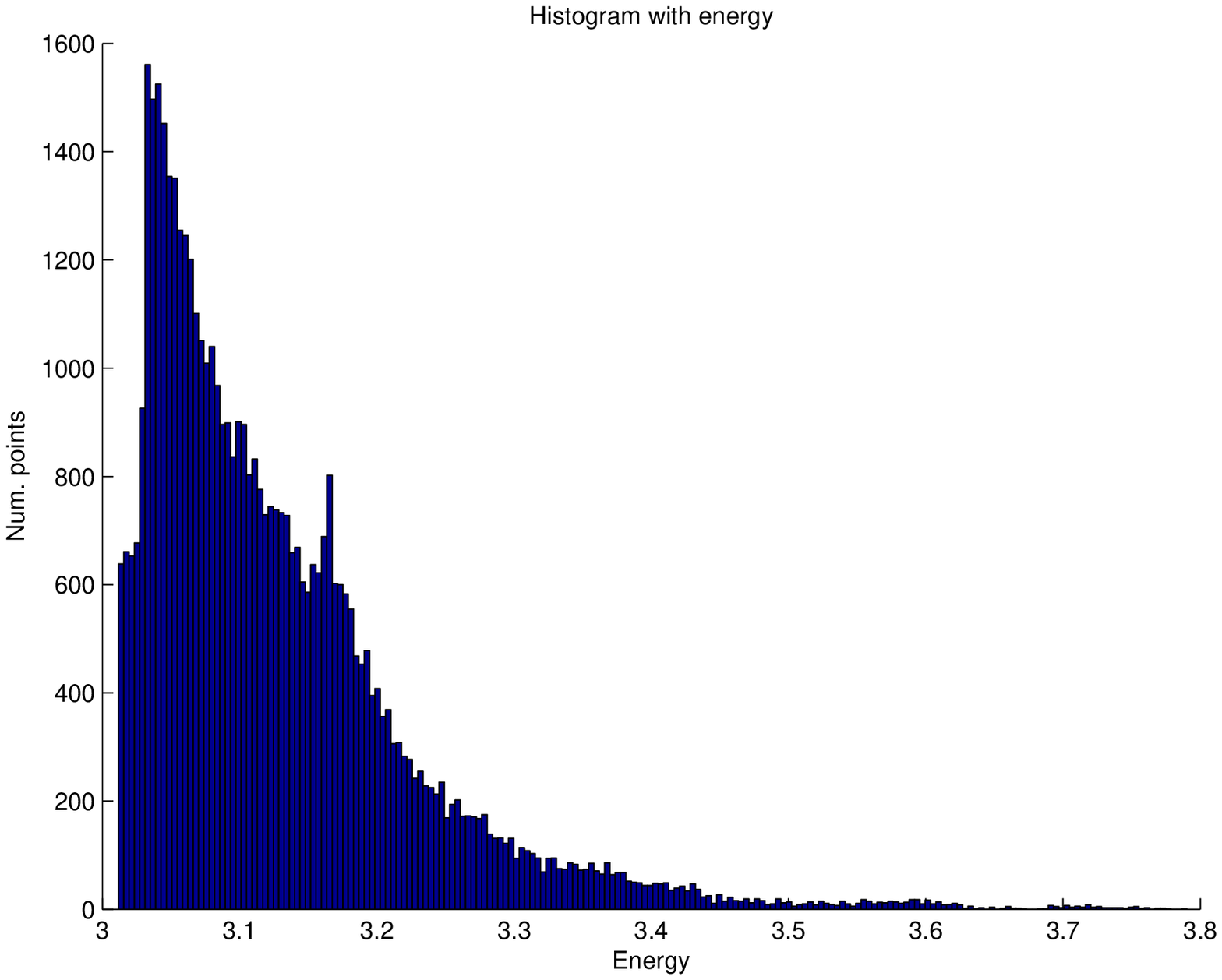}\\
\end{tabular}
\caption{ Brownian motion. Top: time of the reaction coordinate for a system with a free energy defined, by the amino acid PHE van Mises propensity potential, as shown in Fig. \ref{fig10}. Most of the time the reaction coordinate fluctuate near the stable states $A$ and $B$. Bottom: around the energy values $q_A$ and $q_B$ typical for stable states $A$ and $B$, respectively. Rarely, on the timescale of the stable stable fluctuations, the system switches between $A$ and $B$.}\label{fig11-4}
\end{figure}

The second test were performed using the Augmented  Lagrangian Formulation as stabilization methods and using to solve involved linear systems LU-factorization. The energy decrement between iterations are presented on Figure \ref{ALF1}. The model defined using revolute joints (green) outperform the model using cylindrical joints (red). Also in this case the tangential and normal forces have grater intensities for revolute joints, representing a faster path reparametrization and a faster convergence to the solution. The Augmented  Lagrangian Formulation also have better performance on controlling error propagation for revolute joints (Figure \ref{ALF2}).

\section{Recycle}
The conformations are in this work restricted to probability distributions for pairs of angles on joins linking together three consecutive bodies.  For that we must assumed that each pair of joins have an assigned classification and that joins with the same classification have identical propensities. To analyse the kinematic of this type of systems it is sufficient to have a library of probability distributions and/or directional gradients, however its resolution is of fundamental importance to generate the moments needed to achieve stable conformational trajectories, with minimal entropy. 

Assuming that $\{P_{aa}\}_{aa}$ is a library of probability distributions $P_{aa}$ for a  body of type $aa$, such that if $j$ is a body of type $aa$, then $P_{aa}(\phi_{i,j},\phi_{j,k})$ is the probability of the angles for the joint between  bodies $i$ and $j$, and  $j$ and $k$ be, respectively,  $\phi_{i,j}$ and $\phi_{j,k}$, its contribution to the moment to be applied to body $i$, to impose a local propensity describe by distributions $P_{aa}$, is
\begin{equation}
\bm{n}_i=\nu_{i,j}\tilde{\bm{w}_i}\bm{l}_{i,j}/\|\bm{l}_{i,j}\|,
\end{equation}
for the bodies $j$ and $k$ are, respectively
\begin{equation}
 \bm{n}_j=-\nu_{i,j}\tilde{\bm{w}_j}\bm{l}_{i,j}/\|\bm{l}_{i,j}\|+\nu_{j,k}\tilde{\bm{w}_j}\bm{l}_{j,k}/\|\bm{l}_{j,k}\| 
 \end{equation}
and 
\begin{equation}
  \bm{n}_k=-\nu_{j,k}\tilde{\bm{w}_k}\bm{l}_{j,k}/\|\bm{l}_{j,k}\|,
\end{equation} 
in both cases scalars $\nu_{i,j}$ and $\nu_{j,k}$ are defined by the components extracted from the gradient of free energy for the joints propensity space, given by the Boltzmann inverse 
\[
 [\nu_{i,j},\nu_{j,k}]^T=k_bT\nabla(\log(P_{aa}(\phi_{i,j},\phi_{j,k}))
\] 
where $\bm{w}_i$, $\bm{w}_j$ and $\bm{w}_k$ are the bodies rotational coordinates vectors, $\bm{l}_{i,j}$ and $\bm{l}_{j,k}$ are the axis of each joint. The sum moment of all contributions applied to a body $i$ will be denotes by $\bm{n}_i^\ast$, and is defined by the sum of all contribution to the body moments defined by correlation on the dihedral angles of joints where the body is involved. This directional gradient points in the direction of the greatest rate of decrease in the system entropy.

\section{CONCLUSIONS}\label{VIII}

The coarse-grained model for proteins using revolute joints outperforms the model defined by cylindrical joints. However in both case the stopping criteria must be revised. The available  computational power only allowed simulation during short periods of time. The use of gradients of neighbour amino acid conformation directly in the elastic band must be considered, to expedite the convergence to an optimal path. Tests performed using CUDA implementation of algorithms to solve linear systems using generalized inverse  suggest that, in conjunction the Baumargarte Stabilization  Method, can be the best strategy to improve its computational time.

Here we also should note the importance of have trajectories generated using the traditional Monte Carlo Method to evaluate the quality of the optimal path find by the presented methodology.

\end{document}